\documentclass[journal]{IEEEtran}
\ifCLASSINFOpdf
\else
\fi

\usepackage{enumitem}
\setlist[enumerate]{listparindent=\parindent}
\usepackage{CJK}
\usepackage{indentfirst}
\usepackage{amsmath}

\usepackage{cases}
\usepackage{graphicx}
\usepackage{subfigure}
\usepackage{algorithm}
\usepackage{algpseudocode}
\usepackage{amsmath}
\usepackage{tabularx}
\usepackage{float}
\usepackage{color}
\usepackage{fancyhdr}
\makeatletter

\newcommand{\Rmnum}[1]{\expandafter\@slowromancap\romannumeral #1@}
\makeatother

\begin{document}
%
\title{NC-MOPSO: Network centrality guided multi-objective particle swarm optimization for transport optimization on networks}
%
%
%

\author{Jiexin~Wu,~
	Cunlai~Pu,~
	Shuxin~Ding,~
	Guo~Cao,~
	and~Panos M. Pardalos
	\thanks{ This work was supported in part by the National Natural Science Foundation of China under grant U1934220, and in part by the Foundation of China Academy of Railway Sciences Corporation Limited under Grant 2019YJ071. P. M. Pardalos is partially supported by the Paul and Heidi Brown Preeminent Professorship at ISE (University of Florida, USA), and a Humboldt Research Award (Germany). (\textit{Corresponding author: Cunlai Pu.)}}
	\thanks{J. Wu, C. Pu and G. Cao are with the School of Computer Science and Engineering, Nanjing University of Science and Technology, Nanjing 210094, China (e-mail: wujiexin, pucunlai, caoguo@njust.edu.cn).}
	\thanks{S. Ding is with the Signal and Communication Research Institute, China Academy of Railway Sciences Corporation Limited, Beijing 100081, China (e-mail: dingshuxin@rails.cn).}
	\thanks{P. M. Pardalos is with the Center for Applied Optimization, Department of Industrial and Systems Engineering, University of Florida, Gainesville 32608, FL, USA. (e-mail: pardalos@ufl.edu).}}
%
%

\markboth{\tiny{This work has been submitted to the IEEE for possible publication. Copyright may be transferred without notice, after which this version may no longer be accessible.}}%
{Shell \MakeLowercase{\textit{et al.}}: Bare Demo of IEEEtran.cls for IEEE Journals}
%



\maketitle

\begin{abstract}
Transport processes are universal in real-world complex networks, such as communication and transportation networks. As the increase of the traffic in these complex networks, problems like traffic congestion and transport delay are becoming more and more serious, which call for a systematic optimization of these networks. In this paper, we formulate a multi-objective optimization problem (MOP) to deal with the enhancement of network capacity and efficiency simultaneously, by appropriately adjusting the weights of edges in networks. To solve this problem, we provide a multi-objective evolutionary algorithm (MOEA) based on particle swarm optimization (PSO), namely network centrality guided multi-objective PSO (NC-MOPSO). Specifically, in the framework of PSO, we propose a hybrid population initialization mechanism and a local search strategy by employing the network centrality theory to enhance the quality of initial solutions and strengthen the exploration of the search space, respectively. Simulation experiments performed on network models and real networks show that our algorithm has better performance than four state-of-the-art alternatives on several most-used metrics.
\end{abstract}

\begin{IEEEkeywords}
Networks, transport protocols, optimization methods.
\end{IEEEkeywords}

%
\IEEEpeerreviewmaketitle

\section{Introduction}
%
%
%
%
\IEEEPARstart{I}{N} modern society, our life relies so much on various infrastructure networks, including transport, communication and power networks, to deliver either tangible quantities, such as goods and travelers, or intangible quantities, like information and electricity. The performance of transport processes on these infrastructure networks significantly affects the quality of our life. Meanwhile,  traffic load on these networks increases exponentially owing to the rapid development of our society. For example, the latest Cisco traffic report \cite{cisco} revealed that Internet traffic is experiencing an explosion with the  growing number of applications. The increasing traffic poses a great challenge to the efficiency and scalability of these networks, and thus calls for a systematic optimization to address these issues. \\
\indent The performance of network transport can be measured from different aspects. On the one hand, we care about the maximum amount of traffic a network can transport in a period of time, which is usually called network capacity \cite{guimera2002}. Many factors affect network capacity. For instance, generally the larger processing capability of nodes and links, the larger network capacity will be achieved \cite{jiang2019}; homogenous networks have larger transport capacity than heterogeneous ones \cite{yan2006}; the more diverse delivery paths, the larger network capacity \cite{lin2019}. On the other hand, quantities are always expected to be delivered as fast as possible, hence the average number of hops is also a critical metric in transport \cite{danila2007}. This metric is also affected by the network structure, node capability, routing paths, etc. Therefore, we can optimize network transport performance by considering these factors.\\
\indent Optimizing the ``hard'' factors such as network structure and node capability usually results in a large cost and is even impossible in many cases. A more feasible way is to optimize the routing strategies in network transport, which determine how quantities are transmitted from their sources to destinations. One of the most representative routing strategies is the shortest path (SP) routing \cite{sp}, where all edges are treated  equivalently and the paths with the minimum number of edges are selected to transport quantities. However, in real-world networks, edges are different as demonstrated by their weights of various forms. For example, the weight of an edge in computer networks can be defined according to its actual bandwidth \cite{internet}. In air traffic networks, the weight of an edge can be quantified by the number of passengers on the flights  passing through it. Based on our needs, we can even manually allocate the edge weights to influence the selection of transport paths in network transport optimization.
\subsection{Our Motivation}
\label{motivation}
In the SP routing strategy, highly connected (hub) nodes are prone to be congested first because they are usually the intersections of many shortest paths \cite{congestion}. The congestion will then spread to other nodes. This cascading congestion problem significantly affects  network capacity. Therefore, to increase network capacity, we can modify the SP routing strategy to avoid the hub nodes. However, any modification will increase the average number of hops in principle. As a result, it is contradictory to enhance the network capacity and minimize the average number of hops at the same time. Our motivation is then to find the best edge weight allocation scheme, which will achieve the optimal balance between the average number of hops and network capacity. The joint optimization of these two metrics is NP-hard \cite{danila2007}. Therefore, multi-objective evolutionary algorithms (MOEAs) are applicable to our problem. \\
\indent A number of  MOEAs have been developed  in the field of evolutionary computing, nonetheless only a few of them have been tailored for the scenario of network transport problem with yet unsatisfactory performances, such as the Non-dominated Sorting Genetic Algorithm \uppercase\expandafter{\romannumeral2} (NSGA-\uppercase\expandafter{\romannumeral2}) \cite{naga2} and multi-objective particle swarm optimization (MOPSO) \cite{mopso}. The advantage of MOPSO mainly lies in that it requires very few assumptions or mathematical conditions about the problem. It,  however, has two main flaws. One is the early convergence to a local optimum, and the other is the loss of the diversity of particles during iterations. Network centrality has the potential to be applied to countering these flaws, when MOPSO is deployed in the network related applications \cite{ncnetwork}. There are many different definitions of network centrality in network science \cite{nc}. Tailoring the existing network centrality for MOEAs in network applications is nontrival, and more likely  we have to  give  new forms of network centrality based upon the specific optimization problems.   \\
\subsection{Contributions of This Paper}
\begin{enumerate}
	\item We formulate a network transport optimization problem, which involves two correlated objectives and thus belongs to multi-objective optimization problems (MOPs); one of the objectives is to maximize the network capacity, and the other is to minimize the average number of hops. By solving this MOP, we can obtain the optimal tradeoff between network capacity and efficiency in network transport.
	\item We propose a MOEA called  network centrality guided MOPSO (NC-MOPSO)  to deal with the network transport optimization problem. In this algorithm, we develop a hybrid swarm initialization strategy, in which a heuristic approach based on a new edge centrality metric is proposed to produce a fraction of high-quality initial solutions. Furthermore, we  develop a node-centrality guided local search strategy to expand the search space of the swarm. We also analyze the space and time complexity of the proposed algorithm.
	\item We conduct extensive comparative experiments on both network models and real-world networks to validate the performance of NC-MOPSO. Specifically, we compare it with four state-of-the-art MOEAs, i.e., NSGA-\uppercase\expandafter{\romannumeral2}, MOPSOCD, MOPSOCDELS and GSPSO,  in terms of three most popular performance metrics. The experimental results show that NC-MOPSO outperforms these competitors in all cases. Moreover, the convergence of these algorithms is also analyzed.
\end{enumerate}
\quad The remainder of this paper is organized as follows. Section \uppercase\expandafter{\romannumeral2} provides an overview of related works. In Section \uppercase\expandafter{\romannumeral3}, we present the traffic model and network transport optimization problem. In Section \uppercase\expandafter{\romannumeral4}, we introduce the framework of NC-MOPSO. The experimental results and performance analysis are provided in Section \uppercase\expandafter{\romannumeral5}. Finally, this paper is concluded in Section \uppercase\expandafter{\romannumeral6}.
\section{Related works}
The related works of our paper are presented in three parts. One is about network transport optimization from the perspective of network science, and the other two are respectively multi-objective evolutionary algorithms and their applications in complex networks. We also discuss the connection and difference between our work and the related works.\\
\subsection{Transport Optimization on Complex Networks}
In the past few decades, transport processes in various complex networks have aroused great attention. The biggest concern is traffic congestion problem. Many methods have been put forward to alleviate traffic congestion and enhance  transport capacity of networks. Essentially, the capacity of a network depends mainly on its topological structure. Networks with scale-free topologies tend to cause an uneven distribution of traffic load, which limits the network capacity. Many attempts have been made to optimize network topological structures, such as removing some bottleneck links or nodes and adding some links between nodes with long distance \cite{zhang2019}.\\
\indent Another effective way to improve  network capacity is optimizing resource allocation in complex networks. Network resources, including link bandwidth and node processing capability, are usually finite. Thus, it is necessary to optimize the usage of these limited resources. Wu \textit{et al}. \cite{wu2013} provided an allocation strategy for the processing capacity of nodes, based on the node usage probability. A global dynamic bandwidth allocation strategy is proposed in Ref. \cite{jiang2013}, which sets the bandwidth of each link to be proportional to the length of  real-time queue. Liu \textit{et al}. \cite{liu2015} studied the joint optimization of traffic flow rate and node capacity in complex communication networks.\\
\indent The above strategies, however, are often costly or not practical in real applications, since they are devoted to manipulating the hard factors, such as network topology and resource. A more practical direction is to optimize the routing strategy of the traffic. The traditional SP strategy is vulnerable to traffic congestion because it only considers  path length. Other enhanced routing protocols consider more factors, such as node degree \cite{ma2018}, node load \cite{li2019}, memory information \cite{kirst2016}, and next-nearest neighbors \cite{yang2005}. Generally, for any routing protocol, more information used in routing decisions will contribute to better performance, nevertheless, the related computational cost will also be larger. \\
\indent The aforementioned works mainly focus on the optimization of network capacity, which may be at the cost of degenerating other important transport performances, e.g., the average number of hops. Our work considers the optimization of these two objectives, i.e., network capacity and  average number of hops, simultaneously. Furthermore, different from the methodologies in the aforementioned works, we employ evolutionary algorithms to optimize network transport performance.
\subsection{Multi-objective Evolutionary Algorithms}
 Multi-objective optimization is an effective framework to deal with plenty of real-world problems involving multiple conflicting optimization objectives \cite{moop,ntmoea}. The solution of MOPs is known to be a Pareto front (PF), which consists of nondominated solutions. MOEAs are population-based algorithms and have been successfully applied to deal with MOPs. During the past few decades, various MOEAs have been developed, which show  performance on many complex multi-objective benchmark functions. Zitzler \textit{et al}. \cite{spea} proposed the strength Pareto evolutionary algorithm (SPEA), where the diversity of the population is preserved by a novel niching method. SPEA2 \cite{spea2}, an updated version of SPEA, further adopts three mechanisms, namely a fine-grained fitness assignment, a density estimation, and an enhanced archive truncation. The nondominated sorting genetic algorithm (NSGA) \cite{nsga}, a more promising type of MOEAs, integrates the classification of nondominated fronts into the genetic algorithm. The main defects of NSGA are the high computational complexity of nondominated sorting and the lack of elitism. To address these defects, an improved version called NSGA-II \cite{naga2} was devised, which supports the elitism of the swarm and fast computation of nondominated sorting.\\
\indent MOPSO \cite{mopso} is a mainstream algorithm of MOEAs. It is essentially a multi-objective extension of Particle Swarm Optimization (PSO) that itself models a simplified version of bird flocking.   Recently, many improved versions of MOPSO have been proposed. Raquel \textit{et al}. \cite{mopsocd} provided the so-called MOPSOCD, in which the diversity of nondominated solutions is measured by a metric named crowding distance. Zheng \textit{et al}. \cite{zheng2013} used the comprehensive learning strategy to enhance the diversity of the swarm in MOPSO. Yuan \textit{et al}. \cite{han2017} proposed a geometric structure-based MOPSO, which improves the effectiveness and efficiency of the search by exploiting the geometric structure of solutions. To enhance the convergence and efficiency, Ding \textit{et al}. \cite{ding2018} considered multiple improvements in the framework of MOPSOCD,  such as global best selection and perturbation, and  further proposed an  algorithm named MOPSOCDELS, in which an elitist learning strategy is involved to add more diversity of the particles. \\
\subsection{Applications of MOEAs in Complex Networks}
\label{moeacom}
\indent In recent years, MOEAs have stepped into the field of network science. Various optimization problems on complex networks can be solved with MOEAs.  In the task of  network vulnerability assessment, Zhang et. al. \cite{zhang2020} formulated  a bi-objective optimization problem, which simultaneously maximizes the destructiveness and minimizes the cost of an attack on the network. Zhou et. al. \cite{zhou2017} investigated the problem of improving the robustness of scale-free networks against both node-based and link-based attacks, which is finally treated in a  multi-objective framework,  and they further proposed a two-phase MOEA for solving this problem. In  the complex network clustering problem, Gong et. al. \cite{gong2014} provided a multi-objective discrete PSO algorithm to minimize two  evaluation objectives termed as kernel k-means and ratio cut.  In addition, a problem-specific MOEA/D was developed to solve the so-called tradeoff barrier coverage problem in wireless sensor networks \cite{zhang2016}. \\
\indent Enlightened by the aforementioned works, we study the application of  MOEAs  in a so-far untouched  problem, namely, transport optimization on complex networks with multiple objectives. In particular, we optimize  two conflicting  objectives, network capacity and average number of hops, in the framework of MOPSO. Furthermore,  we provide an improved version of MOPSO, which employs the theory of network centrality to enhance the initialization and local search. We have also considered the state-of-the-art MOEAs in the network transport optimization, which nonetheless  have worse performance than our improved MOPSO. \\
\section{Problem statements}
In this section, we present the network transport optimization problem in detail. Firstly, we illustrate the widely used traffic model. Secondly, we introduce the metrics for evaluating network transport processes. Finally, we formulate a MOP in network transport.
\subsection{Traffic Models}
\label{tm}
\begin{figure}[htpb]
	\centering
	\includegraphics[width=0.5\textwidth]{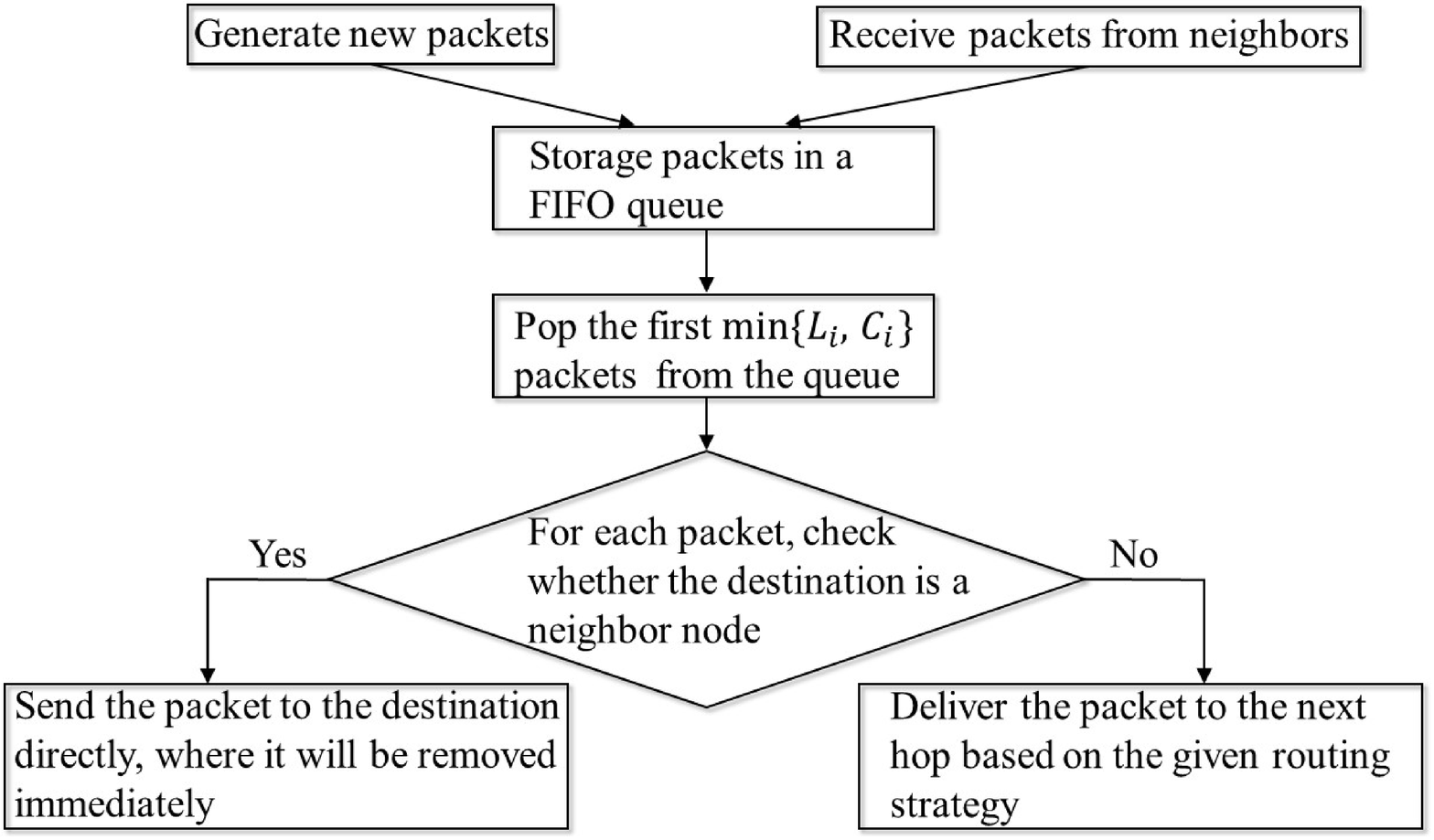}
	\caption{The behavior of node $i$ in a time step.}
	\label{fig1}
\end{figure}

Network transport is a fundamental and ubiquitous process in reality, e.g., packet transmission in the Internet and passenger travel in transport networks. Despite the different nature of the  transport processes in various networks, they have four common essential elements: generation, storage, forwarding, and routing of quantities of various kinds. A traffic model is thus required to involve these critical elements to characterize the essence of network transport. For the sake of concreteness, we take packet transport in communication networks as an instance to elaborate the frequently used traffic model. The edges are assumed to have weights in communication networks, which can be associated with bandwidth of edges in practice. The traffic model of communication networks  \cite{danila2007} is given as follows:\\
\indent \emph{Generation}: Each node generates packets with a rate of $ \lambda $, thus at each time step an average of $ N\lambda $ packets will be inserted into the network, where $N$ represents the number of nodes. The destination node of each packet is randomly selected from the network excluding the source node.\\
\indent \emph{Storage}: Every node has an infinite queue for buffering packets abiding by  the first-in-first-out (FIFO) rule, which is a representative way to schedule packets. When a node generates or receives a packet, it will append the packet to the tail of the queue.\\
\indent \emph{Forwarding}: The processing capability of a node is usually limited. Without loss of generality, at each time step, a node $i$ is assumed to be able to deliver at most $C_i$ packets. When a node forwards a packet, it will first check the corresponding destination node; if the destination is one of its neighbors, the packet will be delivered directly to the destination and then be removed immediately; otherwise, the node will send the packet to the next hop determined by the given routing strategy.\\
\indent \emph{Routing}: There could be multiple paths between the source and destination, and the routing strategy  selects the optimal path for delivering packets. In our work, we use a generalized version of the SP routing, i.e., the smallest-weight path (SWP) routing, which selects the path with the least sum of edge weights.\\
\indent Fig.~\ref{fig1} is an illustration of the whole operation of a node $i$ in one step of the packet transport process, where $L_i$ is the number of packets in the queue of node $i$.
\subsection{Metrics of Network Transport }
We employ two representative metrics to evaluate network transport processes: network capacity and average number of hops. These two metrics are both determined by the routing betweenness centrality \cite{betweenness}, which is a basic measurement for network components in network science. The routing betweenness centrality of node $i$ is defined as
\begin{equation}
B_i = \sum_{s\ne i \ne t} \frac{n_{st}^i}{n_{st}},
\end{equation}
\\where $ n_{st} $ is the number of routing paths from node $ s $ to node $ t $, and $ n_{st}^i $ is the number of routing paths from node $s$ to node $t$ that also pass through node $i$. According to this definition, we can infer that the nodes with large betweenness bear more traffic load than the nodes with small betweenness. Thus, the onset of traffic congestion is often observed in large-betweenness nodes. Generally, once the packet generation rate $\lambda$ surpasses a critical value $\lambda_c$, traffic congestion begins to happen; hence, the critical packet generation rate is usually taken as a measurement of network capacity. Liang \textit{et al}. \cite{congestion} derived the critical packet generation rate
\begin{equation}
\lambda_c = \frac {\left( N-1 \right) C_{L_{max}}}{B_{L_{max}}},
\label{capacity}
\end{equation}
\\where $L_{max}$ represents the index of the  node  with the largest routing betweenness centrality.  For simplicity, at each time step, each node is further assumed to deliver at most one packet, i.e., $C_i=1$. Then, (\ref{capacity}) can be simplified as
\begin{equation}
\lambda_c = \frac {N-1}{B_{L_{max}}}.
\label{ca2}
\end{equation}
\indent In addition to network capacity, the average number of hops is also a main indicator for the performance of network transport. A small average number of hops means a fast delivery of packets when delivery delay in the queue is not considered. The average number of hops can be calculated as
\begin{equation}
H_{avg}=\frac {B_{avg}}{N-1},
\label{havg}
\end{equation}
\\where $B_{avg}$ represents the average routing betweenness centrality of nodes in the network.
\subsection{Multi-objective Optimization Formulation}
We consider optimizing the routing strategy to obtain the best compromise between two metrics in network transport, i.e., network capacity and  average number of hops. This optimization problem is however very challenging since these two metrics are contradictory in that enhancing one suppresses the other. Moreover, it has been pointed out in \cite{danila2007} that optimizing  network capacity is already  NP-hard, indicating that the time needed to calculate the exact solution will increase with the number of network nodes faster than any polynomial. Collectively, these difficulties call for an evolutionary computation framework to handle the optimization of network transport.\\
\indent The dual optimization problem is formally defined as follows. Given an undirected weighted network $G=(V,E,X)$, where $V$ is the node set with $ N=|V|$ nodes, $E$ is the edge set with $M=|E|$ edges, and $ X=(x_1, x_2 , ... , x_M)^T $ is a $M$-dimensional vector, in which the element $x_e$ represents the weight of the $e$-$th$ edge. In the SWP routing, whether a path is used to transport packets is determined by the weights of edges in the path.  The weight setting of the edges directly affects the routing paths and eventually the routing betweenness of each node. In other words, the node routing betweenness is essentially a function of the edge weight. Considering the weight vector $X$ as a decision variable and formulating the optimization problem as
\begin{equation}
\left\{
\begin{aligned}
&\max\quad \lambda_c  = \frac{N-1}{\max\limits_{i \in N}B_i(X)}  \\
&\min\quad H_{avg}  = \frac{\sum_{i=1}^N B_i(X)}{N\left(N-1\right)}
\end{aligned}
\right.
\end{equation}
\begin{equation}
\begin{aligned}
\mbox{s.t.}\quad
& x_e \in (0,1],\quad\forall x_e \in X,
\end{aligned}
\end{equation}
where $B_i(X)$ is the routing betweenness of node $i$ under the weight vector $X$. The first objective maximizes the network capacity, and the second one minimizes the average number of hops. Note that we constrain the weight of each edge in $(0,1]$ to avoid too many redundant solutions and accelerate the convergence of the evolutionary computation. Therefore, our optimization problem is essentially a constrained MOP, whose solution is a set of optimal compromises.
\section{Algorithm specification and implementation}
In this part, a new algorithm called NC-MOPSO is presented to solve the defined transport optimization problem. Specifically, we first provide the framework of NC-MOPSO. Then, we give a detailed description of its two key mechanisms: hybrid initialization and  local search. Finally, we discuss the complexity of NC-MOPSO.
\subsection{NC-MOPSO for Network Transport Optimization }
PSO is a popular single-objective optimization algorithm with fast convergence, simple framework and easy implementation. In PSO, a swarm of particles seek the optimal solution by exploring a given space, each of which has a position and velocity. A candidate solution is expressed by the position vector of a particle. In our problem, we let the position of a particle $k$ be $P_k=(p_1, p_2, ... , p_M)^T$, where $p_e\in(0,1]$ is the weight of edge $e$, i.e., $p_e=x_e$, and $M$ is the number of edges in the network. The position and velocity are updated in the following way:
\begin{equation}
\left\{
\begin{aligned}
v_{k}^{t+1} & = \omega\times v_{k}^{t}+c_1 \times r_1 \times \left(P^{t}_{kpbest}- P_{k}^{t}\right)\\
 &+c_2 \times r_2 \times \left(P^{t}_{gbest}- P_{k}^{t}\right)  \\
P_{k}^{t+1} & = P_{k}^{t}+v_{k}^{t+1},
\end{aligned}
\label{pso}
\right.
\end{equation}
where ${v_k}^t$ and ${P_k}^t$ are the velocity and position of particle $k$ at iteration time $t$, respectively; $ \omega $ is an inertia weight; $ c_1 $ and $ c_2 $ are two acceleration coefficients; $ r_1 $ and $ r_2 $ are two learning coefficients, which are randomly and independently selected from interval $[0,1]$; $P^{t}_{kpbest}$ is the best personal solution of particle $k$ until time $t$, while $P^{t}_{gbest}$ is the best global solution in the entire swarm until time $t$.\\
\indent MOPSOCD \cite{mopsocd} is the multi-objective version of PSO. In MOPSOCD, the particles update their personal best solutions  at each iteration. Specifically, if the current solution dominates the  so-far best solution, i.e., the former is better than the latter for all the objectives considered, then the latter will be updated with the former. The personal best solution of each particle is  put into an external archive, where the comparisons between the solutions are further conducted, and the dominated ones are deleted accordingly. The relative density of non-dominated solutions in the external archive is measured by the crowding distance mechanism. Then, the global best solutions are randomly selected among those with large crowding distances, which together with the personal best solutions guide the movement of particles (see (\ref{pso})). Note that the number of non-dominated solutions grows with iterations until reaching the given upper bound. When the archive is not large enough, the solutions with the smallest crowding distance will be discarded with priority. After the maximum number of iteration is reached, the solutions in the archive are the desired results of MOPSOCD. \\
\indent In our work, we propose a new MOEA, named as NC-MOPSO,  to solve the bi-objective network optimization problem, which is essentially an improvement of MOPSOCD. Particularly, the theory of network centrality is employed to enhance the quality of solutions for the defined transportation problem. In NC-MOPSO, we develop novel strategies for population initialization and local search, which cooperate with each other to strengthen this algorithm. The framework of NC-MOPSO is given in Algorithm \ref{alg:Framework}, which consists of two parts: initialization (line 1 to line 6) and updating (line 7 to line 15). First, an initial swarm is produced by the hybrid initialization strategy (Algorithm \ref{alg:ini}). Afterwards, each particle will update its position and velocity according to the given rules. The fitness and the personal best position will update subsequently. The local search (Algorithm \ref{alg:local}) performs to explore the neighborhood of a solution randomly selected from the current swarm. Finally, the global best solution is selected from the external archive, where all non-dominated solutions are stored. This cycle is repeated until the stopping criterion is reached. The algorithm returns the external archive that contains the final non-dominated solutions.
\begin{algorithm}[htb]
	\caption{Framework of NC-MOPSO Algorithm.}
	\label{alg:Framework}
	\begin{algorithmic}[1]
		\Require
		The adjacency matrix of a network: $A$; swarm size: $ pop $; maximum generation: $ maxgen $.
		\Ensure
		PF solutions. Each solution corresponds to a weight allocation scheme.\\
		Position initialization according to the edge-centrality guided hybrid initialization (\textbf{Algorithm 2}): $ S=\left\{P_1,P_2,...,P_{pop}\right\}^T $;\\
		Velocity initialization: $ V=\left\{v_1,v_2,\dots,v_{pop}\right\}^T $;\\
		Evaluate each  particle in the swarm;\\
		Personal best position initialization: $S_{pbest}=\left\{pbest_1,pbest_2,\dots,pbest_{pop}\right\}^T$, $ pbest_k=P_k $;\\
		Initialize the external archive by non-dominated solutions;	\\	
	    Set $ t=0 $;\\
	    \textbf{For} $ k=1 $ to $ pop $, \textbf{do}\\
		\qquad Calculate new velocity $ v_k^{(t+1)} $ ;\\
		\qquad Calculate new position $ P_k^{(t+1)} $ ;\\
		\qquad Evaluate $ P_k^{(t+1)} $;\\
		\qquad Update $ pbest_k $;\\
		\textbf{End for}\\
		Improve a solution randomly selected  from the current swarm using node-centrality guided local search (\textbf{Algorithm 3});\\
		Update the external archive;\\
		Control the  size of the external archive and select $ gbest $  from it according to the crowding distance mechanism;
		Stopping criteria: If $ t<maxgen $, then $ t++ $  and go to line 7; otherwise, stop the algorithm and output the result.  		
	\end{algorithmic}
\end{algorithm}
\subsection{Edge-centrality Guided Hybrid Initialization}
\label{echi}
A good initialization mechanism  can reduce the search space and accelerate the algorithm for  finding the global optimal solution. Instead of purely random initialization, we propose an edge-centrality guided  hybrid initialization, named as ECHI,  to provide better initial solutions.

 In our optimization problem, defined in (5), network capacity is a function of the largest node betweenness, while average number of hops is dependent on the average node betweenness.
Since both of the two objectives have strong correlation with node betweenness, we define the centrality of an edge as the arithmetic mean of the betweenness of its two end nodes with
normalization. We name it by node-betweenness based edge centrality (NBEC).  Specifically, the NBEC of an edge $e(i,j)$ with two end nodes $i$ and $j$ is
\begin{equation}
{EC}_{e\left(i,j\right)}=\frac{B_i+B_j}{2\sum_{t=1}^N B_t},
\label{imp}
\end{equation}
where $ \sum_{t=1}^N B_t $ represents  sum of the routing betweenness of nodes in the network. Note that there exist several distinct  definitions of edge centrality in network science. We have tested the other definitions and found that (\ref{imp}) is the best form in our problem (see \cite{sup}).

 In the initialization, we redistribute the values of elements in a randomly selected solution according to the edge centrality, so that an edge of larger centrality will has a larger weight. This redistribution intentionally arranges large centrality edges, which are prone to be congested during transportation,  with large weights. Therefore, their centrality values decrease, and  the SWP routing strategy will reduce the dependence of these edges, which is in favour of the transport process.  The details of our initialization strategy ECHI  are shown in Algorithm \ref{alg:ini}.
\begin{algorithm}[htb]
	\caption{Edge-centrality Guided Hybrid Initialization.}
	\label{alg:ini}
	\begin{algorithmic}[1]
		\Require  The adjacency matrix of a network: $A$; swarm size: $ pop$; heuristic initialization rate: $HIR \left(0\textless HIR\leq 1 \right )$.
		\Ensure  Initial swarm $ S $. \\
		Initialize the swarm with random initialization:\quad$ S=\left\{P_1,P_2,...,P_{pop}\right\}^T$, $P_k$ is a particle in the initial swarm; \\
		Calculate the number of heuristic initialization particles: $n_{hir}=HIR*pop$;\\
		\textbf{For} $k=1$ to $n_{hir}$, \textbf{do} \\
		\qquad Calculate the edge centrality (defined in (8)) of the network with the adjacent matrix $A$ and the current edge weights, which is given by $P_k$;\\
		\qquad Reorder the element values in solution $P_k$ that an element with larger edge centrality will has a larger  value, and thus generate a new solution ${P_k}^{\prime}$;\\
		\textbf{End for}\\
		\textbf{Return} the hybrid initial swarm $S$.
	\end{algorithmic}
\end{algorithm}
\subsection{Node-centrality Guided Local Search}
\label{ncls}
 The technique of local search (LS) has been increasingly used to hybridize the evolutionary algorithms, which can find better PF by exploring the neighborhood of current solutions. In this paper, we propose a node-centrality guided local search (NCLS) to strengthen NC-MOPSO, which executes in each iteration of NC-MOPSO (line 13 of Algorithm 1).

  In the NCLS, we first randomly select a solution from the current swarm. Then, we generate a certain amount of neighbors of the solution. Specifically, based on the selected solution, we calculate the node routing betweenness of all nodes, and pick  the node with largest betweenness. We add a random decimal value to the current weight of  each edge of the node. Through this, we obtain a new solution, which is taken as a neighbor of the selected solution. Similarly, we can generate another neighbor based on the new solution, and  so on so forth. For all the generated neighbors, we select the nondominated solutions to replace the original solution, which will enhance the diversity of current solutions. Algorithm \ref{alg:local} depicts the pseudocode of NCLS.

  The idea of NCLS is that by suppressing the maximum node betweenness in an iterative way, the traffic becomes more evenly distributed  on the nodes, which increases the threshold of  traffic congestion and thus the transport performance.
\begin{algorithm}[htb]
	\caption{ Node-centrality Guided Local Search.}
	\label{alg:local}
	\begin{algorithmic}[1]
		\Require The adjacency matrix of a network $A$; a solution randomly selected from the current swarm: $ z $; the given number of the generated neighbors of a solution: $n$; the dimension of a particle: $M$.
		\Ensure  New solutions $ z_{new} $.\\
		\textbf{For} $i=1$ to $n$, \textbf{do}\\
		\qquad Calculate the routing betweenness of each node in the network with the adjacency matrix $A$ and solution $z$, which stores the weight of each edge;\\
        \qquad Find the node with the largest routing betweenness, $L_{max}$;\\
		\qquad Add a random decimal to the weight of each edge connecting to node $L_{max}$, and thus generate a new solution: $z^{\prime}$, which is a neighbor of particle $z$;\\
	    \qquad Update the adjacency matrix according to $z^{\prime}$;\\
	    \textbf{End for}\\
	    \textbf{Return} the nondominated solutions in all neighbors: $ z_{new} $.	   		
	\end{algorithmic}
\end{algorithm}
\subsection{Complexity Analysis}
\begin{enumerate}
	\item \emph{Space complexity}: Three main memorizers are needed for NC-MOPSO. The first one is for storing the adjacency matrix of a network, which has a space complexity of $O(N^2)$, where $N$ is the number of network nodes. The second one is to store particles; its space complexity is  $O(M \cdot pop)$, where $pop$ and $M$ are the number and dimension of particles, respectively. The last one is for the external archive, which has $O(s\cdot M)$ space complexity, where $s$ is the size of the external archive. Note that the sizes of all these memorizers will not increase with iterations since they are determined in the initialization.
	\item  \emph{Time complexity}: The time complexity mainly lies in the updating process of NC-MOPSO, which consists of objective function computation, local search, crowding distance computation, and nondominated comparison in the population and external archive. The objective function computation has $O(m \cdot pop)$ time complexity, where $m$ and $pop$ are the numbers of objectives and particles, respectively. The time complexity of the local search is $O(n)$, where $n$ is the number of neighbors. Sorting the solutions in the external archive with crowing distance has $O(m\cdot s\cdot \log s)$ time complexity.  The time complexity for the nondominated comparison is generally $O(m \cdot pop \cdot s)$. Assuming that the size of the external archive is proportional to the number of particles, the time complexity for the nondominated comparison will be $O(m \cdot pop^2)$. Collectively, the overall time complexity for each generation is $O(m\cdot pop^2)$.
\end{enumerate}
\section{Experimental Results}
In this section, we assess the performance of NC-MOPSO in network transport optimization. First, we introduce the baseline algorithms and network data. Then, we present three popular performance metrics and  parameter settings of  MOEAs. Finally, the results of the comparative experiments are provided.
\subsection{Baseline Algorithms and Network Data}
To show the performance of NC-MOPSO, we compare it with the following baseline algorithms:
\begin{enumerate}
	\item MOPSOCD \cite{mopsocd}, a popular MOEA, which employs the crowding distance mechanism to deal with the global best selection and the filtering of nondominated solutions when the external archive is full.
	\item MOPSOCDELS \cite{ding2018}, an improved version of MOPSOCD, which employs the elitist learning strategy (ELS) to avoid early convergence and improve  optimization efficiency. The ELS adopts a Gaussian perturbation to maintain the diversity of particles and is  fairly effective for global optimization.
	\item NSGA-\uppercase\expandafter{\romannumeral2} \cite{naga2}, one of the most popular MOEAs, which selects individuals according to Pareto dominance relation and propagates offspring in an iterative way. The main feature of this algorithm is that it adopts the elitist nondominated sorting with the crowding distance as a ranking criterion.
	\item GSPSO \cite{han2017}, one of the state-of-the-art MOEAs, which enhances the PSO with the geometric structure of PF.  In this MOEA, the inherent geometric structure of the current generation's PF is exploited to guide particles' flying directly.
\end{enumerate}
\par For fair comparison, comparative experiments will be conducted on different types of networks. The dataset contains eight undirected weighted networks; four of them are generated by network models and the others are real-world networks. Specifically, we employ two widely used network models, i.e., the Barabasi-Albert (BA) model \cite{ba} and Watts-strogatz (WS) model \cite{ws}, to generate the scale-free networks and small-world networks, respectively. The scale-free networks exhibit the power-law degree distribution, in which a small percent of nodes have much larger degree than the others. While the small-world networks have approximately exponential degree distribution and high clustering, which are different from the scale-free networks. In addition, we select four different types of real-world networks, i.e., the power grid (118-bus), email network (email-enron-only) \cite{network}, road network (chesapeake) and Internet (uninett) \cite{zoo}. Table \ref{test data} characterizes these networks with three quantities:  number of nodes $N$,  number of edges $M$ and  average degree $\left \langle D\right \rangle =2M/N$.\\
\begin{table}[H]
	\centering
	\caption{Main characteristics of eight test instances.}
	\label{test data}
	\begin{tabular}{c|c|c|c}
		\hline
		Instance & $ N $  & $ M $ & $\left \langle D\right \rangle$ \\
		\hline
		BA100  & 100 & 196  & 3.92	\\
		\hline
		BA300 & 300 & 597  &  3.98\\
		\hline
		WS100 & 100 & 200 & 4\\
		\hline
		WS300 & 300 & 600 & 4\\
		\hline
		118-bus & 118 & 179 & 3.03\\
		\hline
		email-enron-only & 143 & 623 & 8.71\\
		\hline
		chesapeake & 39 & 170 & 8.72\\
		\hline
		uninett & 74 & 101 & 2.73\\
		\hline
	\end{tabular}
\end{table}
\subsection{Performance Metrics and Parameter Settings}
Three popular metrics are employed to evaluate the performance of different algorithms, namely hypervolume (HV) \cite{chv}, inverted generational distance (IGD) \cite{igd} and set coverage (C-metric) \cite{chv}. HV and IGD measure the quality of PF solutions in terms of convergence and diversity; C-metric compares two solution sets from the point of Pareto dominance.\\
\indent The HV metric gives the volume that is calculated by the solution set $P$ and the reference point $r$, as shown in (\ref{hv}). The reference point adopted here is $(\min(\lambda_c),\max(H_{avg}))$. The larger value of  HV means that the algorithm  produces better solutions.
\begin{equation}
HV\left\{P,r\right\}=Volume\left( \bigcup_{i=1}^{|P|} v_{i}\right).
\label{hv}
\end{equation}
\indent The definition of IGD is provided in (\ref{igd}), where $P^*$ denotes the true PF and $P$ is an obtained solution set in a MOP. $d(p,P)$ is the smallest Euclidean distance from $p\in P^*$ to all points in $P$. The smaller value of IGD, the better solution. Note that the true PFs of real-world optimization problems are usually unknown, while they can be  approximated by selecting non-dominated solutions from all the solutions obtained by different MOEAs \cite{is2014}.
\begin{equation}
IGD\left\{P,P^*\right\}=\frac{ \sum_{p\in P^*}  d\left(p,P \right)}{|P^*|}.
\label{igd}
\end{equation}
\indent The C-metric $C(P,Q)$ is defined as the percentage of the solutions in $Q$ dominated by at least one solution in $P$ (see (\ref{c})), where $P$ and $Q$ are two PFs. Assuming $P^*$ is the true PF, the lower the value of $C(P^*,P)$, the better $P$ is.
\begin{equation}
C\left(P,Q\right)=\frac{|\left\{ q \in Q| \exists p \in P: p \quad dominates \quad q\right\}|}{|Q|}.
\label{c}
\end{equation}
\begin{table}[htpb]
	\centering
	\caption{Parameter settings of the algorithms.}
	\label{parameters}
	\begin{tabular}{c|c|c}
		\hline
		Algorithms & Parameters  & References \\
		\hline
		MOPSOCD \& MOPSOCDELS \& GSPSO   & $ pop=200 $ & \cite{mopsocd} \\
		& $ c_1=1.5 $ & \\
		& $c_2=2$   &\\
		& $\omega=0.4$ &\\
		& $maxgen=500$ &\\		
		\hline
		NC-MOPSOCD    & $ pop=200 $ &  \\
		& $ c_1=1.5 $ & \\
		& $c_2=2$   &\\
		& $\omega=0.4$ &\\
		& $n=300$ &\\
		&$ maxgen=500 $ \\
		\hline
		NSGA-\uppercase\expandafter{\romannumeral2}  & $ pop=200 $ & \cite{naga2} \\
		& $ p_c=0.9 $ & \\
		& $ p_m=\frac{1}{M} $ & \\
		&$ maxgen=500 $ \\
		\hline		
	\end{tabular}
\end{table}
\begin{table}[htpb]
	\centering
	\caption{HV values of NC-MOPSO for different $c_1$ and $c_2$ values on the BA100 instance.}
	\label{c1c2}
	\begin{tabular}{c|c|c}
		\hline
		$c_1$  & $ c_2 $ & \ HV values \\
		\hline
		1 & 1 & 1.8082\\
		\hline
		1 & 1.5 & 1.7992\\
		\hline
		1 & 2 & 1.8721\\
		\hline
		1.5 & 1 & 1.7936\\
		\hline
		1.5 & 1.5 & 1.8281\\
		\hline
		\textbf{1.5} & \textbf{2} & \textbf{1.8922}\\
		\hline
		2 & 1 & 1.8473\\
		\hline
		2 & 1.5 & 1.8188\\
		\hline
		2 & 2 & 1.8559\\
		\hline 		
	\end{tabular}
\end{table}
\begin{figure}[htpb]
	\centering
	\includegraphics[width=0.3\textwidth]{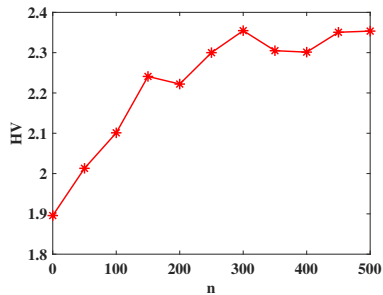}
	\caption{HV values of NC-MOPSO vs.  different $n$ values on the BA100 instance.}
	\label{np}
\end{figure}
\indent The experiments are conducted by using MATLAB R2017a on a computer, which has inter(R) Core(TM) i5-8300H with 2.30 GHz speed and 8 GB of RAM. The parameter settings of all the concerned MOEAs are shown in Table \ref{parameters}. The first column presents the names of the algorithms. The second column provides the values for different parameters, where $pop$ is the population size and $maxgen$ is the maximum number of iterations; $c_1$ and $c_2$ are two acceleration coefficients for PSO-based algorithms; $p_c$ and $p_m$ represent the crossover probability and the mutation probability in NSGA-\uppercase\expandafter{\romannumeral2}, respectively. The references for these parameter settings are given in the last column. \\
\indent For NC-MOPSO, we conduct the sensitivity test to select the optimal values for $c_1$, $c_2$, and $n$, which are the most important parameters in our algorithm. Table \ref{c1c2} shows the performance of NC-MOPSO in terms of HV metric for different $c_1$ and $c_2$ on BA100. It can be observed that the performance of NC-MOPSO is the best when $c_1 = 1.5$ and $c_2 = 2$. In addition, Fig.~\ref{np} shows the HV metric of NC-MOPSO for different $n$ values on BA100. It can be seen that $n = 300$ is the best point corresponding to the largest value of HV metric. \\
\begin{table}[htpb]
	\scriptsize
	\caption{HV (mean(std))  for different HIR (HIR0, HIR50 and HIR100) on each test instance.}
	\label{hir}
	\begin{tabular}{c|c|c|c}
		\hline
		Instance &  HIR0   &  HIR50  & HIR100 \\
		\hline
		BA100  & 0.7063(0.0783) & \textbf{1.7833(0.0824)} &	1.6303(0.1097)	\\
		\hline
		BA300 & 0.2465(0.0293) & \textbf{0.8939(0.0131)}	& 0.7977(0.0137)\\
		\hline
		WS100 & 1.1345(0.0778) & \textbf{1.4727(0.1097)} & 1.2892(0.0352)\\
		\hline
		WS300 & 0.6289(0.0205) & \textbf{0.7426(0.0632)} & 0.6475(0.0442)\\
		\hline
		118-bus & 0.1772(0.0046) & \textbf{0.1821(0.0092)} & 0.1780(0.0095)\\
		\hline
		email-enron-only & 0.9448(0.1675) & \textbf{1.7815(0.1089)} & 1.1880(0.1307)\\
		\hline
		chesapeake  & 1.9334(0.5761) & \textbf{4.3292(0.4083)} & 3.8410(0.1884)\\
		\hline
		uninett & 0.5382(0.0105) & \textbf{0.6010(0.0070)} & 0.5930(0.0112)\\
		\hline
	\end{tabular}
\end{table}
\begin{table}[htpb]
	\scriptsize
	\caption{IGD (mean(std)) for MOPSOCD, MOPSOCD\_in and NC-MOPSO on each test instance.}
	\label{zs}
	\begin{tabular}{c|c|c|c}
		\hline
		Instance &  MOPSOCD   &  MOPSOCD\_in  & NC-MOPSO \\
		\hline
		BA100  & 0.1675(0.0183) & 0.0717(0.0058) & \textbf{0.0032(0.0005)}	\\
		\hline
		BA300 & 0.4072(0.0268) & 0.1077(0.0181)	& \textbf{0.0059(0.0005)}\\
		\hline
		WS100 & 0.0364(0.0051) & 0.0283(0.0029) & \textbf{0.0059(0.0014)}\\
		\hline
		WS300 & 0.1336(0.0028) & 0.0955(0.0101) & \textbf{0.0030(0.0015)}\\
		\hline
		118-bus & 0.0109(0.0081) & 0.0089(0.0043) & \textbf{0.0042(0.0030)}\\
		\hline
		email-enron-only & 0.0979(0.0195) &0.0454(0.0137)& \textbf{0.0154(0.0167)}\\
		\hline
		chesapeake  & 0.1333(0.0435) & 0.0361(0.0099) & \textbf{0.0049(0.0017)}\\
		\hline
		uninett & 0.0136(0.0009) & 0.0031(0.0007) & \textbf{0.0009(0.0003)}\\
		\hline
	\end{tabular}
\end{table}
\begin{figure}[htpb]
	\centering
	\includegraphics[width=0.3\textwidth]{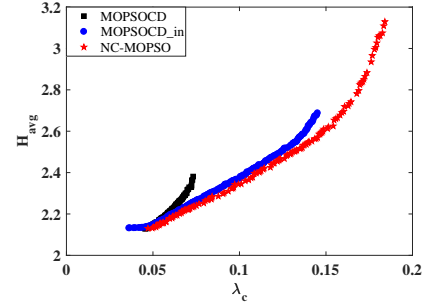}
	\caption{Nondominated solutions (PFs) of MOPSOCD, MOPSOCD\_in and NC-MOPSO on the BA100 instance.}
	\label{zs_1}
\end{figure}
\begin{figure}[htpb]
	\centering
	\subfigure[BA100]{
		\label{fig:subfig:ia} 
		\includegraphics[width=4cm,height=3cm]{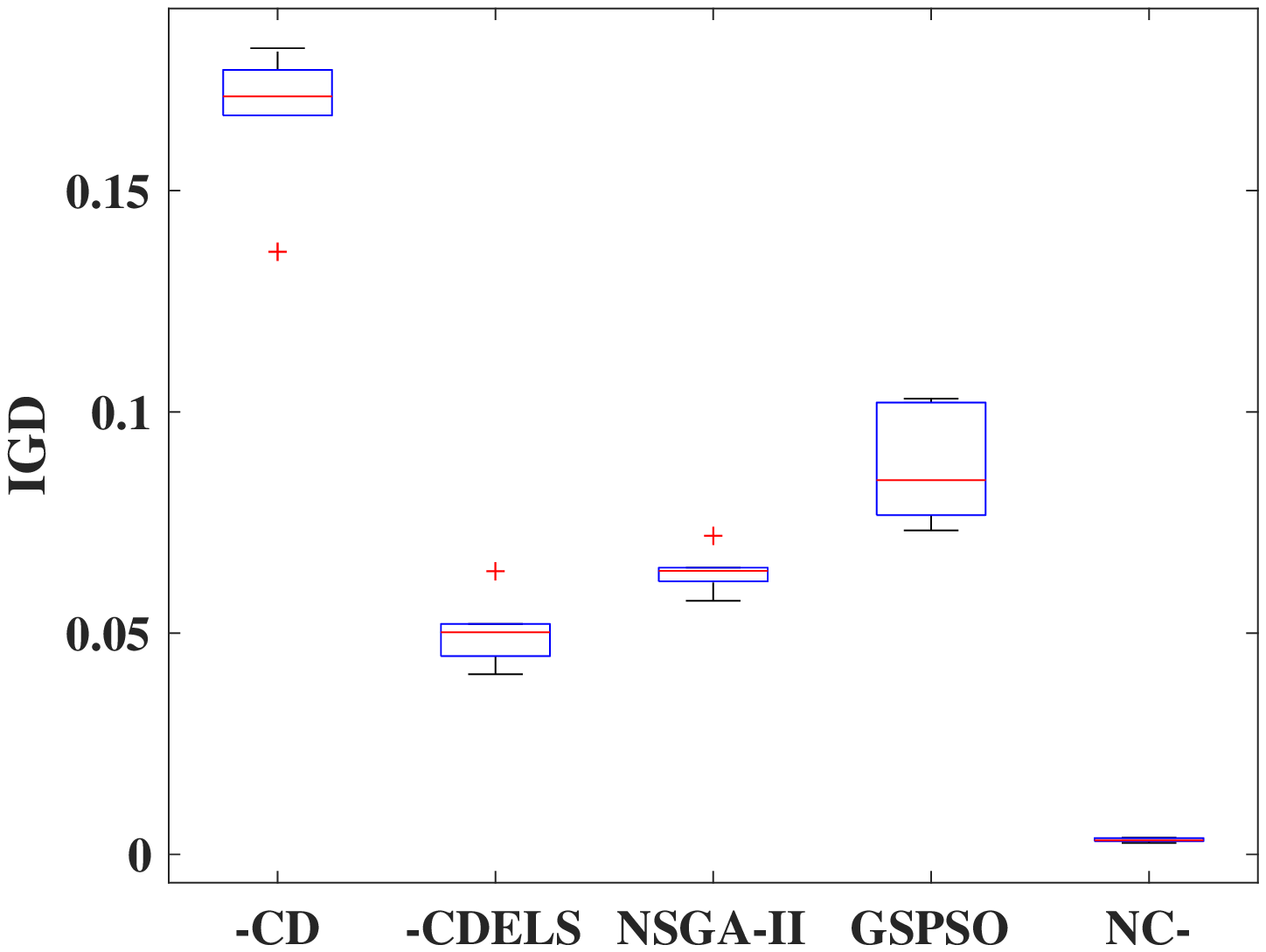}}
	\subfigure[BA300]{
		\label{fig:subfig:ib} 
		\includegraphics[width=4cm,height=3cm]{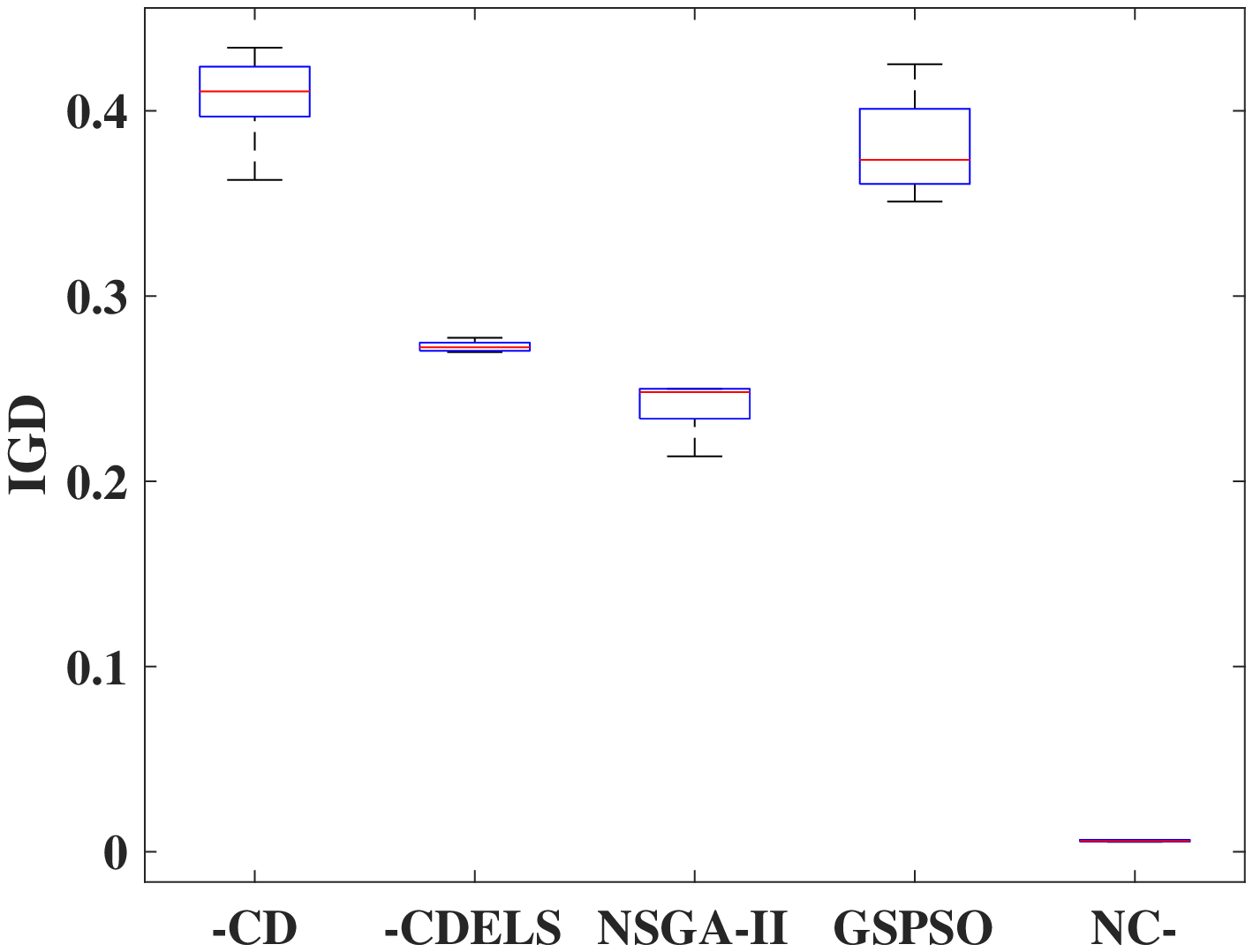}}\\
	\subfigure[WS100]{
		\label{fig:subfig:ic} 
		\includegraphics[width=4cm,height=3cm]{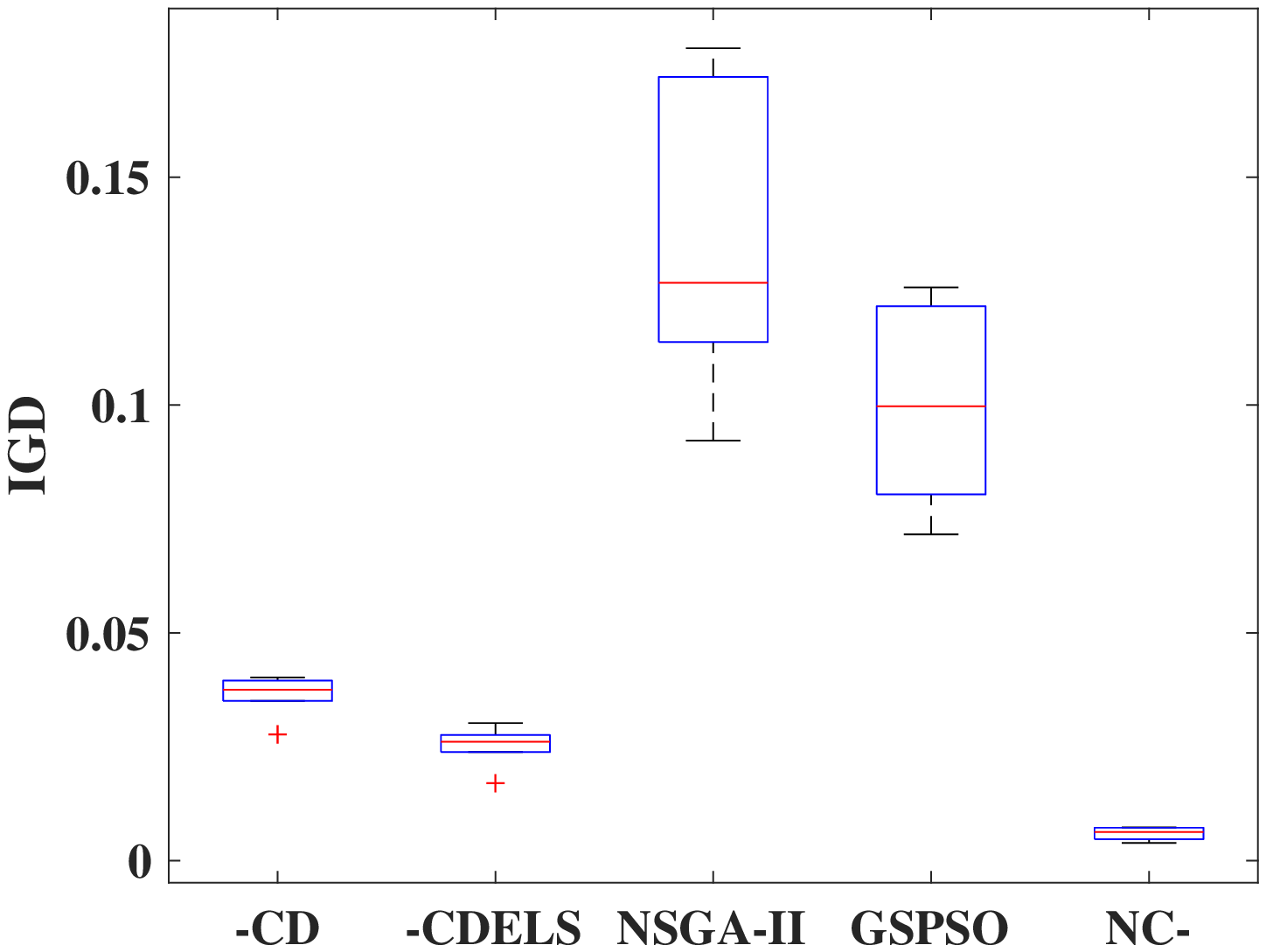}}
	\subfigure[WS300]{
		\label{fig:subfig:id} 
		\includegraphics[width=4cm,height=3cm]{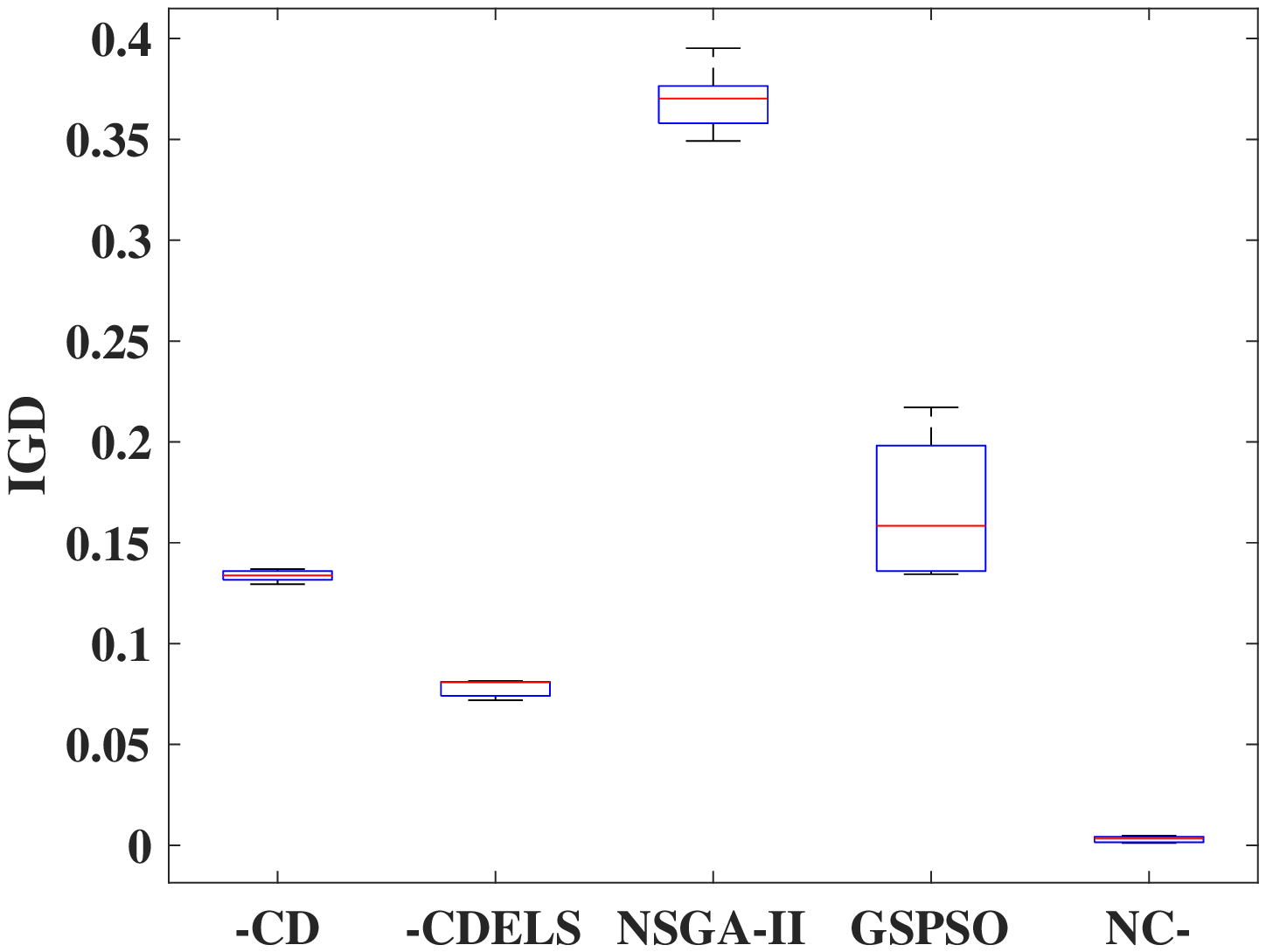}}\\
	\subfigure[118-bus]{
		\label{fig:subfig:ie} 
		\includegraphics[width=4cm,height=3cm]{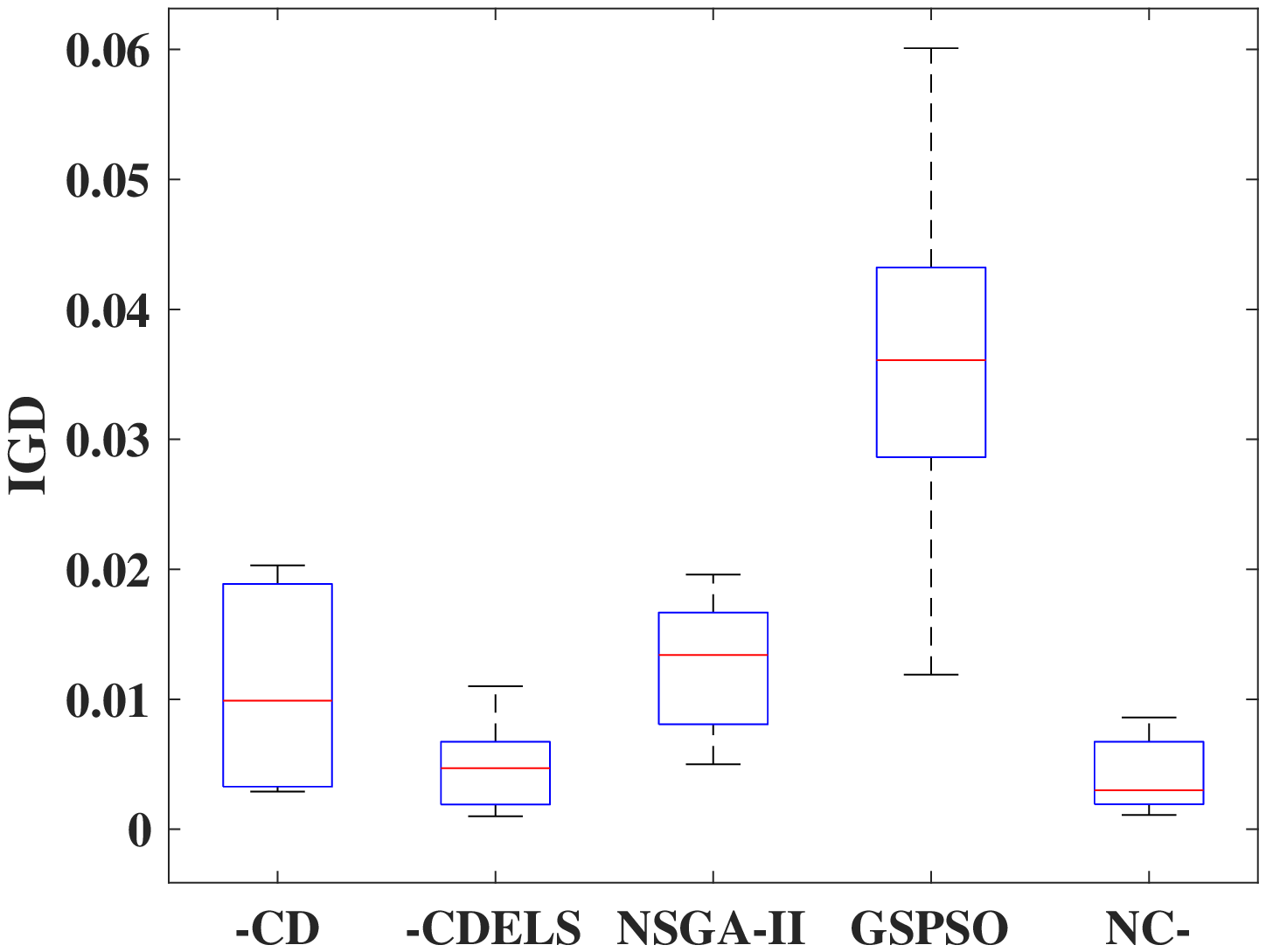}}
	\subfigure[email-enron-only]{
		\label{fig:subfig:if} 
		\includegraphics[width=4cm,height=3cm]{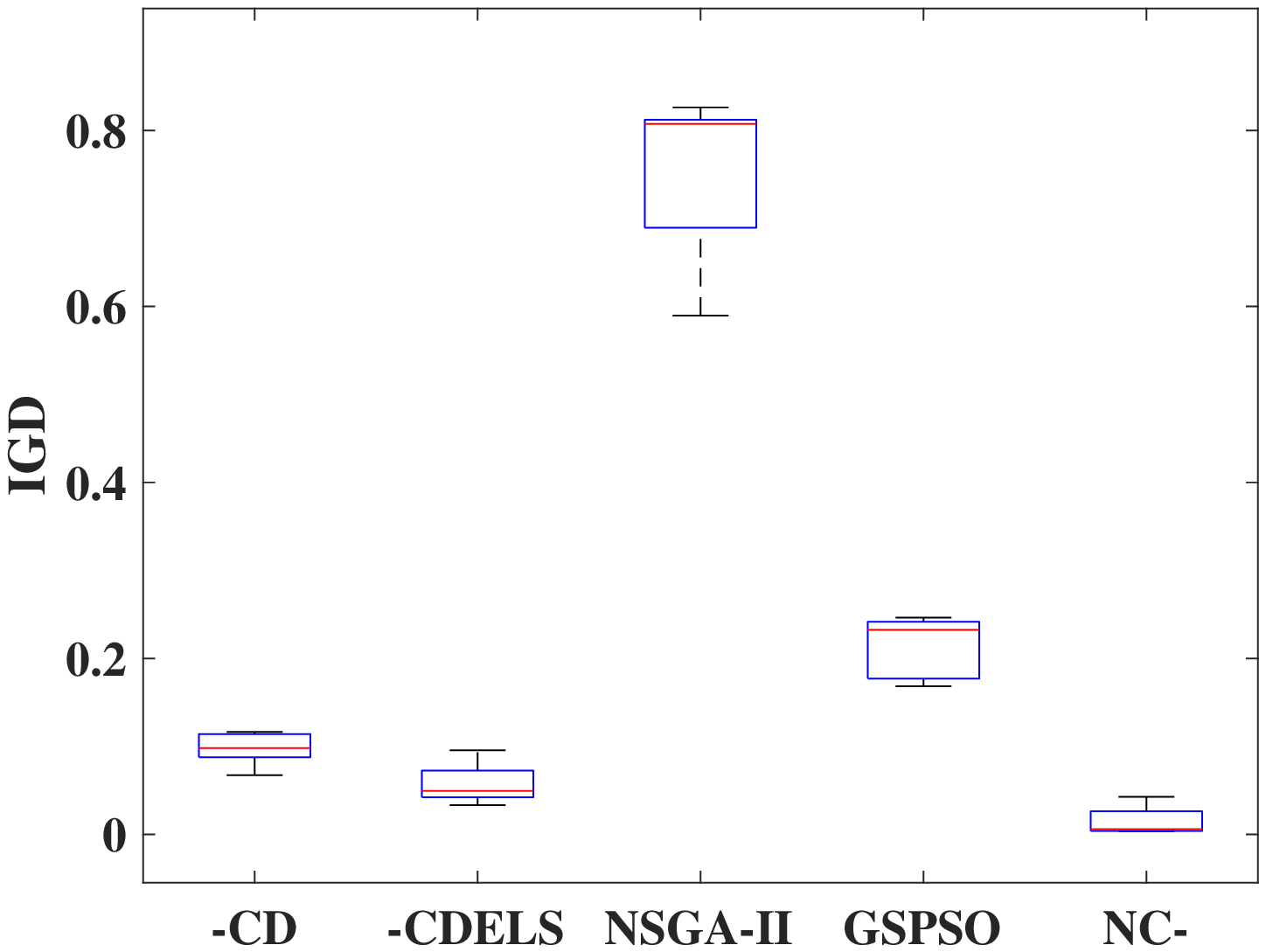}}\\
	\subfigure[chesapeake]{
		\label{fig:subfig:ig} 
		\includegraphics[width=4cm,height=3cm]{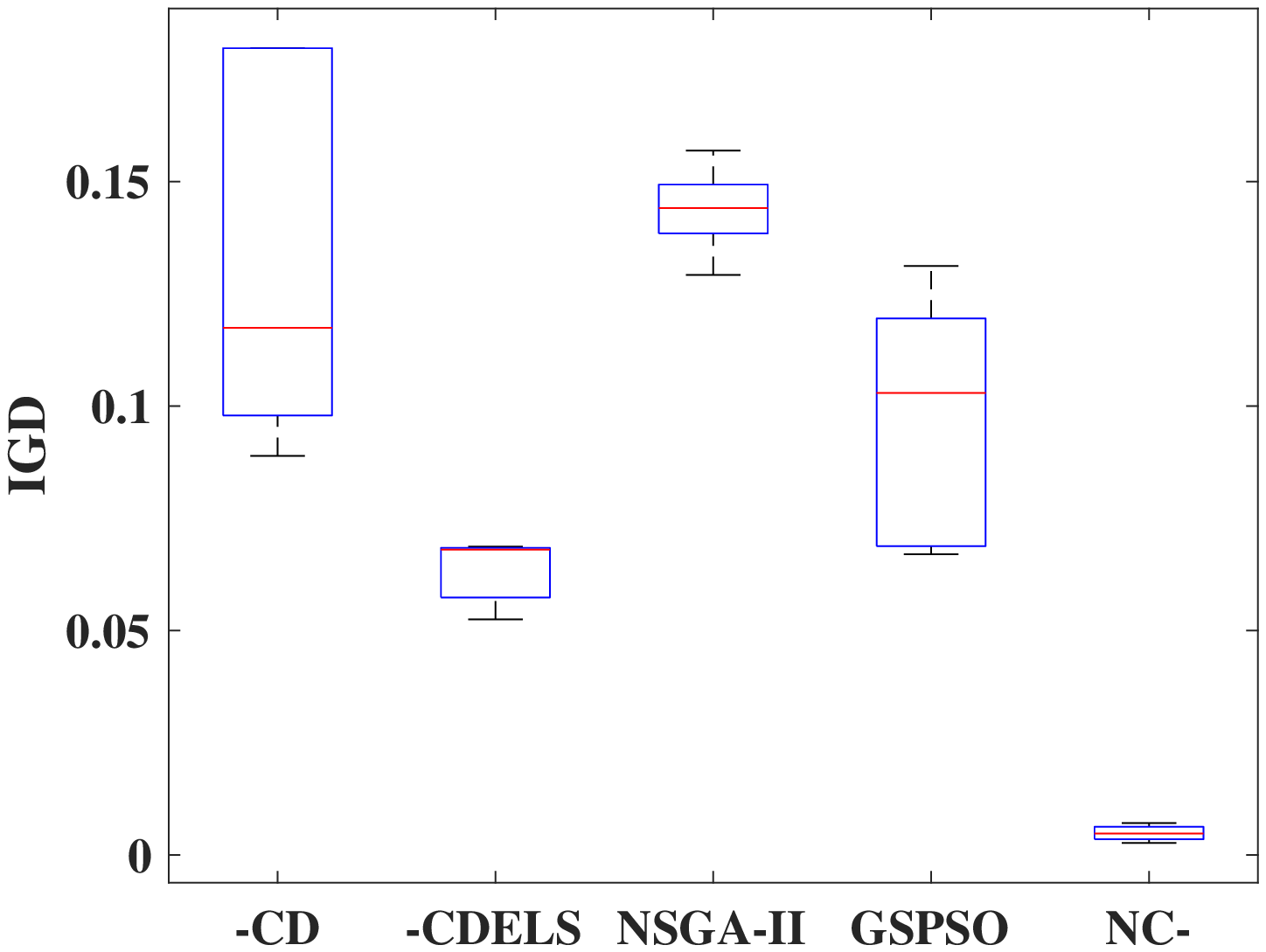}}
	\subfigure[uninett]{
		\label{fig:subfig:ih} 
		\includegraphics[width=4cm,height=3cm]{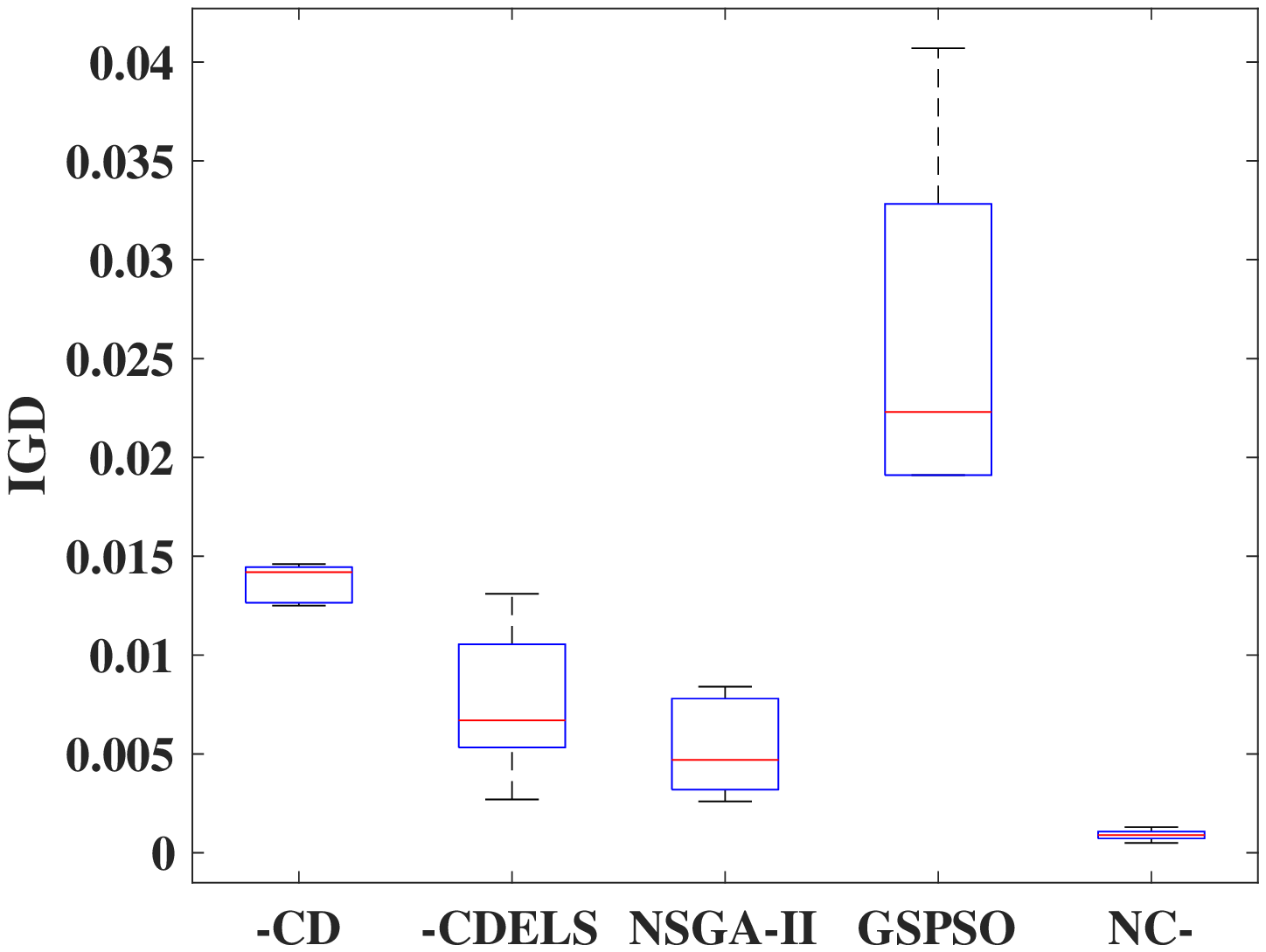}}\\
	\caption{The box-plot of the metric: IGD for all test instances.}
	\label{fig:igd} 
\end{figure}
\begin{figure}[htpb]
	\centering
	\subfigure[BA100]{
		\label{fig:subfig:ca} 
		\includegraphics[width=4cm,height=3cm]{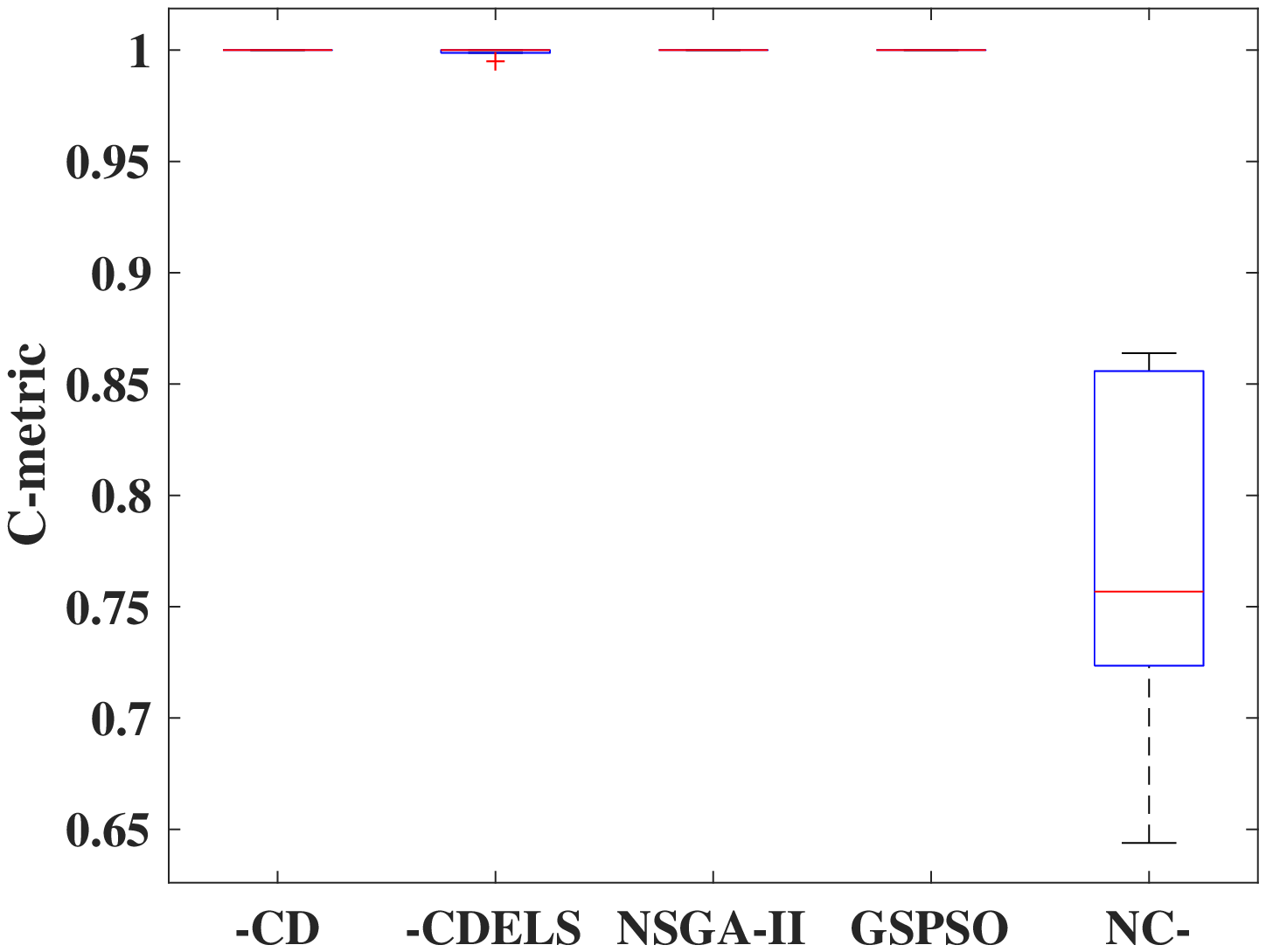}}
	\subfigure[BA300]{
		\label{fig:subfig:cb} 
		\includegraphics[width=4cm,height=3cm]{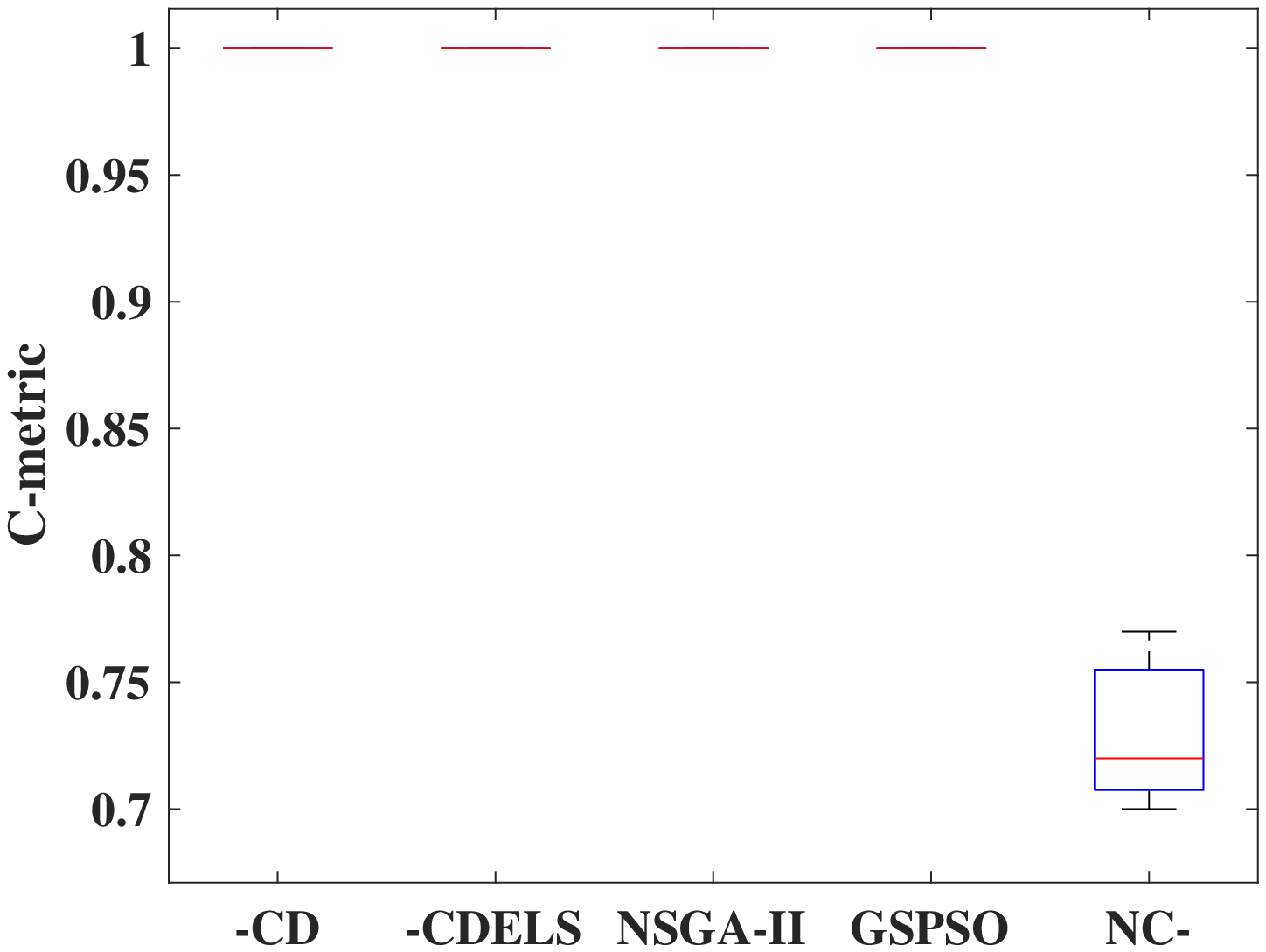}}\\
	\subfigure[WS100]{
		\label{fig:subfig:cc} 
		\includegraphics[width=4cm,height=3cm]{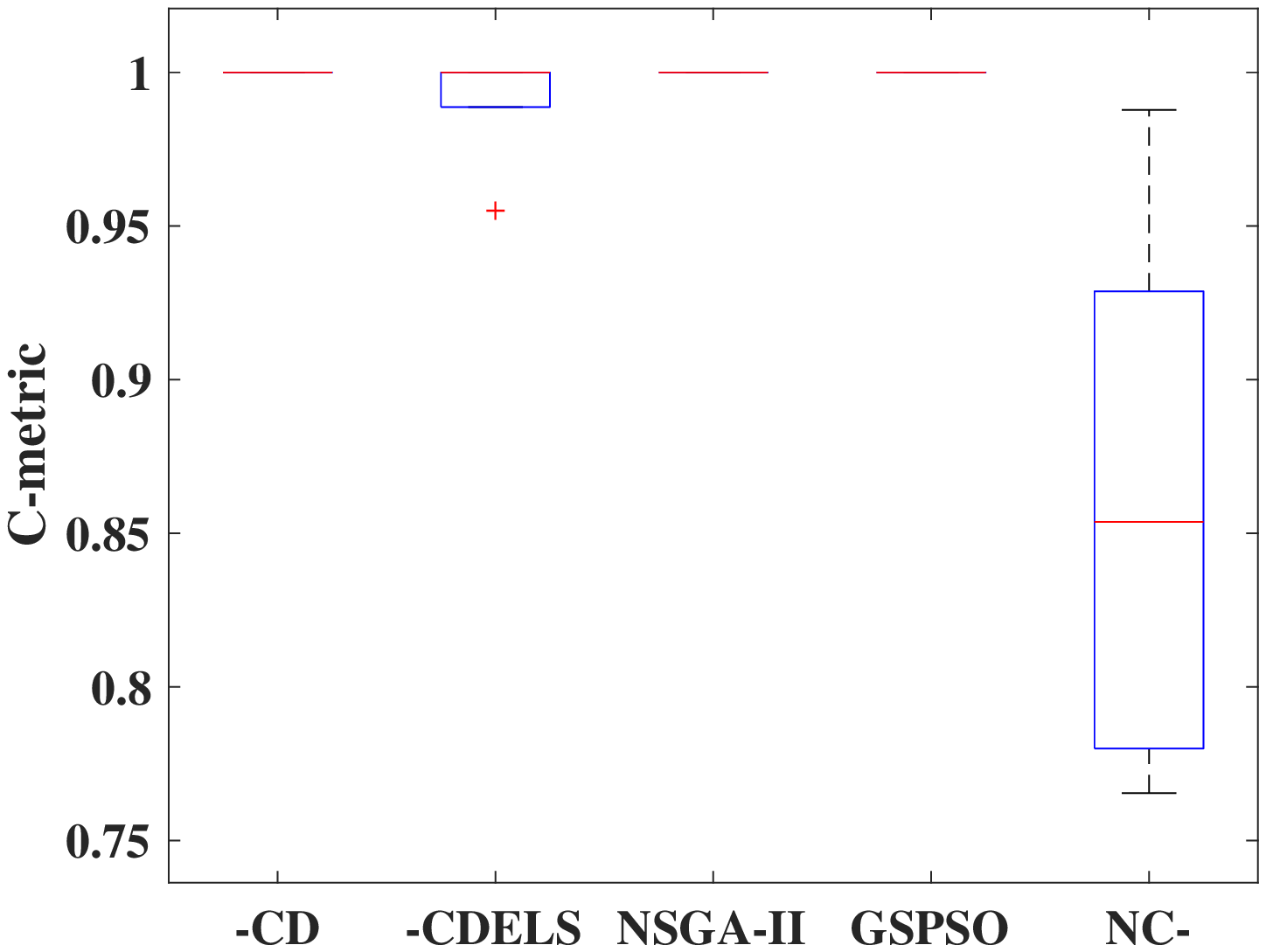}}
	\subfigure[WS300]{
		\label{fig:subfig:cd} 
		\includegraphics[width=4cm,height=3cm]{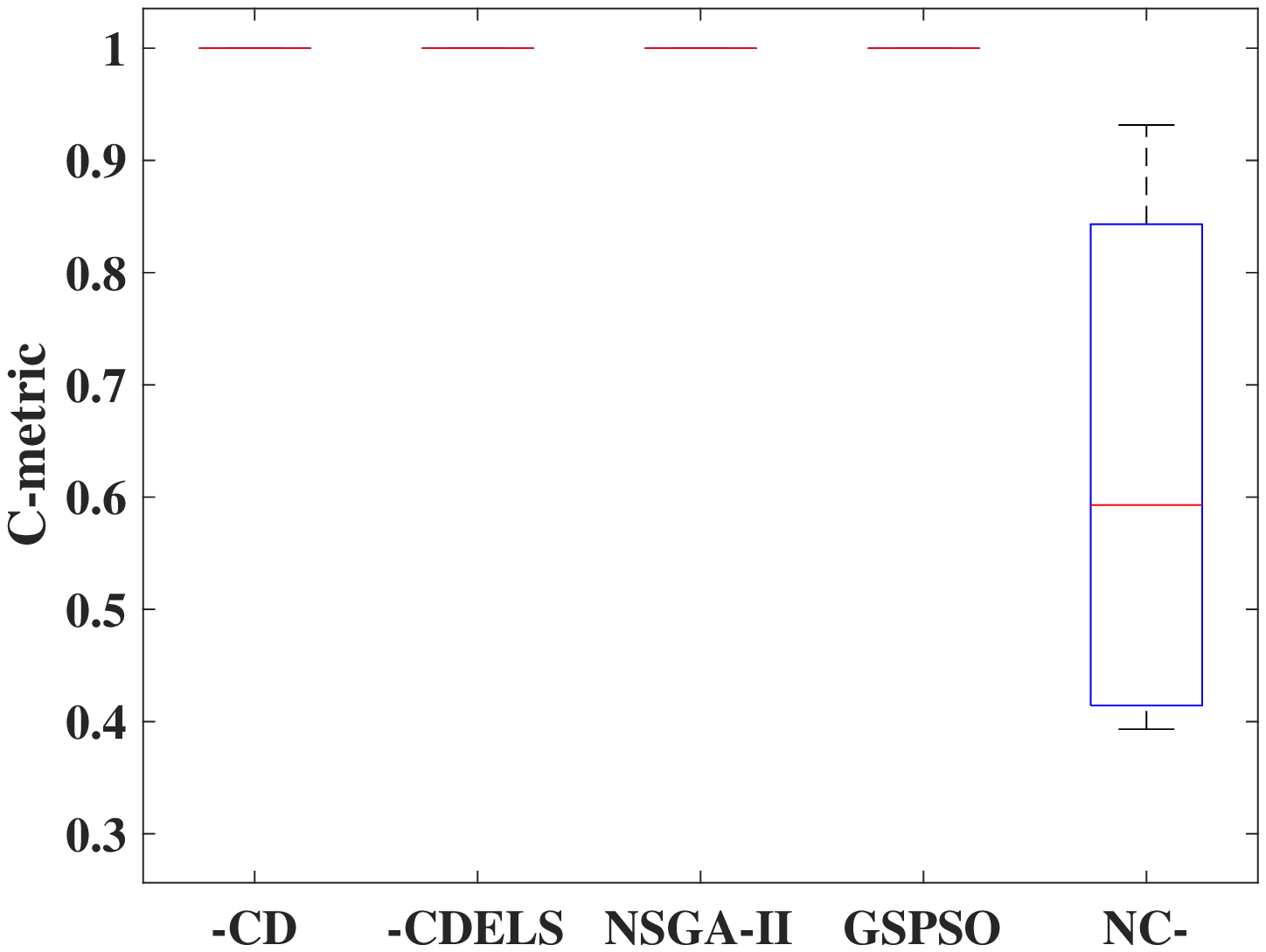}}\\
	\subfigure[118-bus]{
		\label{fig:subfig:ce} 
		\includegraphics[width=4cm,height=3cm]{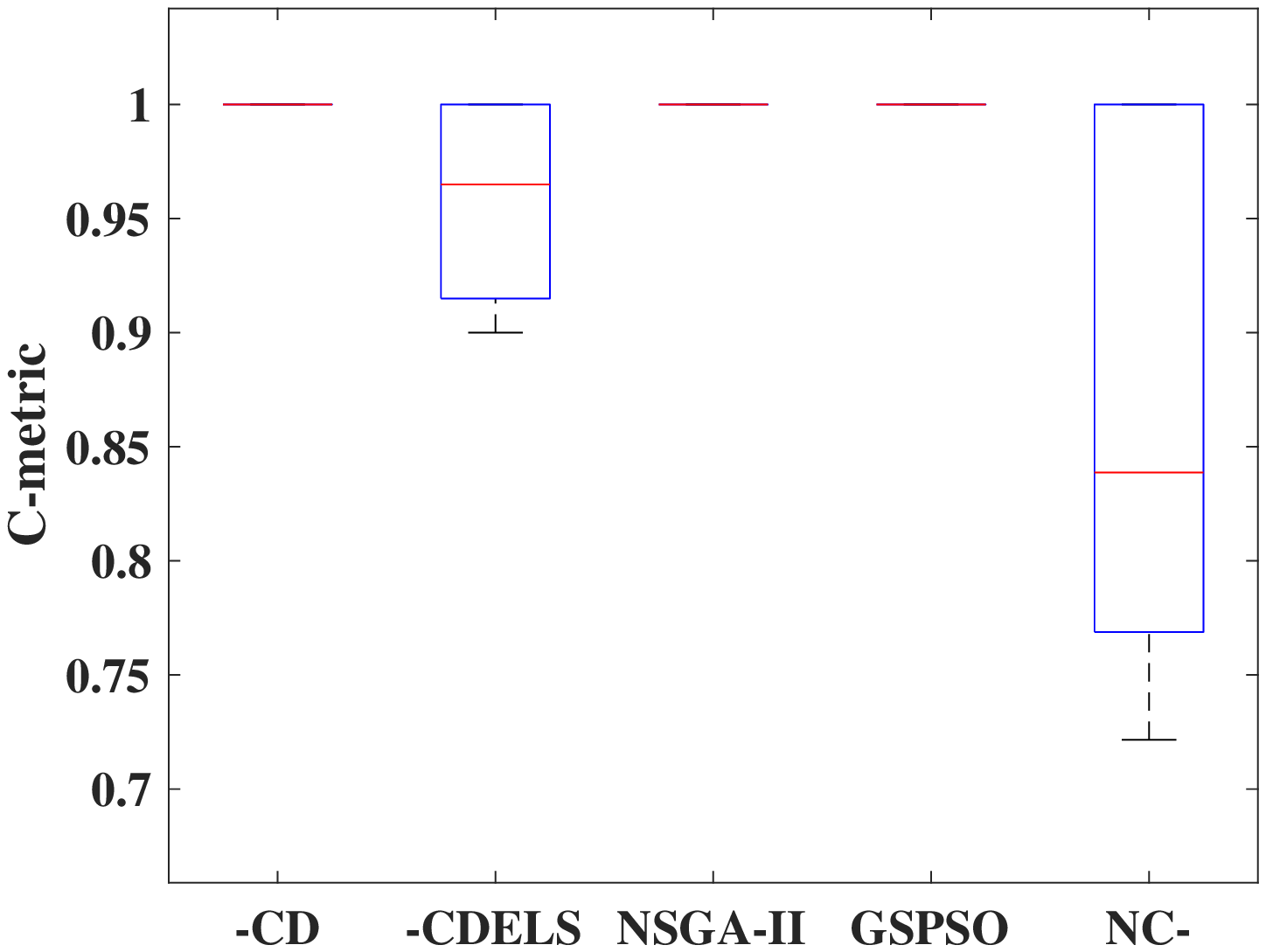}}
	\subfigure[email-enron-only]{
		\label{fig:subfig:cf} 
		\includegraphics[width=4cm,height=3cm]{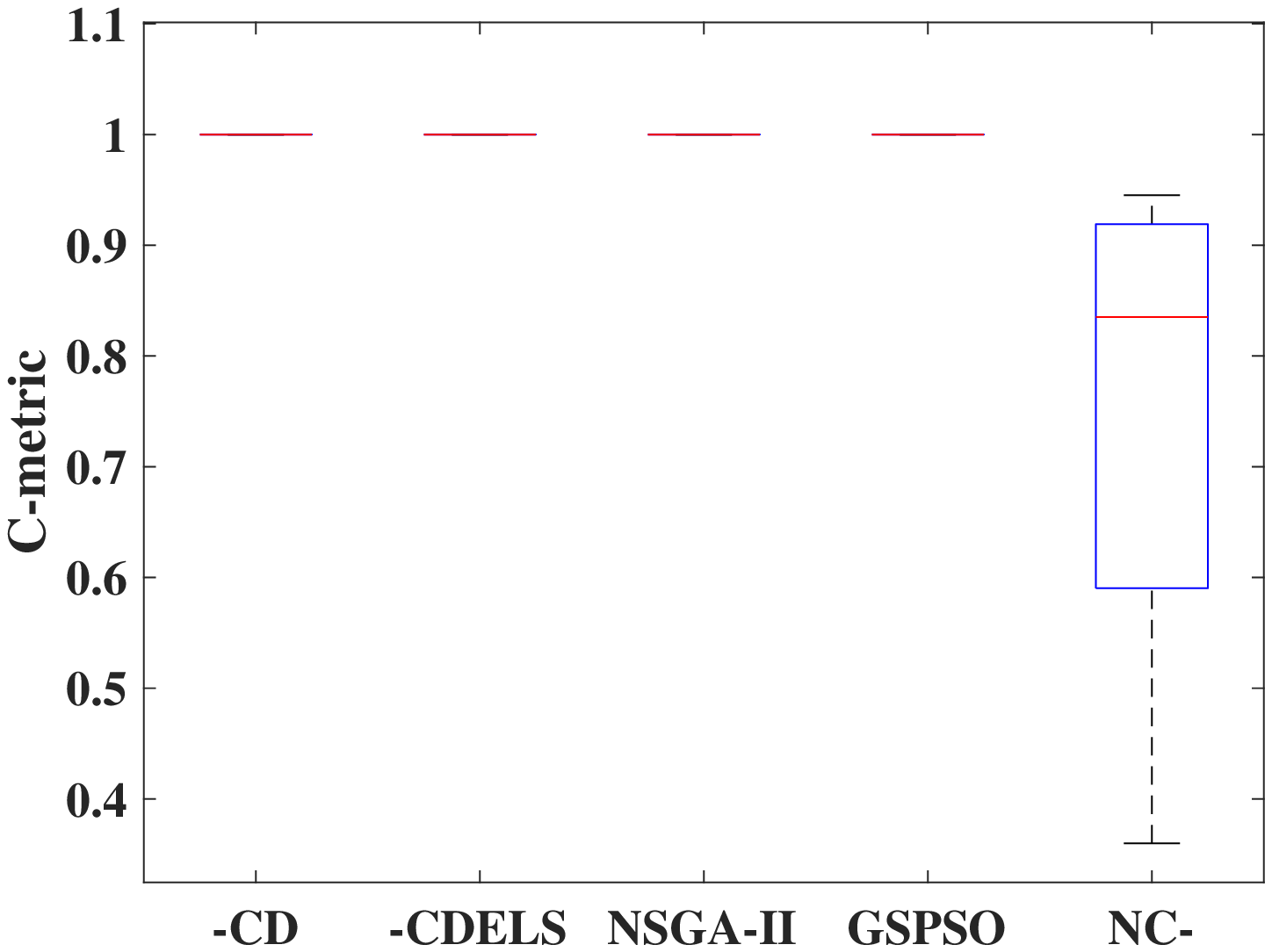}}\\
	\subfigure[chesapeake]{
		\label{fig:subfig:cg} 
		\includegraphics[width=4cm,height=3cm]{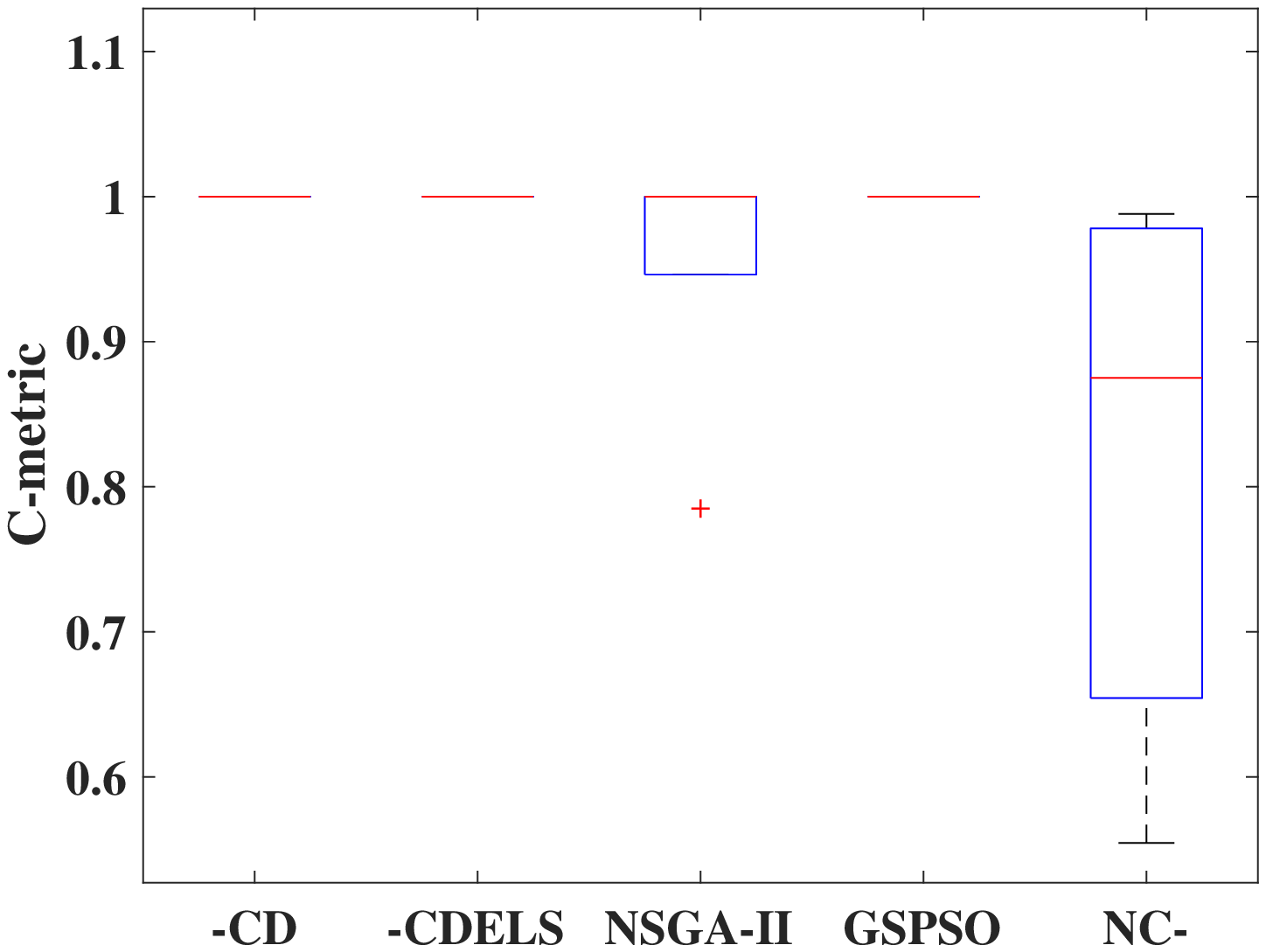}}
	\subfigure[uninett]{
		\label{fig:subfig:ch} 
		\includegraphics[width=4cm,height=3cm]{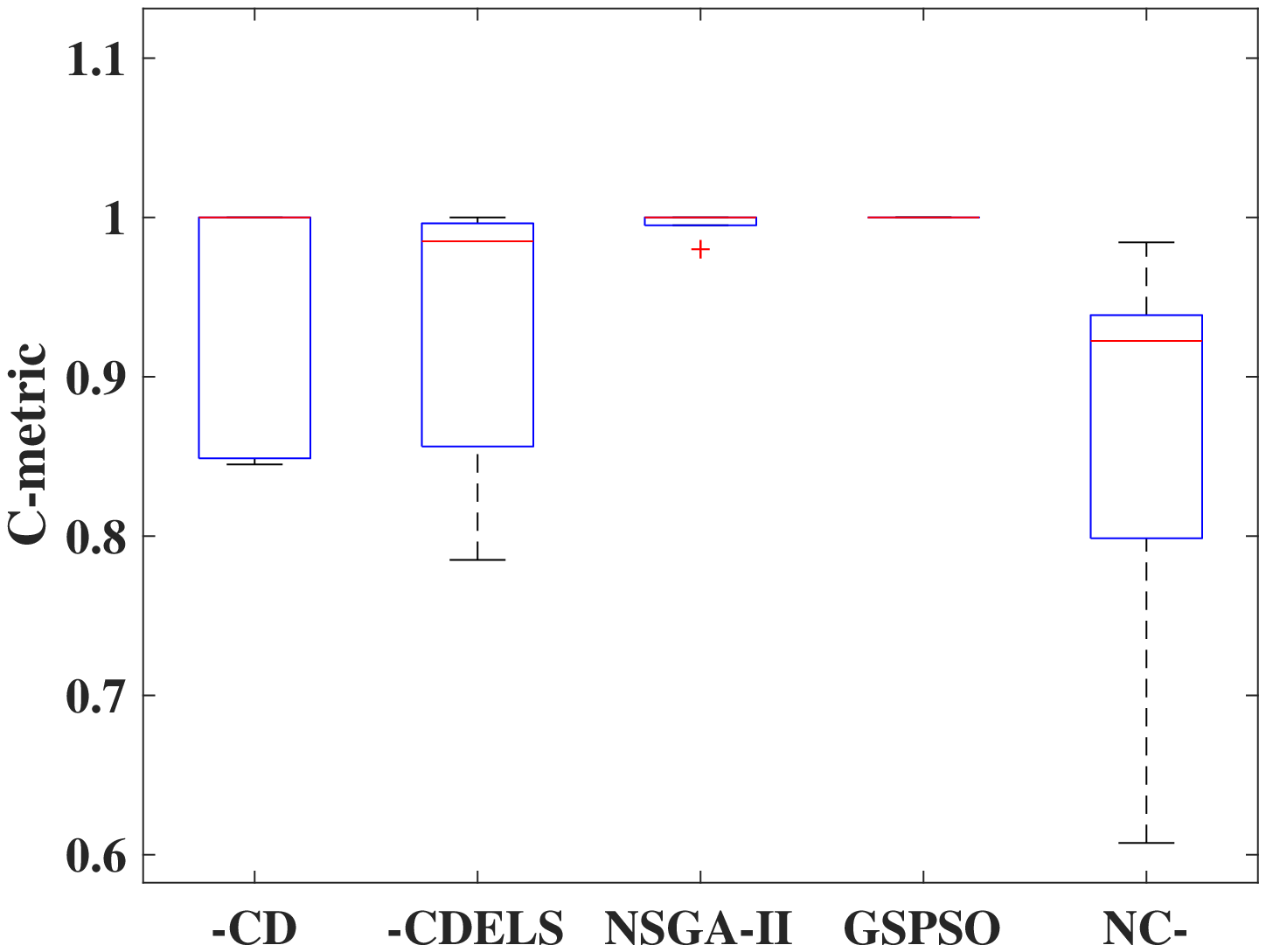}}\\
	\caption{The box-plot of the metric: C-metric for all test instances.}
	\label{fig:c} 
\end{figure}
\begin{figure}[htpb]
	\centering
	\subfigure[BA100]{
		\label{fig:subfig:ha} 
		\includegraphics[width=4cm,height=3cm]{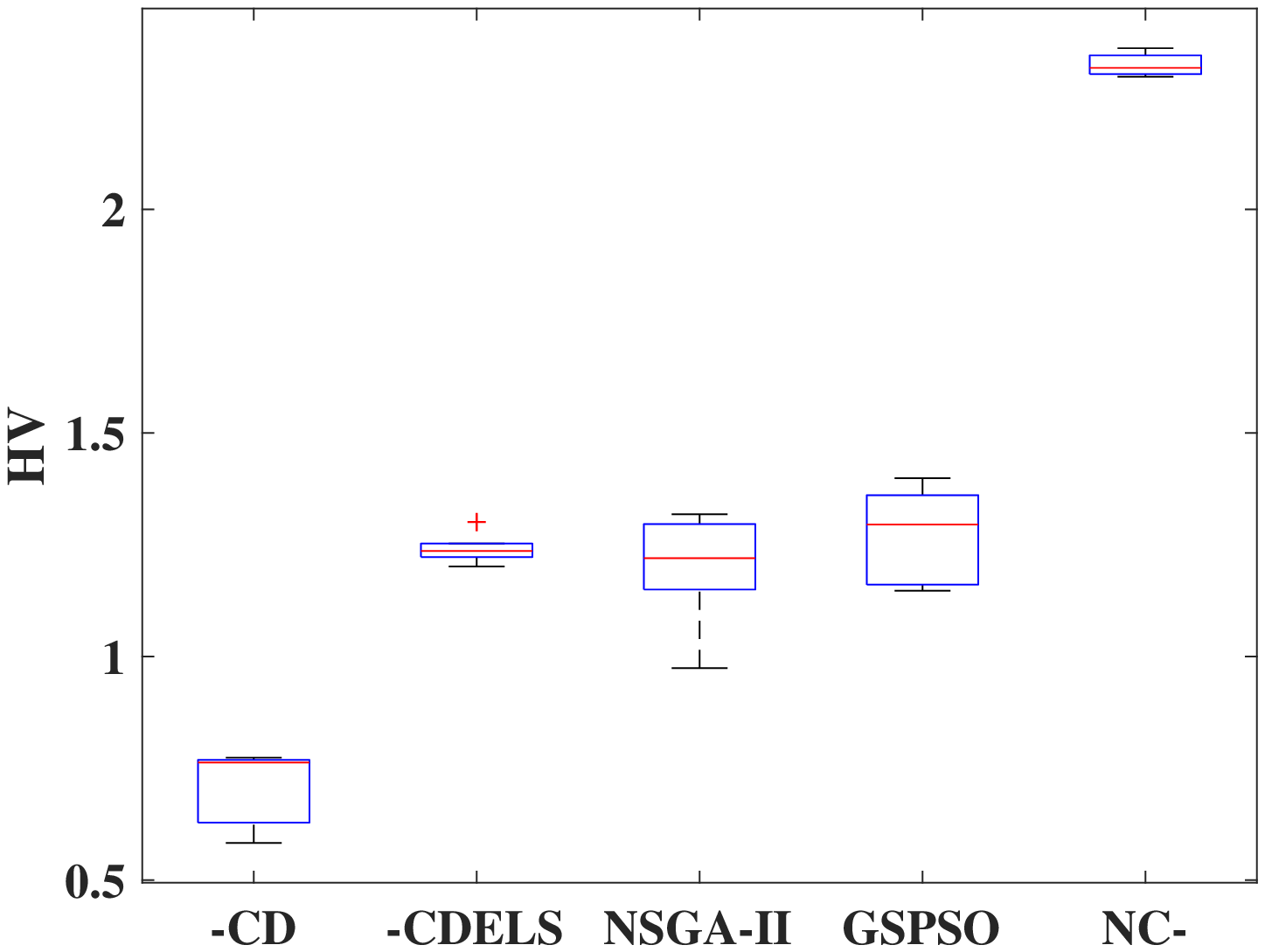}}
	\subfigure[BA300]{
		\label{fig:subfig:hb} 
		\includegraphics[width=4cm,height=3cm]{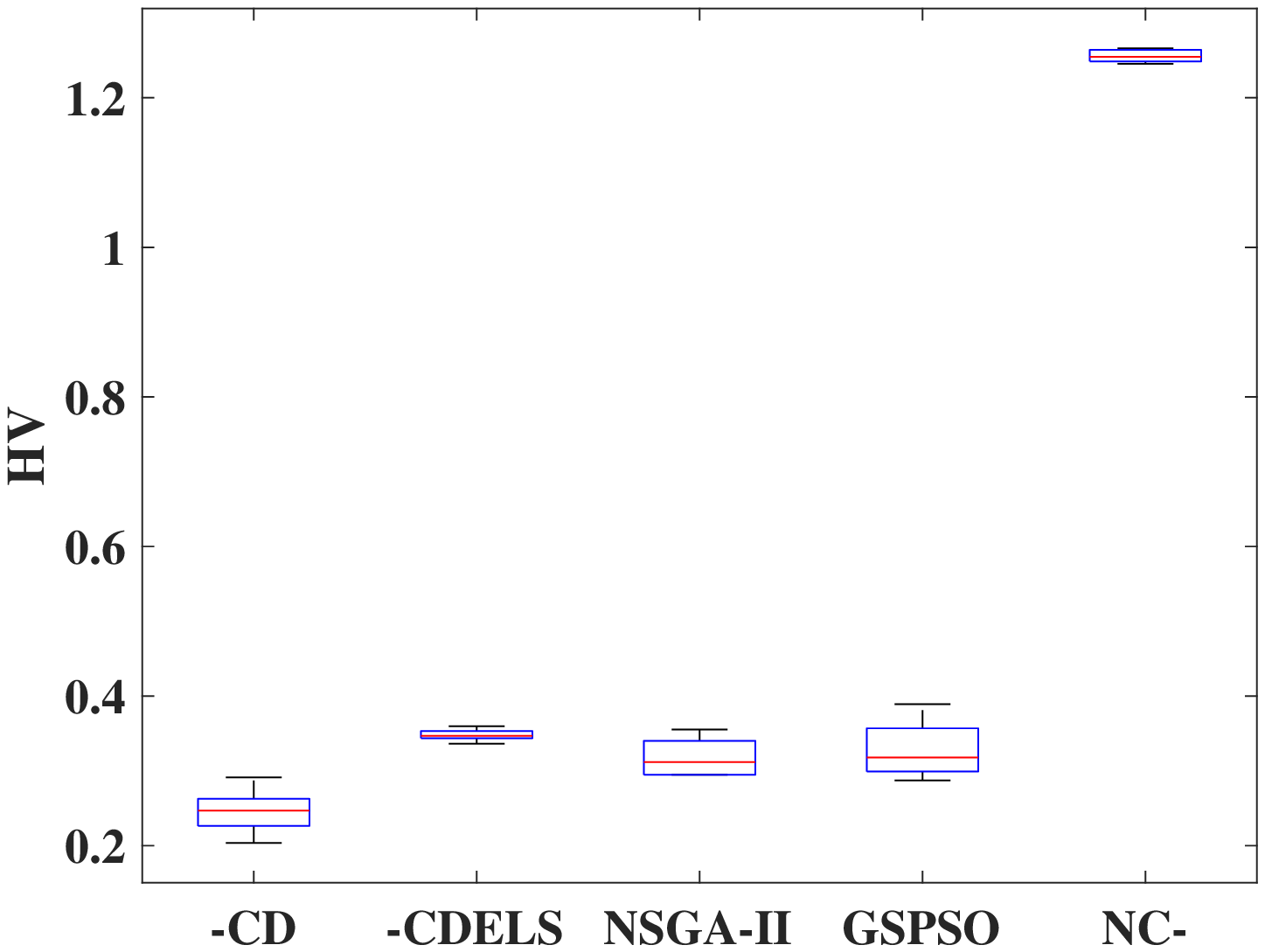}}\\
	\subfigure[WS100]{
		\label{fig:subfig:hc} 
		\includegraphics[width=4cm,height=3cm]{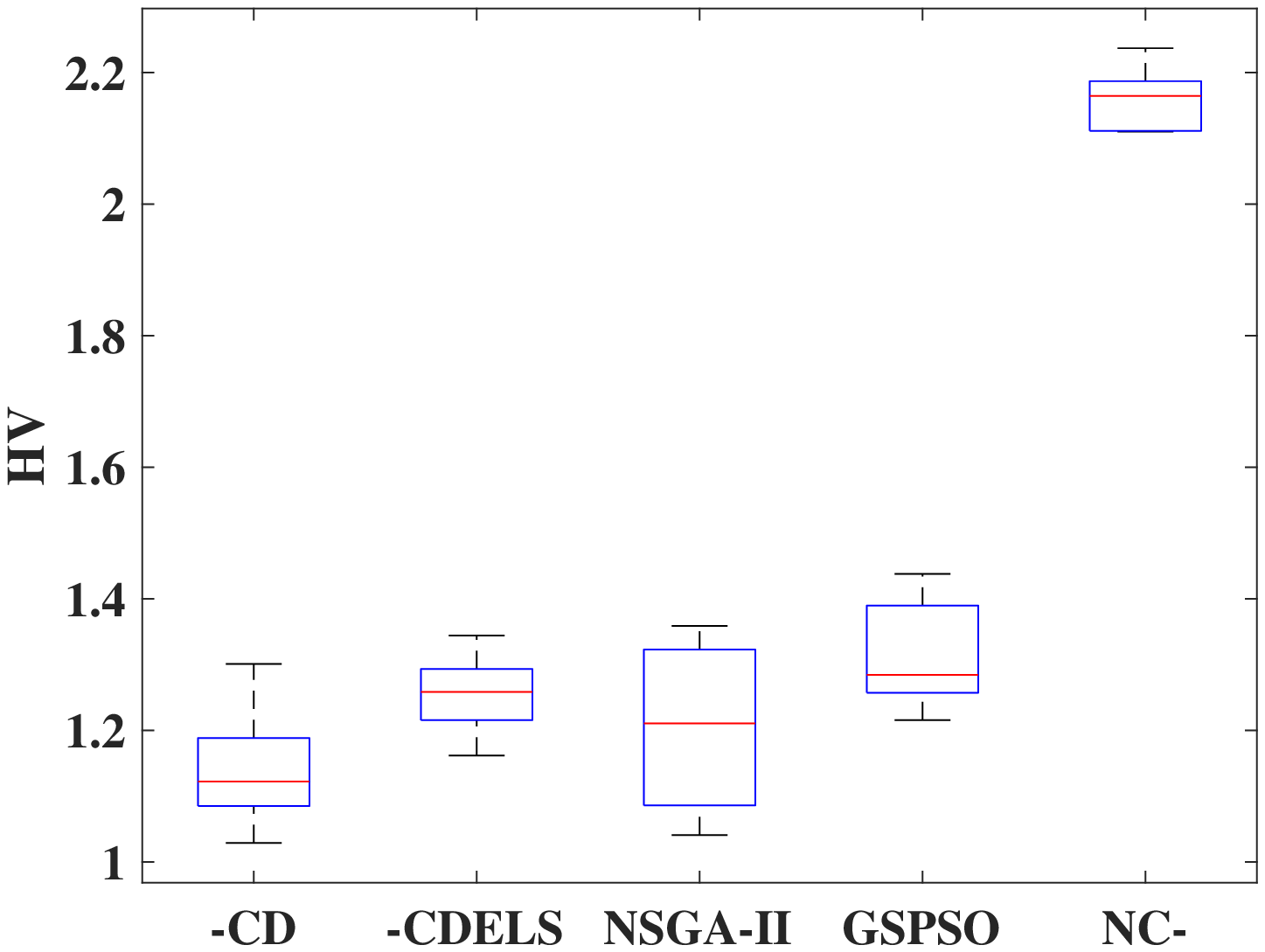}}
	\subfigure[WS300]{
		\label{fig:subfig:hd} 
		\includegraphics[width=4cm,height=3cm]{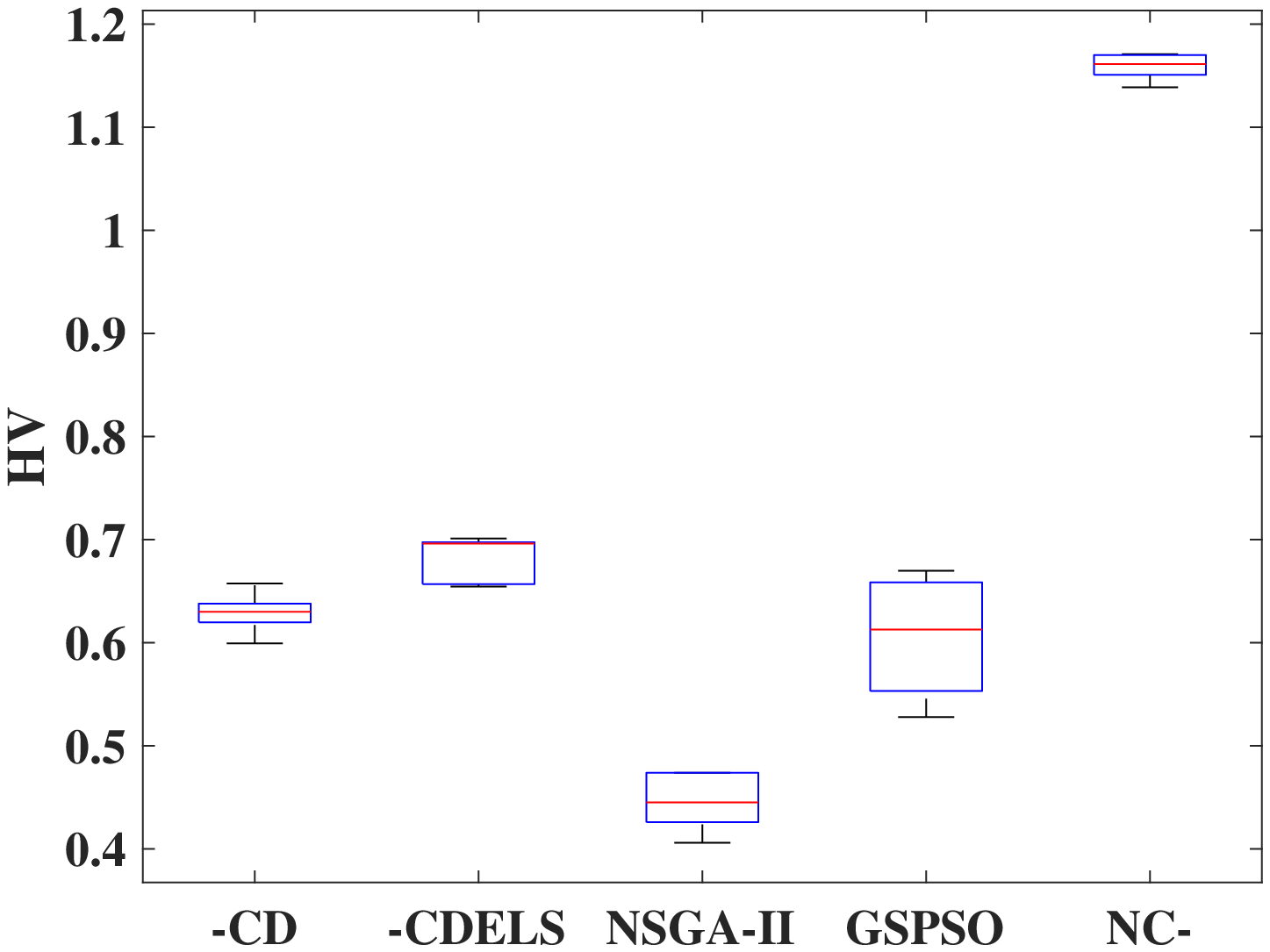}}\\
	\subfigure[118-bus]{
		\label{fig:subfig:he} 
		\includegraphics[width=4cm,height=3cm]{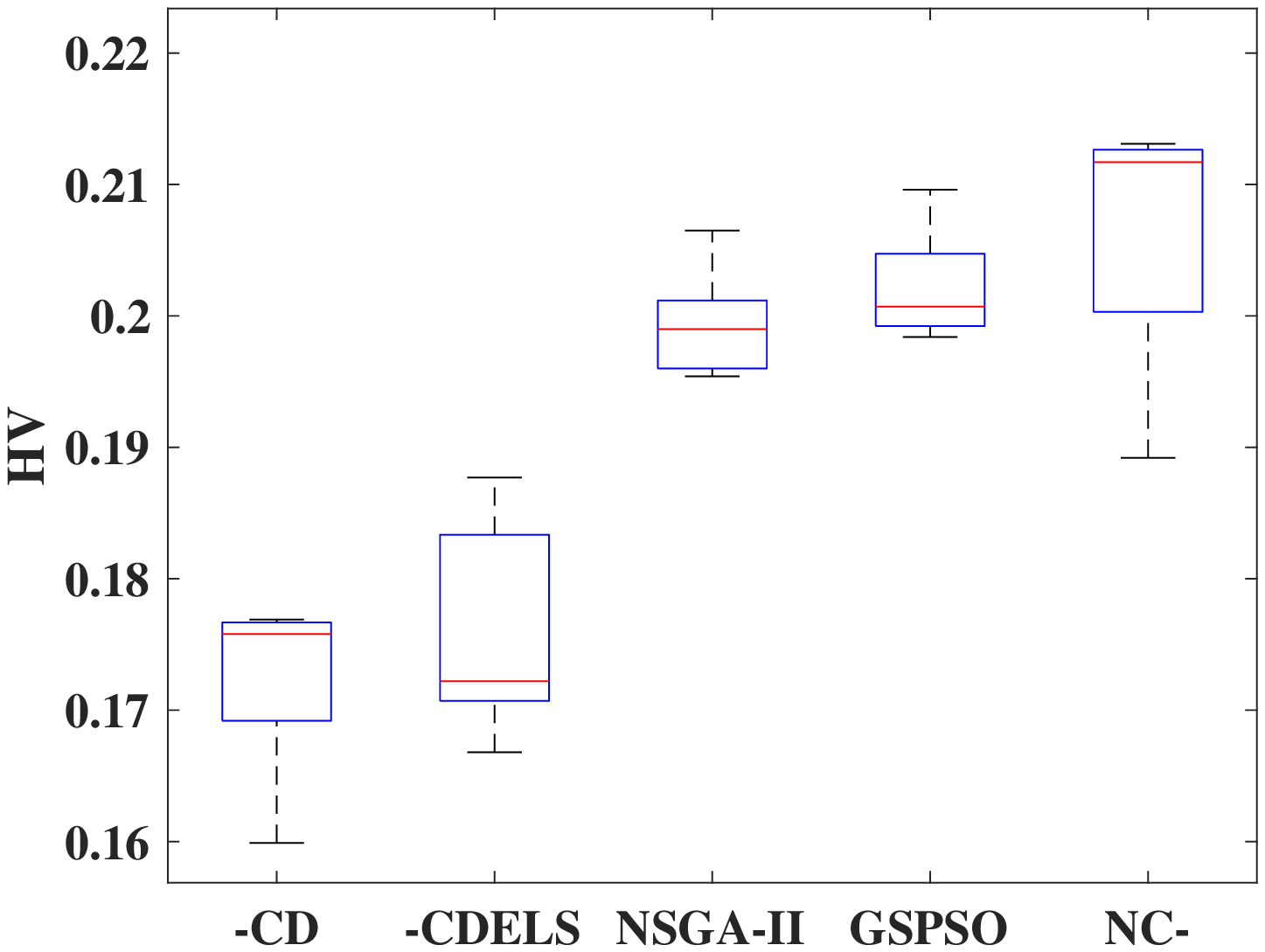}}
	\subfigure[email-enron-only]{
		\label{fig:subfig:hf} 
		\includegraphics[width=4cm,height=3cm]{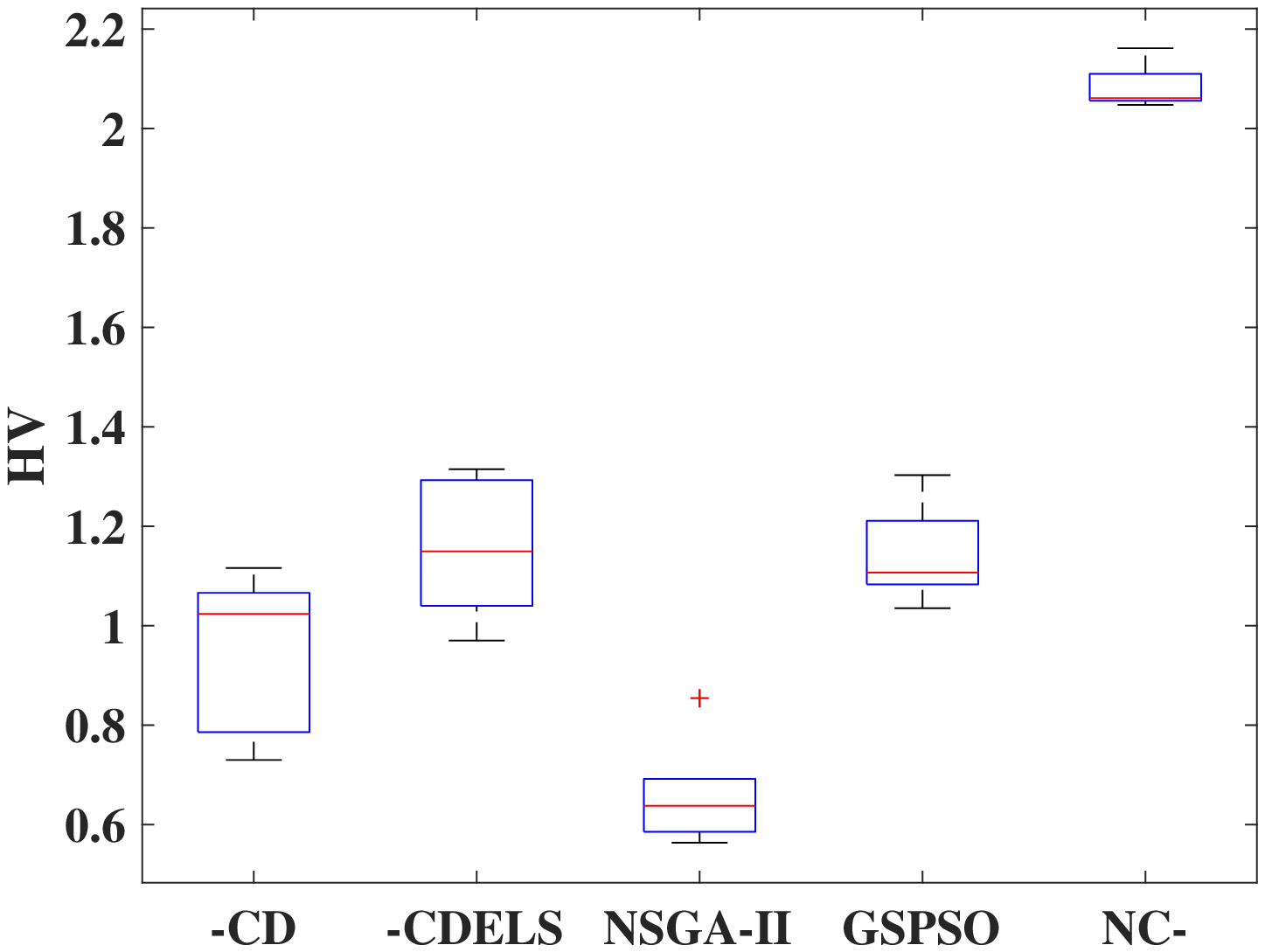}}\\
	\subfigure[chesapeake]{
		\label{fig:subfig:hg} 
		\includegraphics[width=4cm,height=3cm]{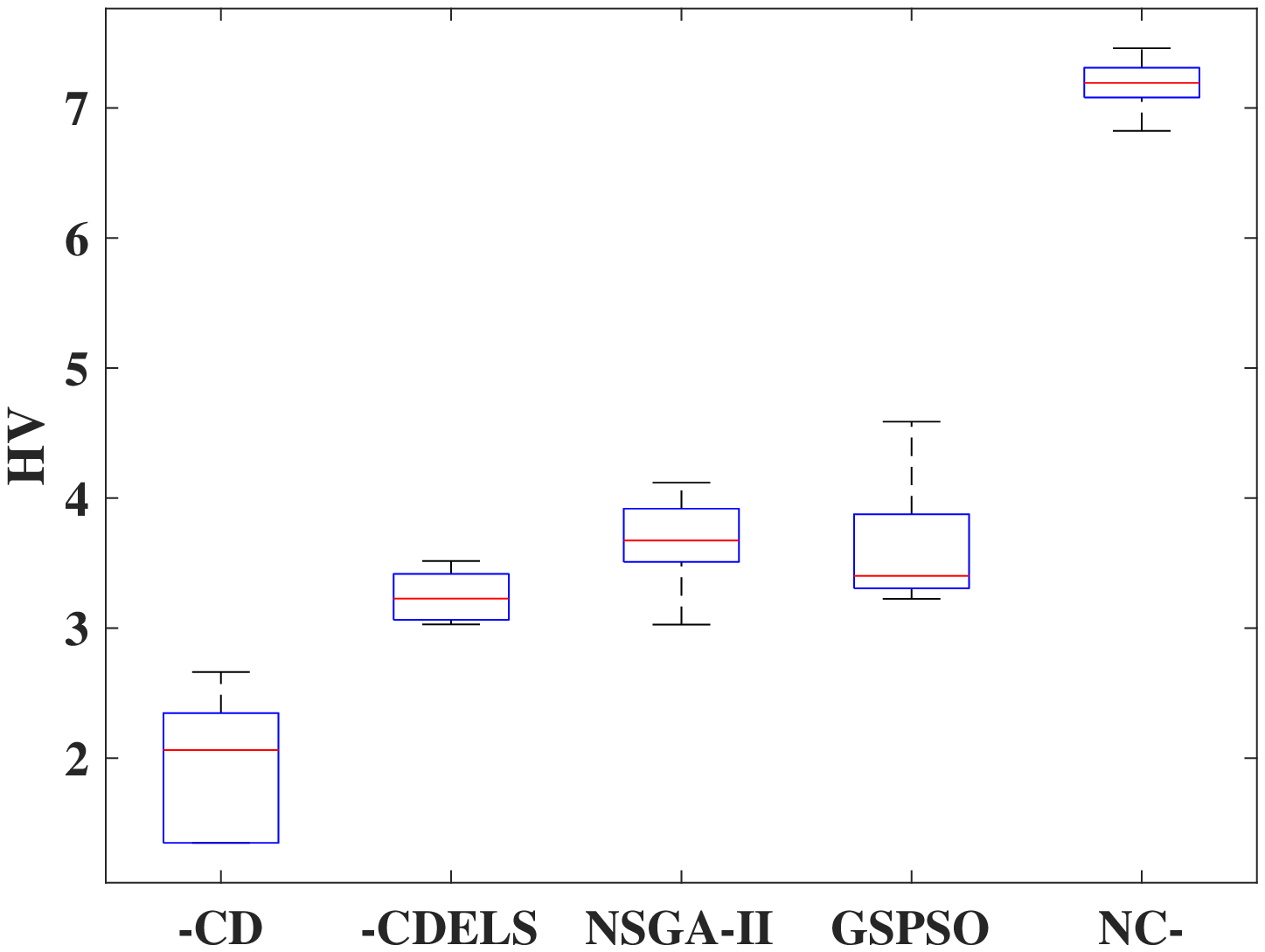}}
	\subfigure[uninett]{
		\label{fig:subfig:hh} 
		\includegraphics[width=4cm,height=3cm]{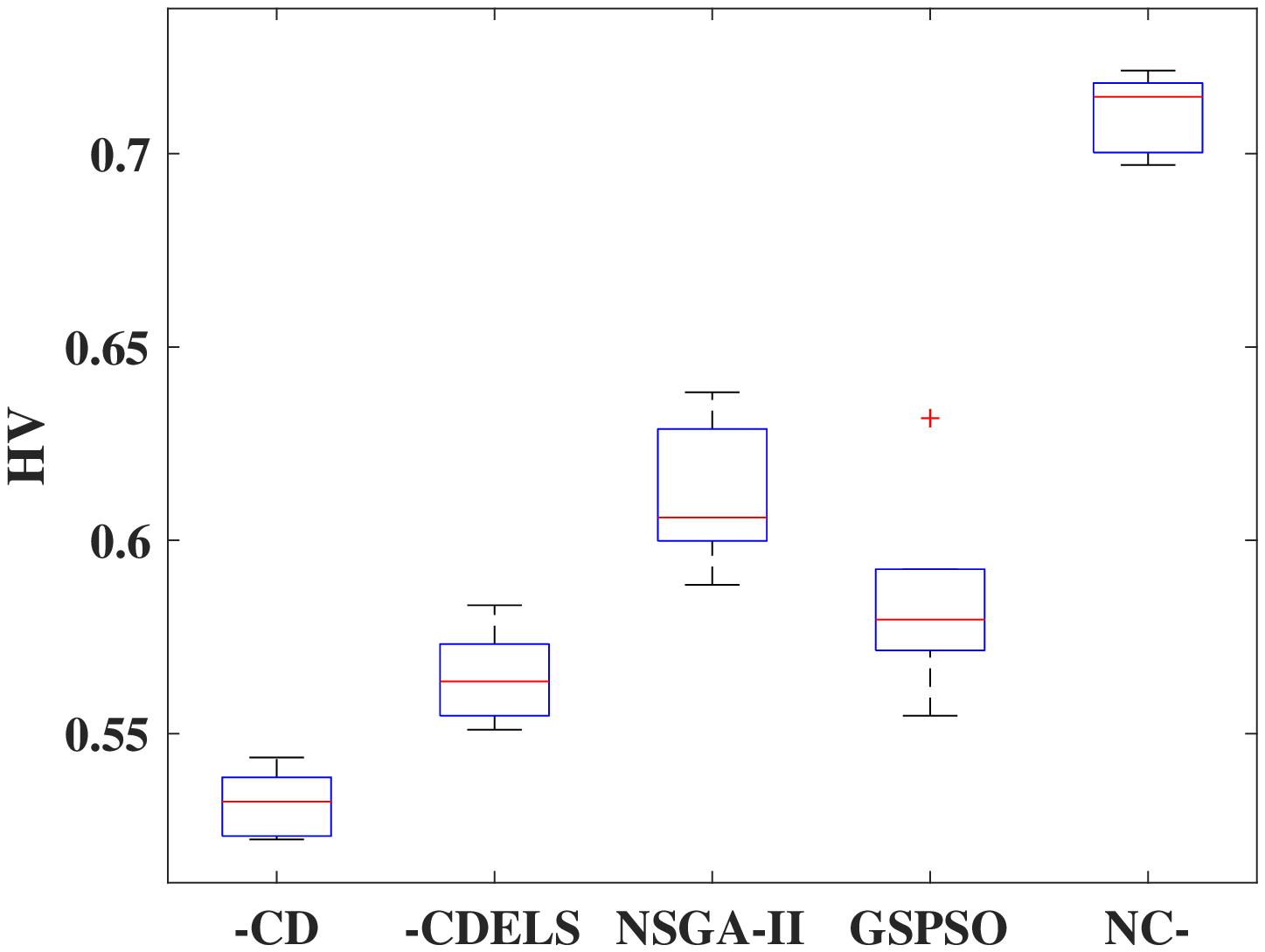}}\\
	\caption{The box-plot of the metric: HV for all test instances.}
	\label{fig:hv} 
\end{figure}
\subsection{Performance Evaluation}
The results of comparative experiments are presented in this part. Specifically, we shall first show the performance of NC-MOPSO for different values of HIR. Next, we present the effectiveness of the two improvement operators: ECHI and NCLS. Finally, we give the performance comparison between NC-MOPSO and four representative MOEAs.\\
\indent With ECHI, the initial population of NC-MOPSO is composed of heuristic and random solutions, the ratio of which is controlled by HIR. We study how the value of HV changes with HIR to obtain an adequate HIR setting. The means and standard deviations of HV over ten runs of NC-MOPSO  on eight instances are shown in Table \ref{hir}, where HIR50 means $50\%$ of initial solutions are generated heuristically and so on. We can see that the means of HIR50 and HIR100 are greater than HIR0 for all instances, which indicates that population with heuristic initialization performs better than with purely random initialization. Furthermore, HIR50 shows advantage over HIR100. For totally randomly initialization, the particles will have difficulties and are not efficient to find the best solution since the initial solution is, with high probability, far from the best one.  While in the case of purely heuristic initialization, the particles are prone to fall into the local optimum. The hybrid initialization  with heuristic and random solutions, such as ECHI, make a necessary compromise between efficiency and diversity. Therefore, we fix HIR value as $50\%$  in the following experiments. \\
\begin{table*}[htpb]
	\caption{Performance comparison among MOPSOCD, MOPSOCDELS, NSGA-\uppercase\expandafter{\romannumeral2}, GSPSO and NC-MOPSO in terms of IGD (mean[ranking](std)).}
	\label{four}
	\begin{center}
		\begin{tabular}{clllll}
			\hline  		
			Instance &  MOPSOCD &	MOPSOCDELS & NSGA-\uppercase\expandafter{\romannumeral2} & GSPSO & NC-MOPSO \\		
			\hline
			BA100  & 0.1675$\dagger$[5](0.0183) & 0.0503$\dagger$[2](0.0088) & 0.0641$\dagger$[3](0.0053) & 0.0873$\dagger$[4](0.0142) & \textbf{0.0032[1](0.0005)}	\\
			\hline
			BA300 & 0.4072$\dagger$[5](0.0268) & 0.2728$\dagger$[3](0.0031) & 0.2404$\dagger$[2](0.0155)  & 0.3808$\dagger$[4](0.0315) & \textbf{0.0059[1](0.0005)}\\
			\hline
			WS100 & 0.0364$\dagger$[3](0.0051) & 0.0252$\dagger$[2](0.0049) & 0.1376$\dagger$[5](0.0014)  & 0.1001$\dagger$[4](0.0232)  & \textbf{0.0059[1](0.0014)}\\
			\hline
			WS300 & 0.1336$\dagger$[3](0.0028) & 0.0779$\dagger$[2](0.0043) & 0.3691$\dagger$[5](0.0169) & 0.1671$\dagger$[4](0.0391)& \textbf{0.0030[1](0.0015)}\\
			\hline
			118-bus & 0.0109$\approx$[3](0.0081) & 0.0048$\approx$[2](0.0038) & 0.0125$\dagger$[4](0.0056) &0.0359$\dagger$[5](0.0171)& \textbf{0.0042[1](0.0031)}\\
			\hline
			email-enron-only & 0.0979$\dagger$[3](0.0195) & 0.0577$\dagger$[2](0.0239) & 0.7506$\dagger$[5](0.9850)&0.2134$\dagger$[4](0.0363) & \textbf{0.0154[1](0.0167)}\\
			\hline
			chesapeake &0.1333$\dagger$[4](0.0435) & 0.0633$\dagger$[2](0.0072)&0.1437$\dagger$[5](0.0099)&0.0972$\dagger$[3](0.0283)&\textbf{0.0049[1](0.0017)}\\
			\hline
			uninett &0.0136$\dagger$[4](0.0009)&0.0076$\dagger$[3](0.0039)&0.0053$\dagger$[2](0.0025)&0.0262$\dagger$[5](0.0092)&\textbf{0.0009[1](0.0003)}\\
			\hline
			$\dagger/\S/\approx$  & 7/0/1 & 7/0/1 &	8/0/0 & 8/0/0  &    \\	
			\hline
		\end{tabular}
	\end{center}
\end{table*}
\begin{table*}[htpb]
	\caption{Performance comparison among MOPSOCD, MOPSOCDELS, NSGA-\uppercase\expandafter{\romannumeral2}, GSPSO and NC-MOPSO in terms  of HV (mean[ranking](std)).}
	\label{five}
	\begin{center}
		\begin{tabular}{clllll}
			\hline  		
			Instance &  MOPSOCD &	MOPSOCDELS & NSGA-\uppercase\expandafter{\romannumeral2} & GSPSO & NC-MOPSO \\		
			\hline
			BA100  & 0.7063$\dagger$[5](0.0783) & 1.2355$\dagger$[4](0.0540) & 1.2365$\dagger$[3](0.1246) &1.2709$\dagger$[2](0.1112)& \textbf{2.3235[1](0.0262)}	\\
			\hline
			BA300 & 0.2465$\dagger$[5](0.0293) & 0.3480$\dagger$[2](0.0085) & 0.3183$\dagger$[4](0.0264) & 0.3281$\dagger$[3](0.0435)& \textbf{1.2558[1](0.0087)}\\
			\hline
			WS100 & 1.1345$\dagger$[5](0.0778) & 1.2504$\dagger$[3](0.0556) & 1.2064$\dagger$[4](0.1188) & 1.3167$\dagger$[2](0.0885)&\textbf{2.1558[1](0.0521)}\\
			\hline
			WS300 & 0.6289$\dagger$[3](0.0205) & 0.6811$\dagger$[2](0.0229) & 0.4462$\dagger$[5](0.0288) &0.6049$\dagger$[4](0.0654)& \textbf{1.1591[1](0.0131)}\\
			\hline
			118-bus & 0.1772$\dagger$[5](0.0046) & 0.1877$\dagger$[4](0.0142) & 0.2073$\approx$[2](0.0166)& 0.2022$\approx$[3](0.0045)&  \textbf{0.2061[1](0.0101)}\\
			\hline
			email-enron-only & 0.9448$\dagger$[4](0.1675) & 1.1566$\dagger$[2](0.1459) & 0.6574$\dagger$[5](0.1145) & 1.1449$\dagger$[3](0.1023)&\textbf{2.0842[1](0.0462)}\\
			\hline
			chesapeake &1.9334$\dagger$[5](0.5761) & 3.2464$\dagger$[4](0.2051)&3.6687$\dagger$[2](0.4027)&3.6380$\dagger$[3](0.5523)&\textbf{7.1809[1](0.2299)}\\
			\hline
			uninett &0.5382$\dagger$[5](0.0105)&0.5651$\dagger$[4](0.0123)&0.6142$\dagger$[2](0.0196)&0.5845$\dagger$[3](0.0.0284)&\textbf{0.7103[1](0.0105)}\\
			\hline
			$\dagger/\S/\approx$  &8/0/0 & 8/0/0 & 7/0/1 & 7/0/1  &	     \\	
			\hline
		\end{tabular}
	\end{center}
\end{table*}
\begin{table}[htpb]
	\caption{Overall performance comparison of the five algorithms on the eight instances.}
	\label{overall}
	\begin{center}
		\begin{tabular}{lll}
			\hline  		
			& 	Mean ranking &	Total $\dagger/\S/\approx$ \\		
			\hline
			MOPSOCD  & 4.1875 & 15/0/1	\\
			\hline
			MOPSOCDELS & 2.6875	& 15/0/1\\
			\hline
			NSGA-\uppercase\expandafter{\romannumeral2} & 3.6250 & 15/0/1\\
			\hline
			GSPSO & 3.5000 &15/0/1 \\
			\hline
			NC-MOPSO & 1.0000 & -\\
			\hline
		\end{tabular}
	\end{center}
\end{table}
\indent In order to enhance the quality of solutions, NC-MOPSO incorporates two improvement operators: ECHI (see Section \ref{echi}) and NCLS (see Section \ref{ncls}). We present the effectiveness of these two operators through comparing NC-MOPSO with MOPSOCD and MOPSOCD\_in.  MOPSOCD is the original algorithm without any improvement, and MOPSOCD\_in is an improved version of MOPSOCD by employing ECHI. NC-MOPSO is also an improved version of MOPSOCD, which utilizes  both ECHI and NCLS.  Table~\ref{zs} shows the means and standard deviations of IGD over ten runs of these three algorithms. We can observe that in all instances,  IGD of MOPSOCD\_in is smaller than MOPSOCD, which demonstrates the effectiveness of ECHI. In other words, the performance of MOPSOCD with ECHI is better than  with random initialization. Furthermore, we  find that  IGD of NC-MOPSO is smaller than MOPSOCD\_in, which validates the effectiveness of NBLS.  Fig.~\ref{zs_1} presents the distribution of nondominated solutions of these three algorithms on instance BA100. It is obviously shown that NC-MOPSO can achieve better solutions than MOPSOCD and MOPSOCD\_in, which further confirms the effectiveness of the two proposed improvement operators.\\
\indent Next, we compare NC-MOPSO with the four baseline algorithms for the three popular metrics: IGD, C-metric and HV. For each algorithm, all results are obtained by averaging  ten independent runs. The statistical results for the metrics IGD and HV are presented in Table \ref{four} and \ref{five}, respectively. For a network instance, the ranking of the  performance of each algorithm is given  in  square brackets, while the standard deviation of the performance is   in  parentheses. We conduct the Wilixon's rank sum test at a $5\%$ significance level to test the significance of the differences between the mean metric values yielded by NC-MOPSO and the other comparison algorithms. The symbols $\dagger$, $\S$ and $\approx$ indicate that the performance of NC-MOPSO is better than, worse than and similar to the baseline algorithms, respectively. Table \ref{overall} shows the overall performance of all algorithms via the Wilixon's rank sum test. Figs.~\ref{fig:igd},~\ref{fig:c} and~\ref{fig:hv} present respectively the box plots of IGD, C-metric and HV of each algorithm for the eight test instances, where ``-CD'', ``-CDELS'' and ``NC-'' represent MOPSOCD, MOPSOCDELS and NC-MOPSO, respectively.\\
\begin{figure}[htpb]
	\centering
	\subfigure[BA100]{
		\label{fig:subfig:a} 
		\includegraphics[width=4cm,height=3cm]{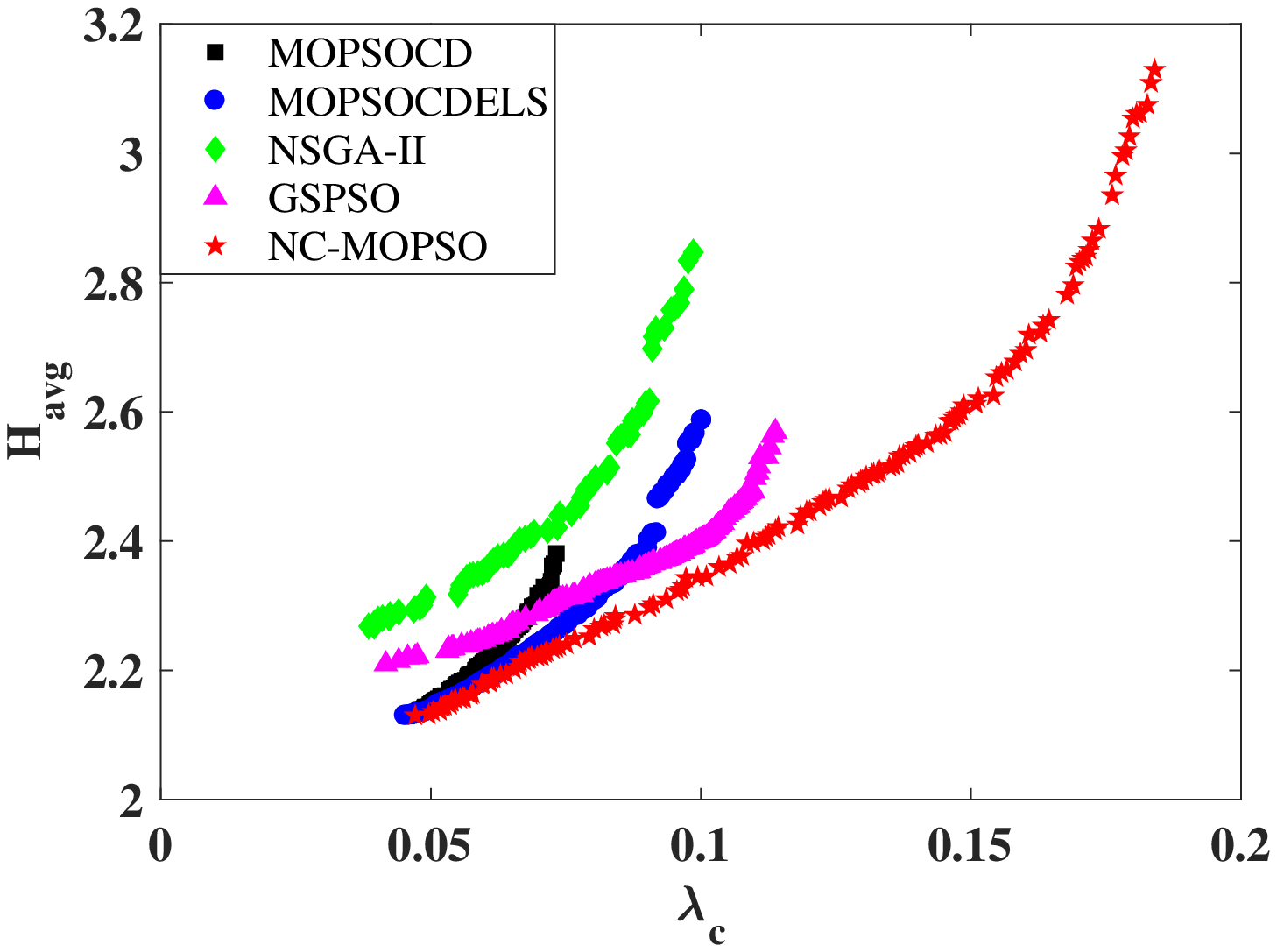}}
	\subfigure[BA300]{
		\label{fig:subfig:b} 
		\includegraphics[width=4cm,height=3cm]{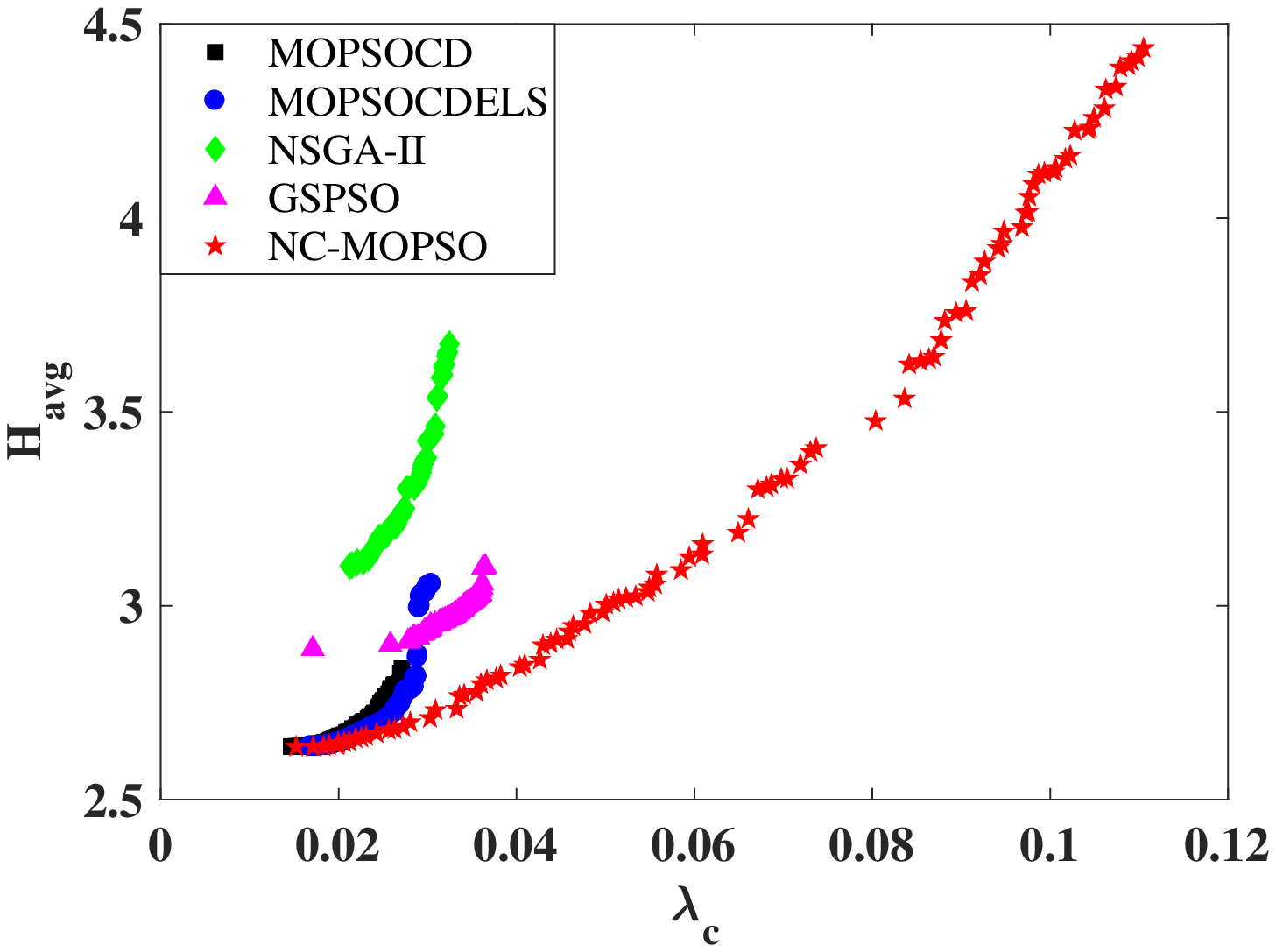}}\\
	\subfigure[WS100]{
		\label{fig:subfig:c} 
		\includegraphics[width=4cm,height=3cm]{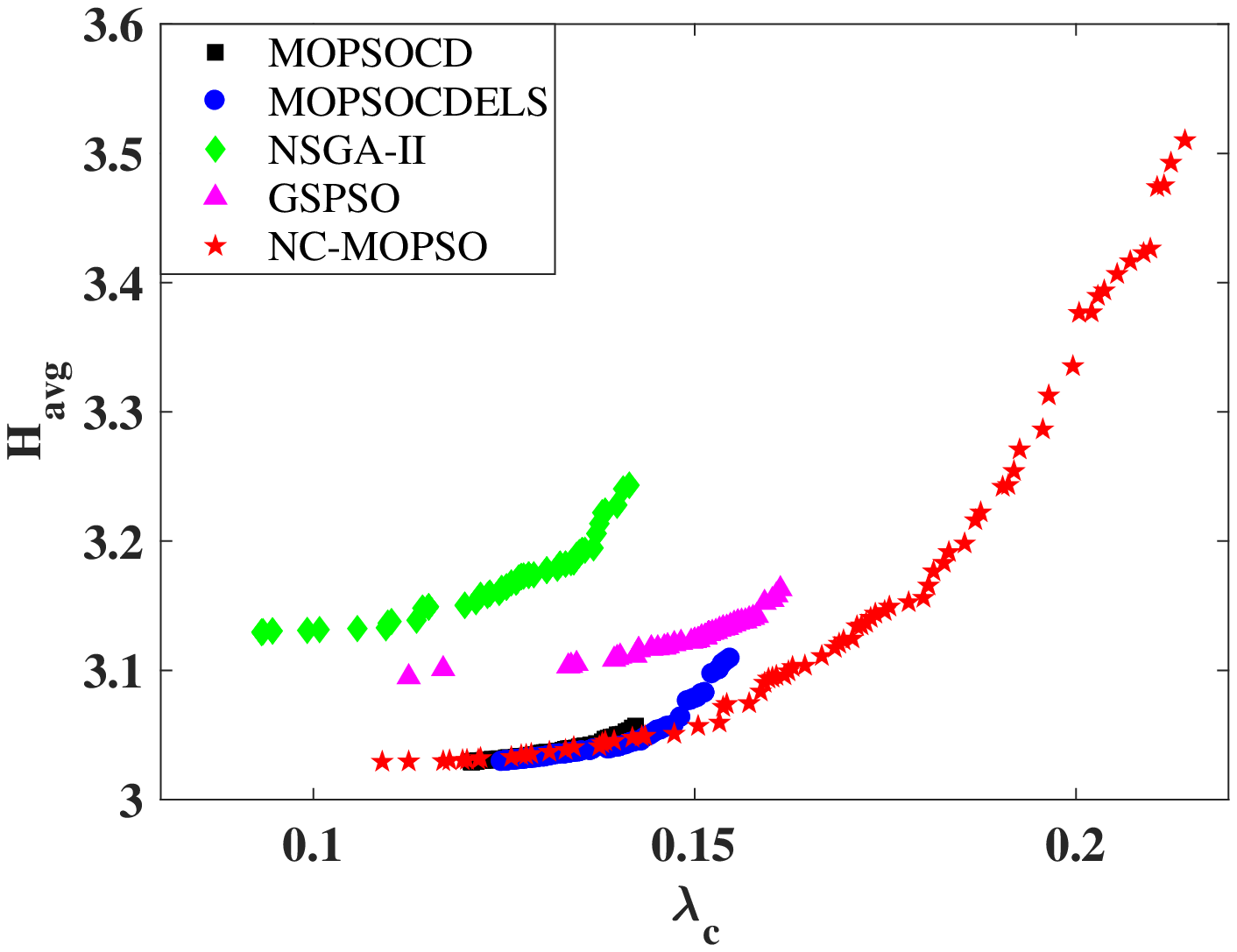}}
	\subfigure[WS300]{
		\label{fig:subfig:d} 
		\includegraphics[width=4cm,height=3cm]{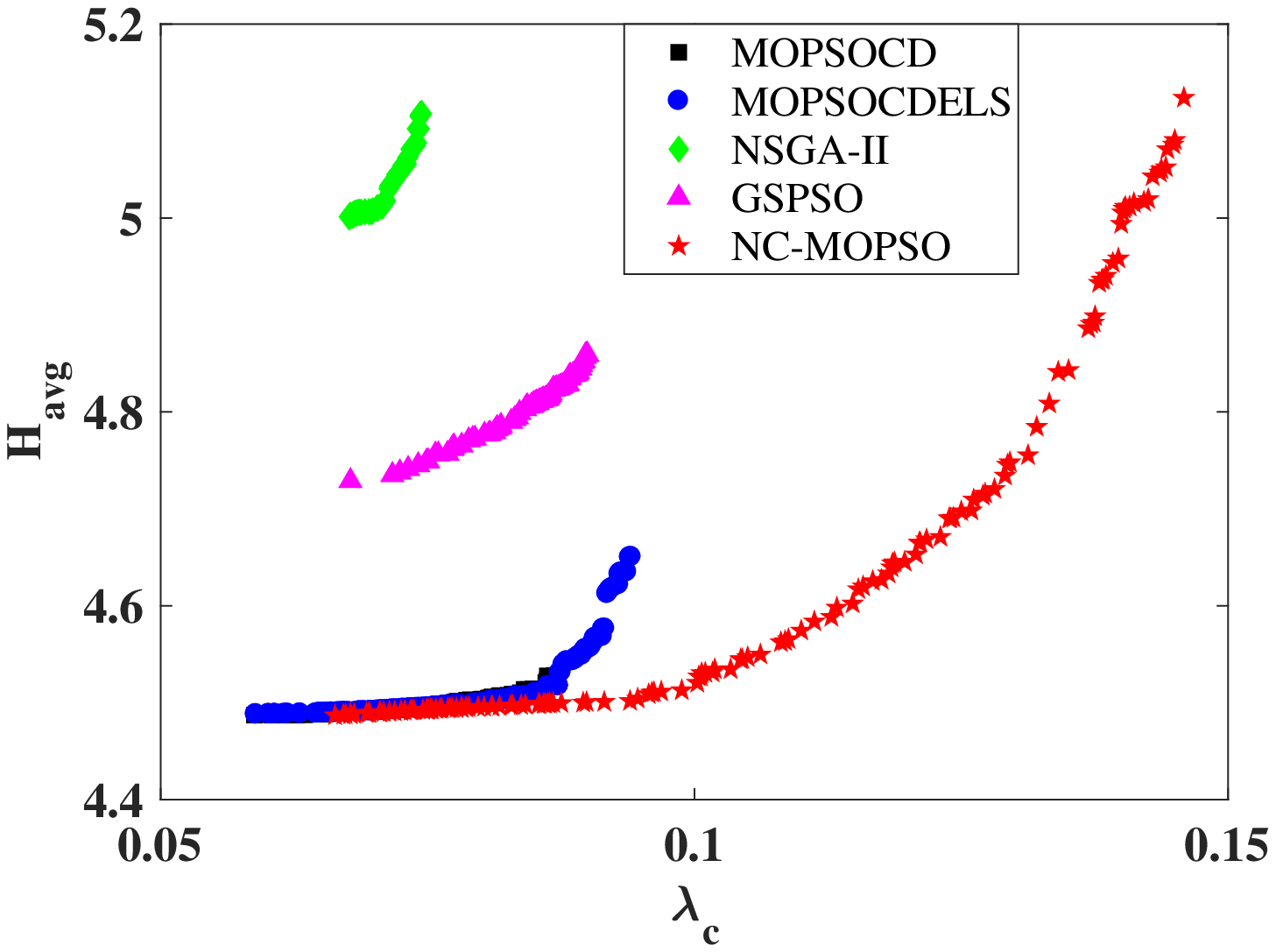}}\\
	\subfigure[118-bus]{
		\label{fig:subfig:e} 
		\includegraphics[width=4cm,height=3cm]{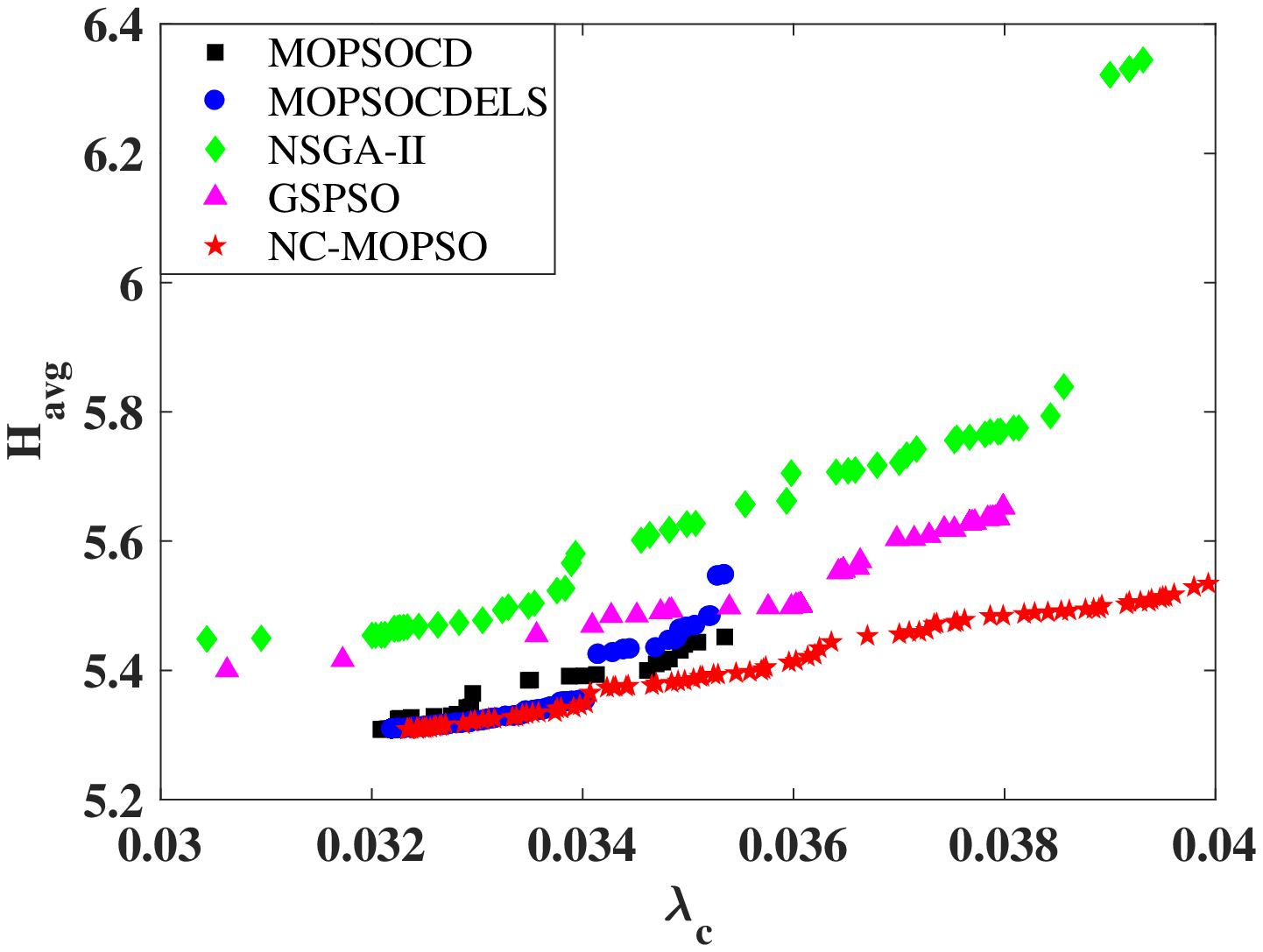}}
	\subfigure[email-enron-only]{
		\label{fig:subfig:f} 
		\includegraphics[width=4cm,height=3cm]{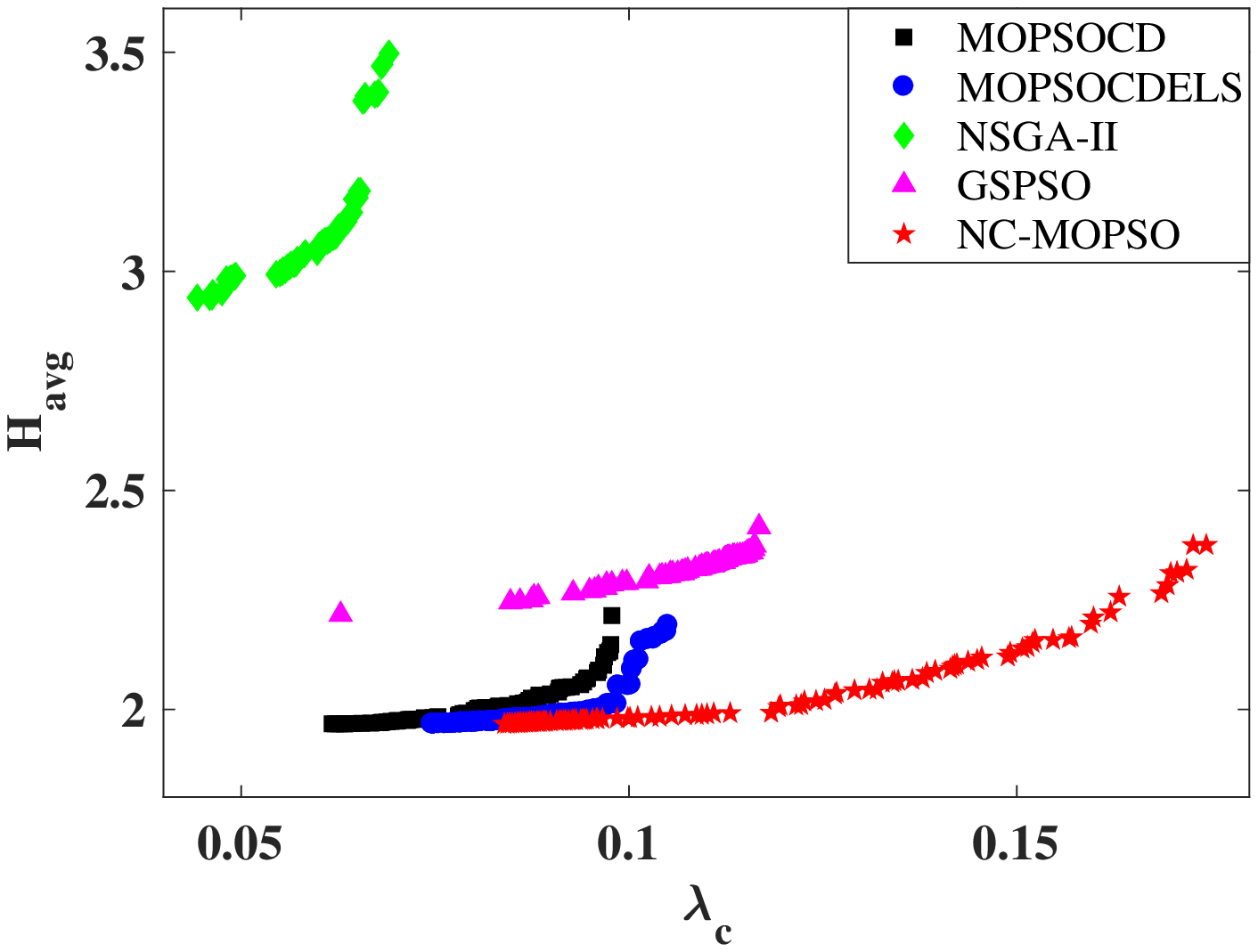}}\\
	\subfigure[chesapeake]{
		\label{fig:subfig:g} 
		\includegraphics[width=4cm,height=3cm]{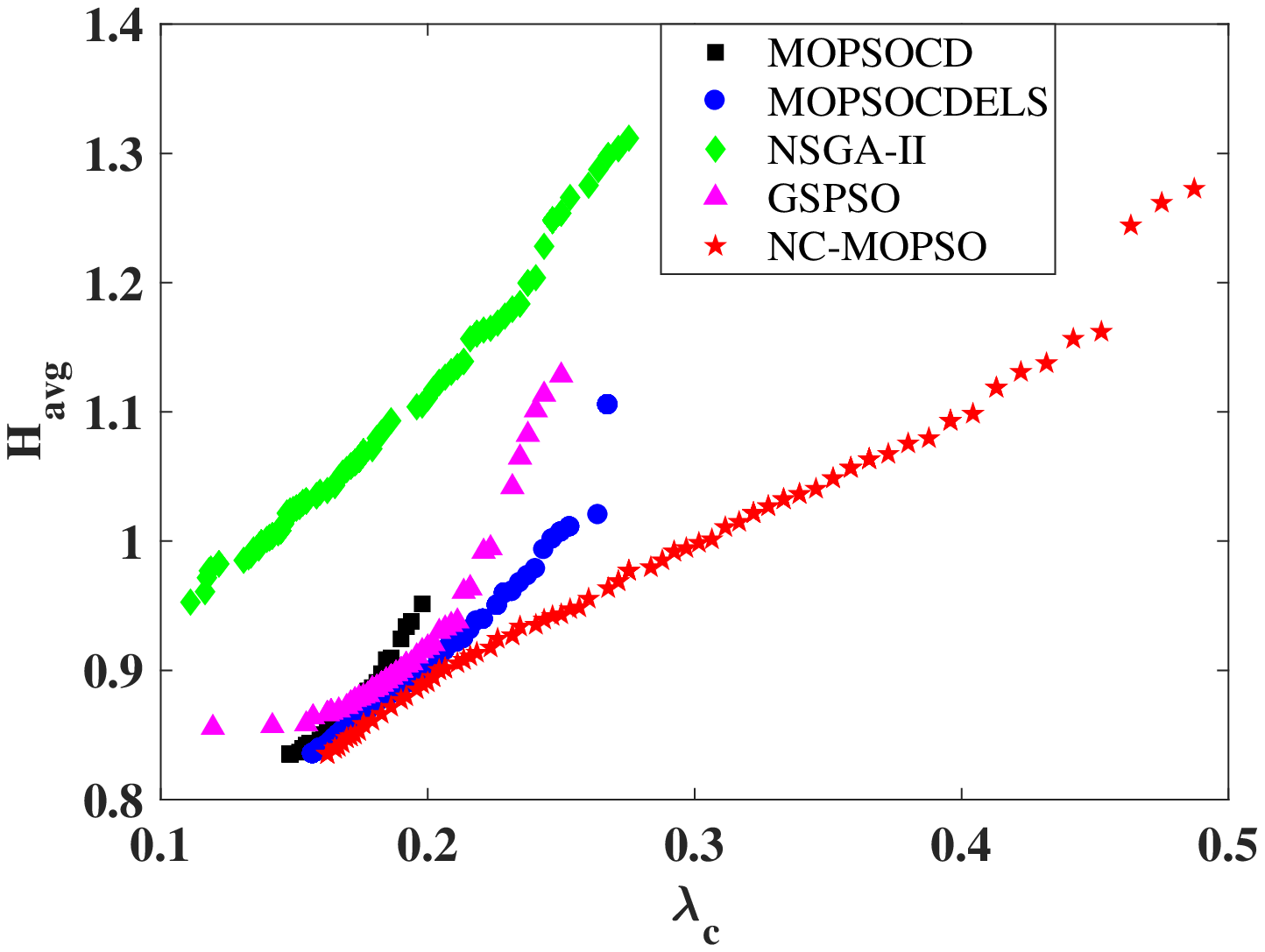}}
	\subfigure[uninett]{
		\label{fig:subfig:h} 
		\includegraphics[width=4cm,height=3cm]{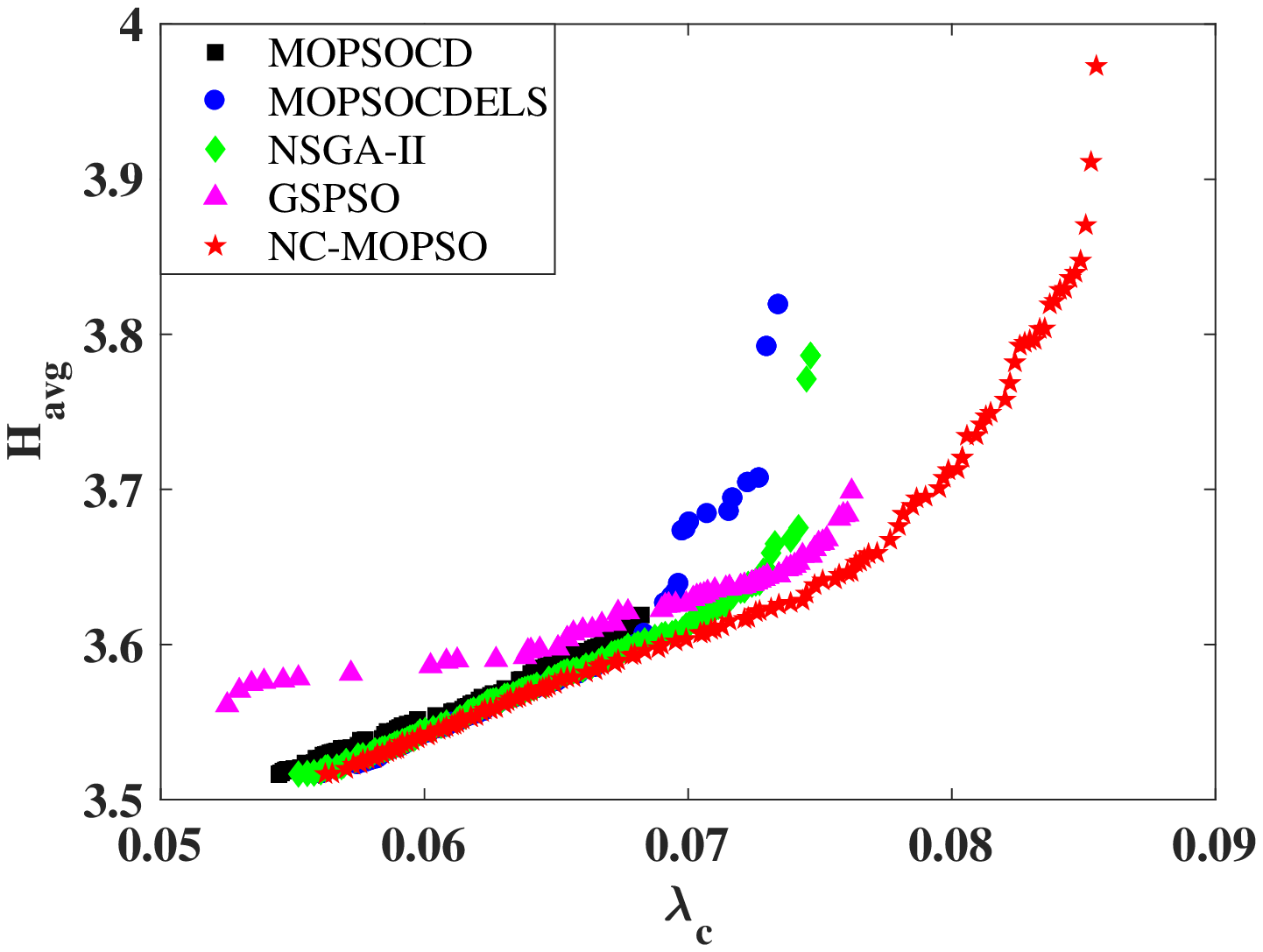}}\\
	\caption{Nondominated solutions (PFs) of all the algorithms for each instance.}
	\label{fig:fig1} 
\end{figure}
\begin{figure}[htpb]
	\centering
	\subfigure[BA100]{
		\label{fig:subfig:3a} 
		\includegraphics[width=4cm,height=3cm]{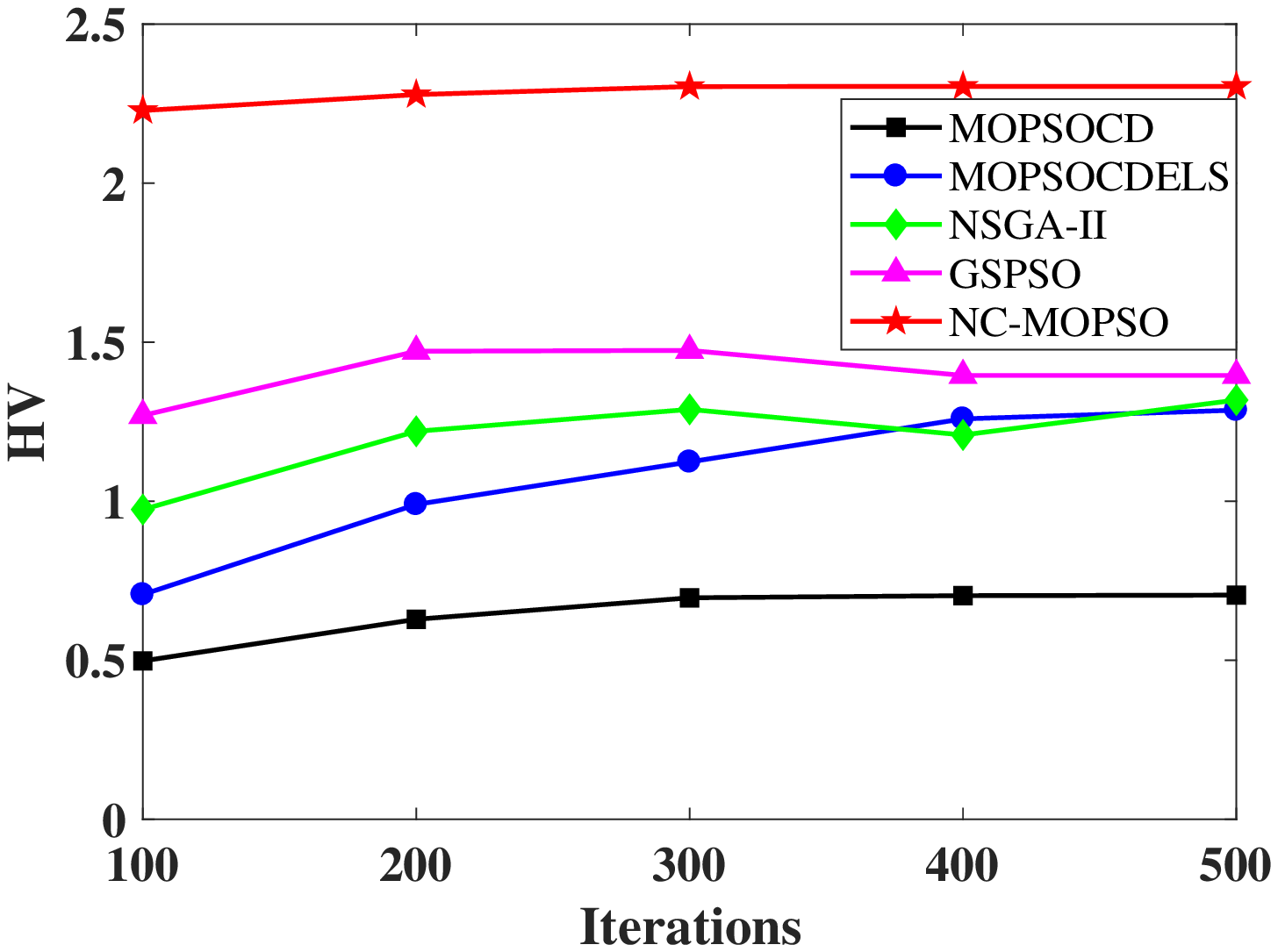}}
	\subfigure[BA300]{
		\label{fig:subfig:3b} 
		\includegraphics[width=4cm,height=3cm]{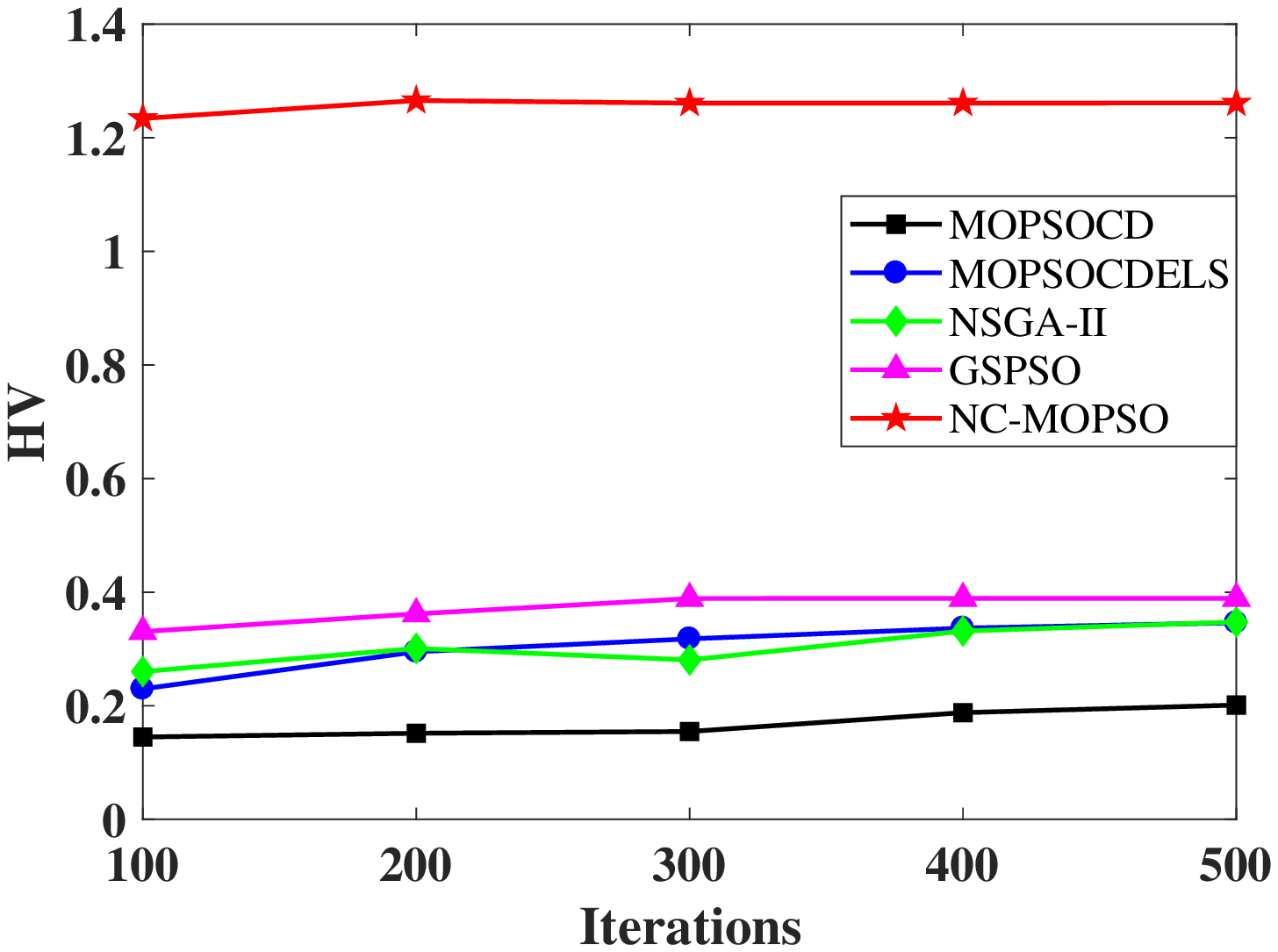}}\\
	\subfigure[WS100]{
		\label{fig:subfig:3c} 
		\includegraphics[width=4cm,height=3cm]{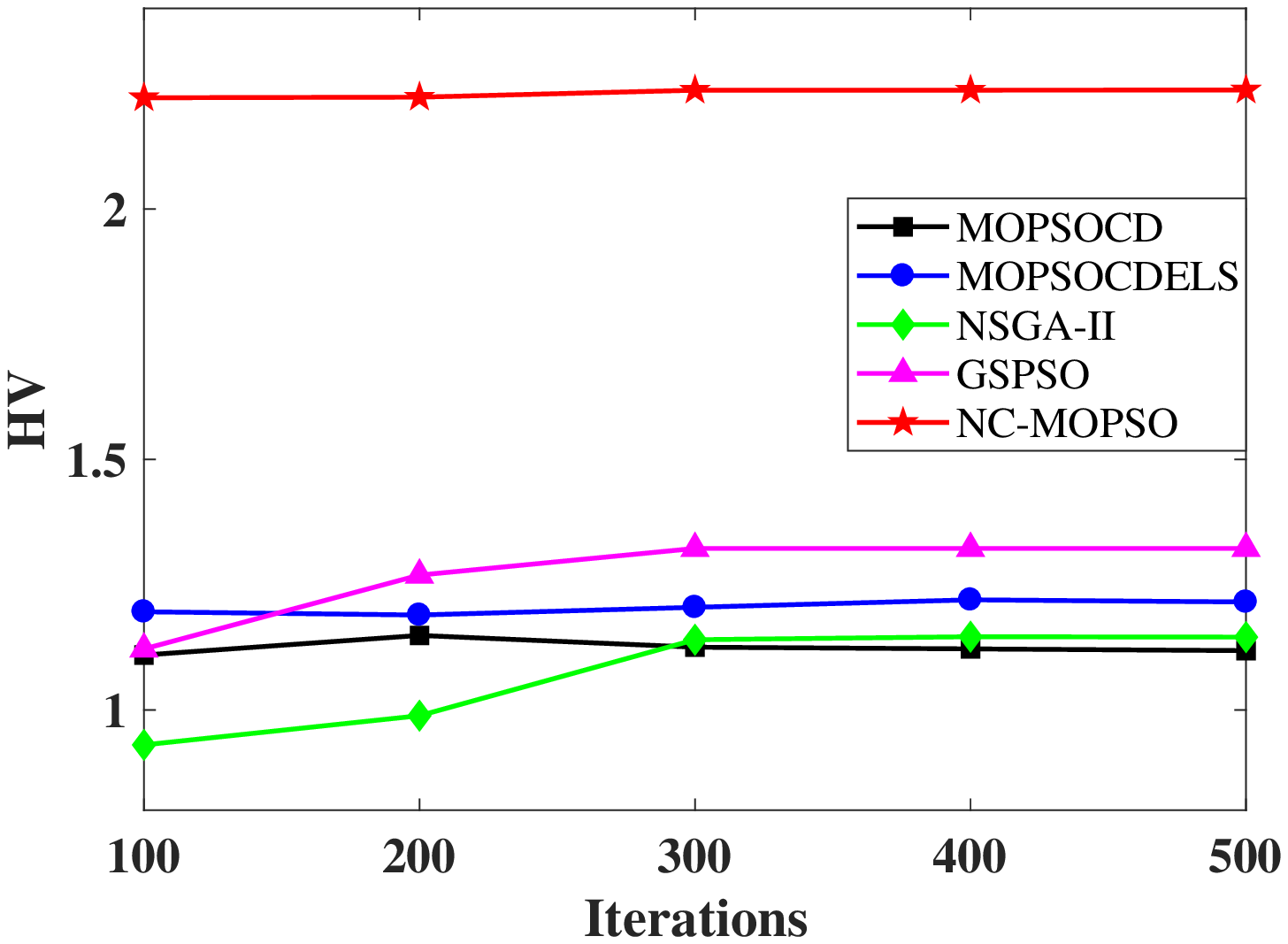}}
	\subfigure[WS300]{
		\label{fig:subfig:3d} 
		\includegraphics[width=4cm,height=3cm]{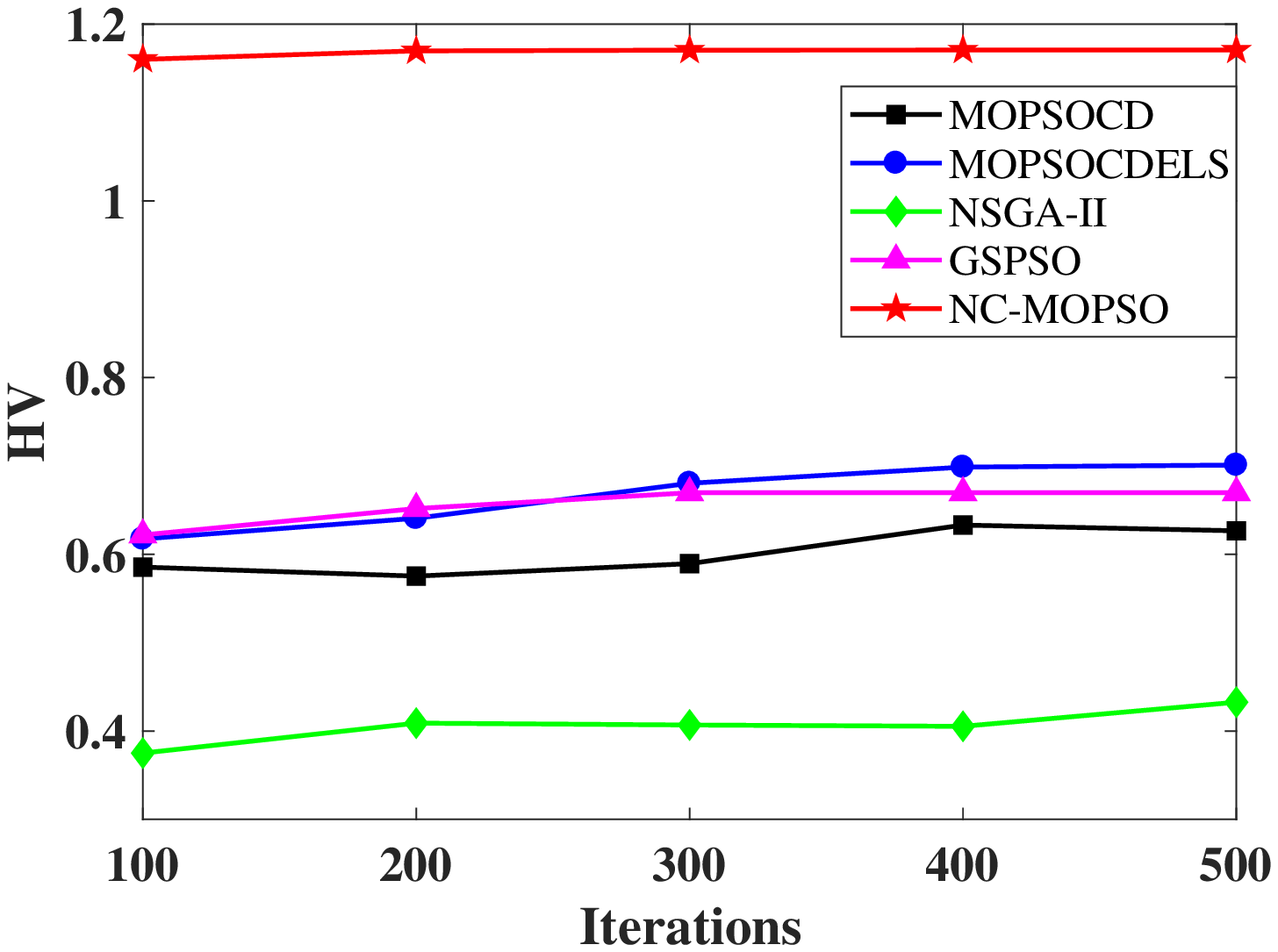}}\\
	\subfigure[118-bus]{
		\label{fig:subfig:3e} 
		\includegraphics[width=4cm,height=3cm]{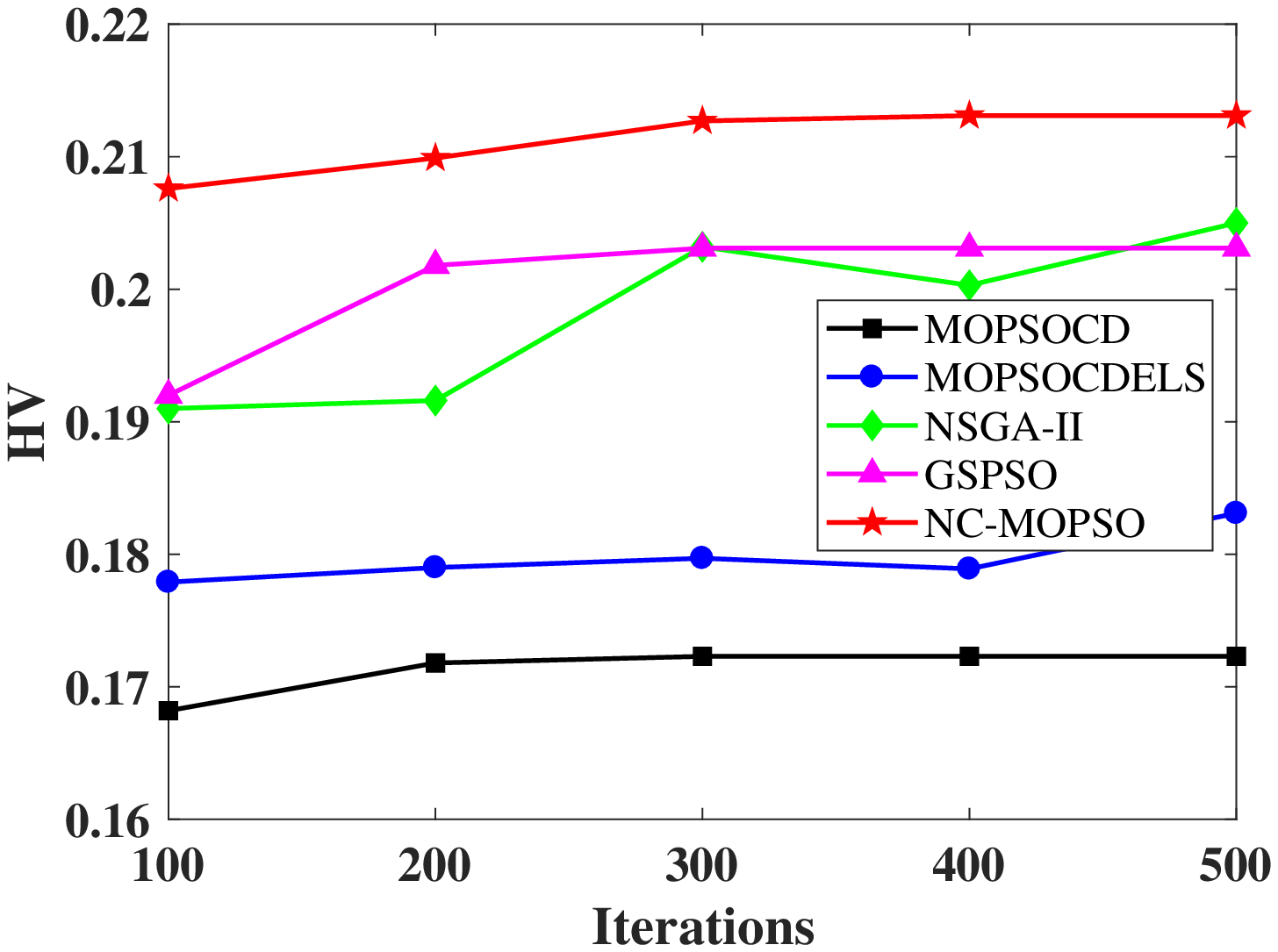}}
	\subfigure[email-enron-only]{
		\label{fig:subfig:3f} 
		\includegraphics[width=4cm,height=3cm]{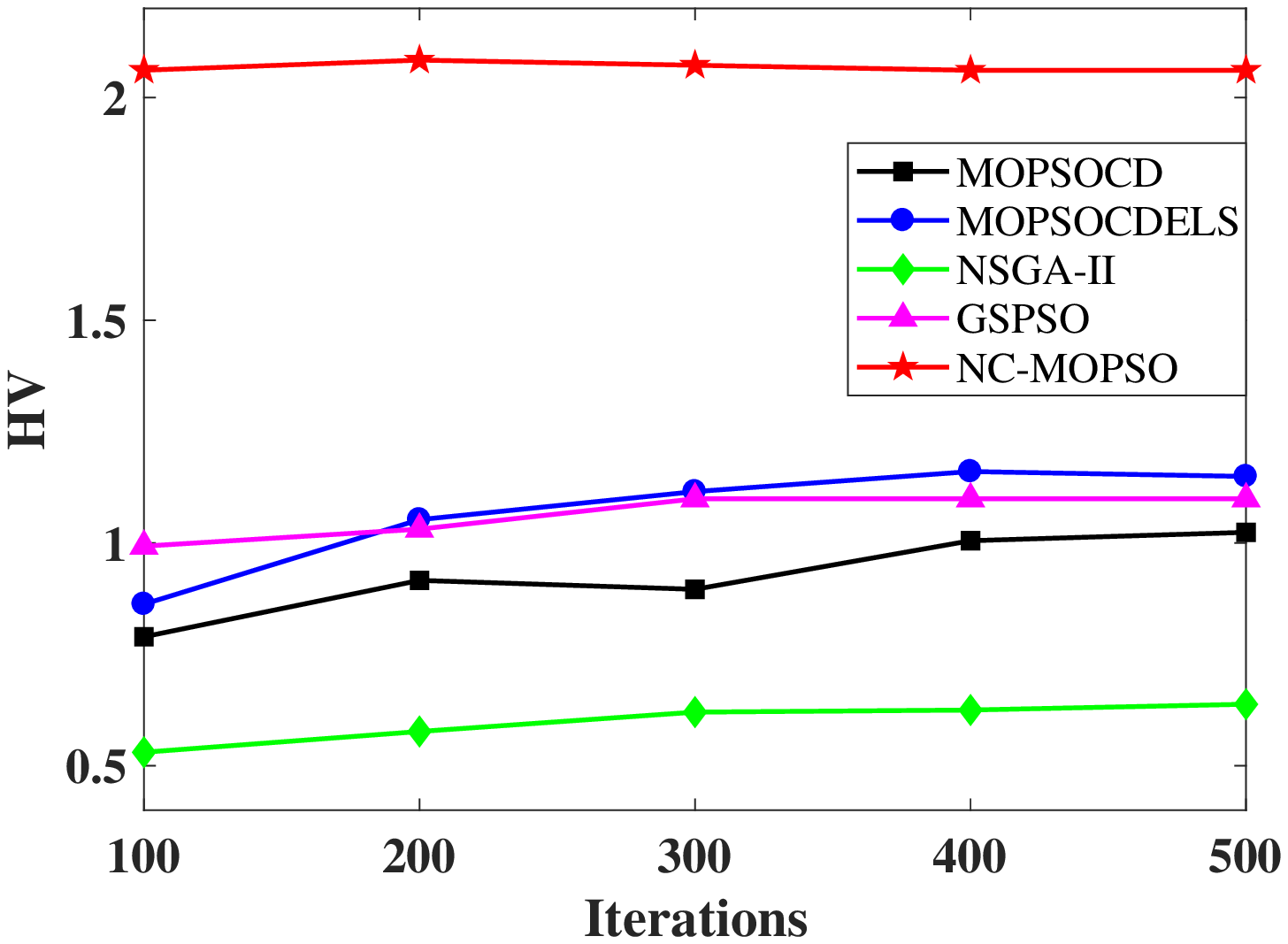}}\\
	\subfigure[chesapeake]{
		\label{fig:subfig:3g} 
		\includegraphics[width=4cm,height=3cm]{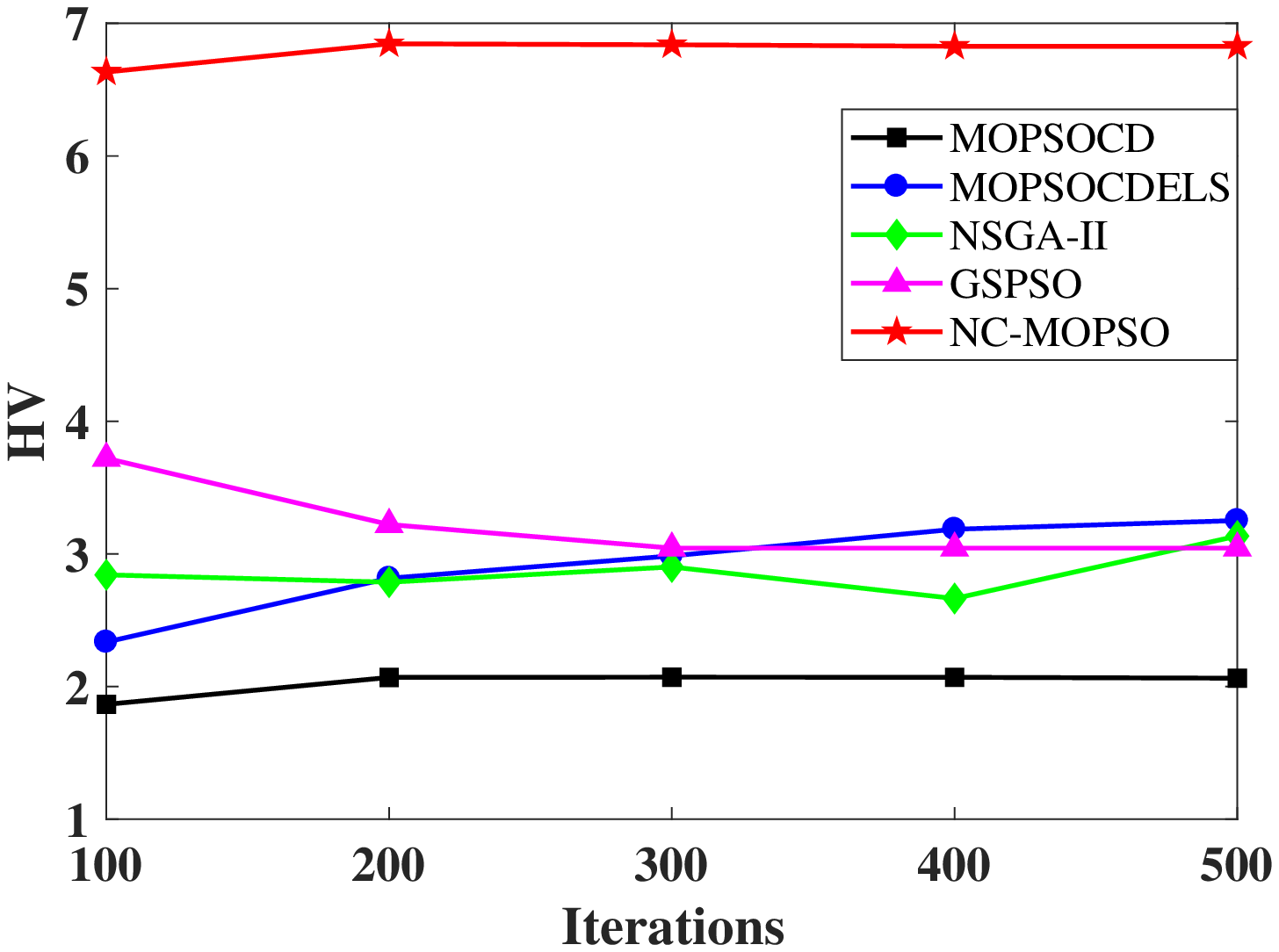}}
	\subfigure[uninett]{
		\label{fig:subfig:3h} 
		\includegraphics[width=4cm,height=3cm]{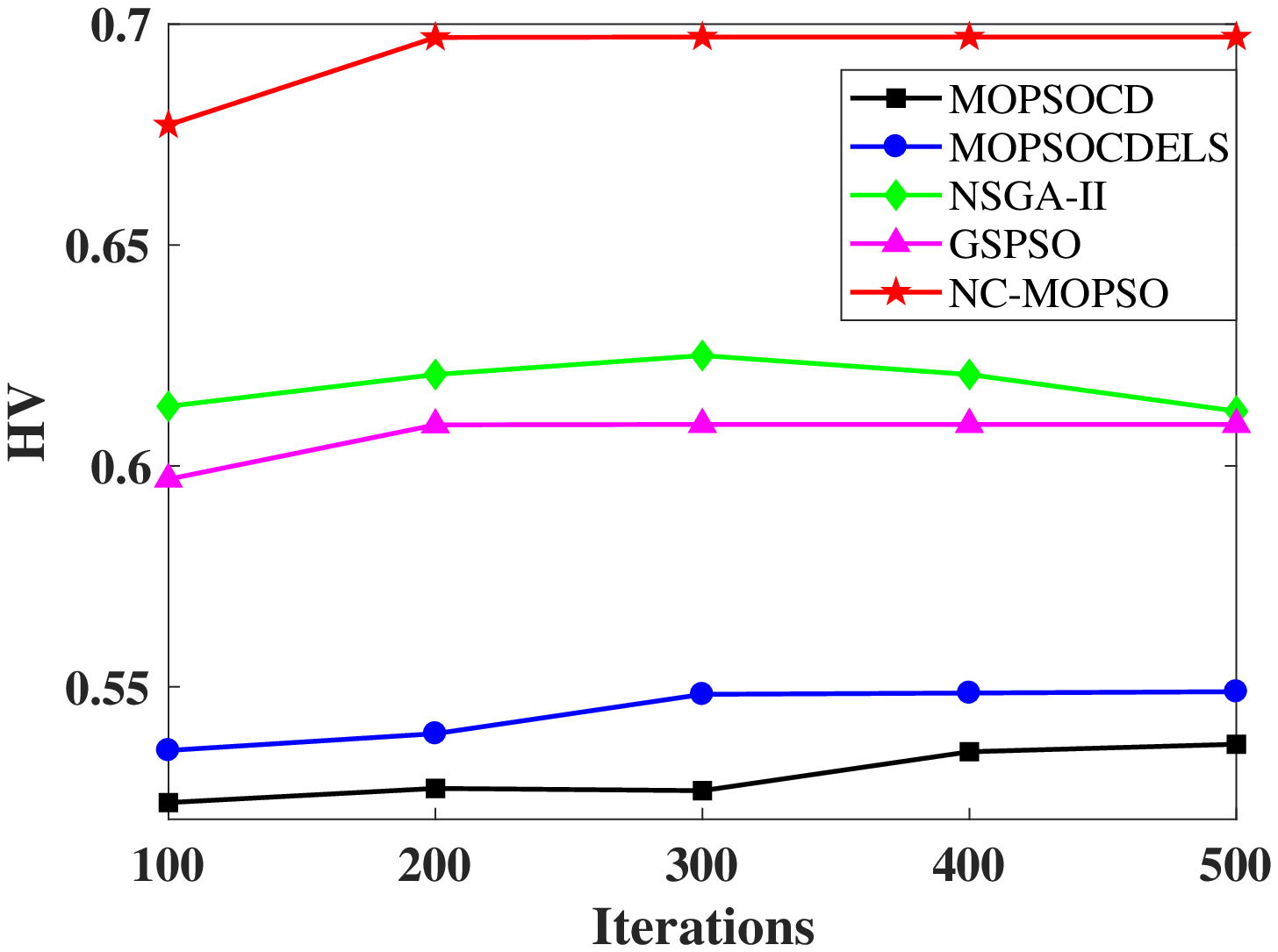}}\\
	\caption{Convergence analysis for all the algorithms on each instance.}
	\label{fig:fig3} 
\end{figure}
\indent We note that smaller values are  better for IGD and C-metric, while the opposite is true for HV. According to the results in Tables \ref{four}-\ref{overall} and Figs.~\ref{fig:igd}-\ref{fig:hv}, we can see that on almost all test instances, NC-MOPSO is superior to the other four competitors. Specifically, in terms of IGD (see Table \ref{four} and Fig.~\ref{fig:igd}), NC-MOPSO has the smallest value on all test instances, which means it has the best performance among all algorithms. The poor IGD results of GSPSO and NSGA-\uppercase\expandafter{\romannumeral2} indicate that their solutions have small diversity, though they have good convergence. For C-metric (see Fig.~\ref{fig:c}), NC-MOPSO has a value smaller than one for all test instances, which means part of the solutions obtained by NC-MOPSO cannot be dominated by the true PF. GSPSO performs relatively poor since the corresponding C-metric values are always one for all instances, which means all the solutions obtained by it will be dominated by the true PF. With regard to the HV metric (see Table \ref{five} and Fig.~\ref{fig:hv}), NC-MOPSO is also superior to the other four algorithms. Note that for the 118-bus network, NSGA-\uppercase\expandafter{\romannumeral2} and NC-MOPSO have close HV values.\\
\indent The mean ranking and the total number of $\dagger$$/$$\S$$/$$\approx$ (Table \ref{overall})  show that NC-MOPSO outperforms all the other comparison algorithms. Moreover, we present the nondominated solutions (PFs)  of all algorithms for each instance  graphically in Fig.~\ref{fig:fig1} to highlight the advantage of NC-MOPSO. We can clearly see from the distribution of the nondominated solutions that NC-MOPSO obtains higher quality solutions than the others.\\
\indent Finally, we analyze the convergence of all MOEAs, and the results  are presented in Fig.~\ref{fig:fig3}, where the HV metric is used as the indicator. It can be seen that all MOEAs generally have good convergence, some of which have occasionally  slight fluctuations due to local minima problem. Overall, the comparative experiments prove that NC-MOPSO can not only obtain high-quality PFs, but also have good convergence  property. The good performance of NC-MOPSO is mainly owing to the following two facts:
\begin{enumerate}
	\item An edge-centrality guided  hybrid initialization, ECHI,  is designed to generate high-quality initial solutions.
	\item A node-centrality guided local search, NCLS,  is adopted to expand the search space and improve the quality of current solutions in each iteration.
\end{enumerate}

\section{Conclusion}
In summary, we formulate a dual optimization problem, which is applicable to a wide range of transport processes in real-word networks. In this problem, we aim at exploring the best compromise of enhancing network capacity and reducing average number of hops. In particular, we propose a network centrality guided MOPSO, i.e., NC-MOPSO, to solve this problem. In this algorithm, an edge-centrality guided hybrid initialization, ECHI,  is proposed to provide high-quality initial solutions, and a node-centrality guided local search, NCLS, is developed to enhance the exploration of search space. We conduct extensive experiments to demonstrate the performance of NC-MOPSO. The results indicate that NC-MOPSO outperforms several state-of-the-art MOEAs in network transport optimization. We believe that NC-MOPSO  can be further improved with additional problem-specific knowledge.


%






\ifCLASSOPTIONcaptionsoff
  \newpage
\fi



%

%
\begin{IEEEbiography}[{\includegraphics[width=1in,height=1.25in,clip,keepaspectratio]{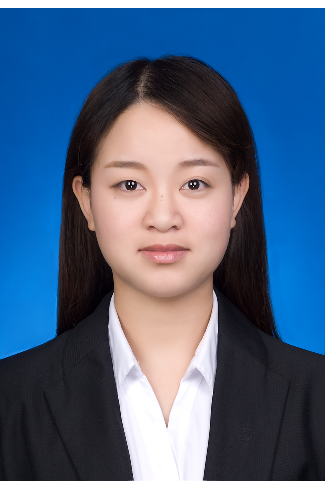}}]{Jiexin Wu}
	
	received the B.E. degree in network engineering from Nanjing University of Science and Technology, Nanjing, China, in 2016. She is currently pursuing the Ph.D. degree in computer science and technology at the School of Computer Science and Engineering, Nanjing University of Science and Technology, Nanjing, China.
	
	Her research interests include network optimization and network science.
	\vspace{-.5cm}
\end{IEEEbiography}
\begin{IEEEbiography}[{\includegraphics[width=1in,height=1.25in,clip,keepaspectratio]{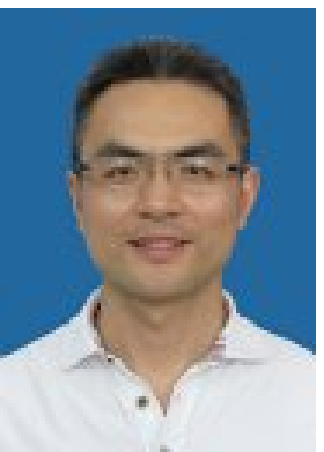}}]{Cunlai Pu}
	received the Ph.D. degree in Information and Communication Engineering from Southeast University, Nanjing, China, in 2012.
	
	He is currently an Associate Professor with the School of Computer Science and Engineering, Nanjing University of Science and Technology, Nanjing, China. His current research interests include network science, communication systems and network optimization.
	\vspace{-.5cm}
\end{IEEEbiography}
\begin{IEEEbiography}[{\includegraphics[width=1in,height=1.25in,clip,keepaspectratio]{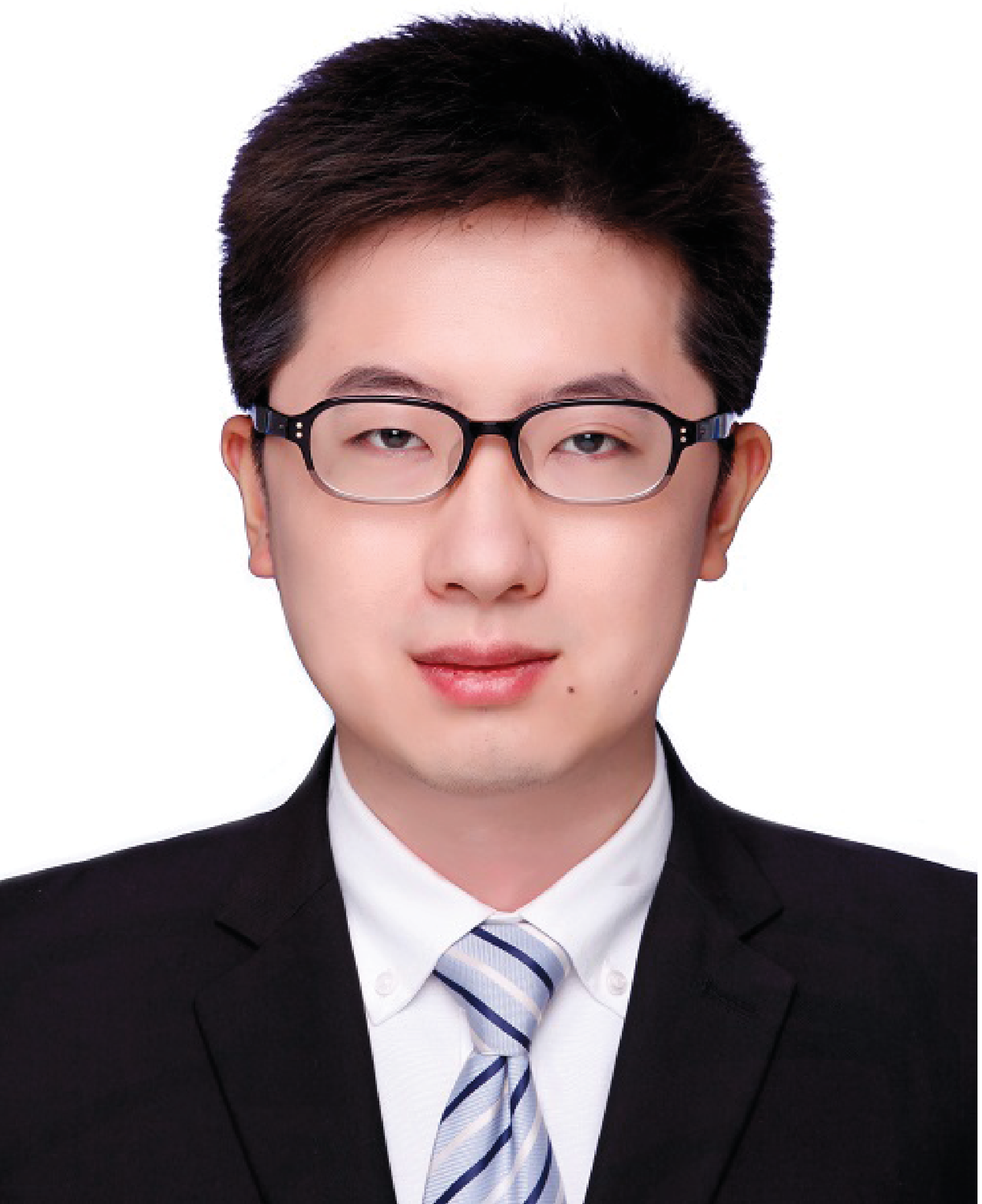}}]{Shuxin Ding}
	received the B.E. degree in automation and the Ph.D. degree in control science and engineering from the Beijing Institute of Technology, Beijing, China, in 2012 and 2019, respectively.
	
	From 2016 to 2017, he was a Visiting Scholar of Industrial and Systems Engineering at the University of Florida, Gainesville, FL, USA. He is currently an Assistant Researcher with the Signal and Communication Research Institute, China Academy of Railway Sciences Corporation Limited. His current research interests include railway scheduling, evolutionary computation, multiobjective optimization, and optimization under uncertainty.
	\vspace{-.5cm}
\end{IEEEbiography}
\begin{IEEEbiography}[{\includegraphics[width=1in,height=1.25in,clip,keepaspectratio]{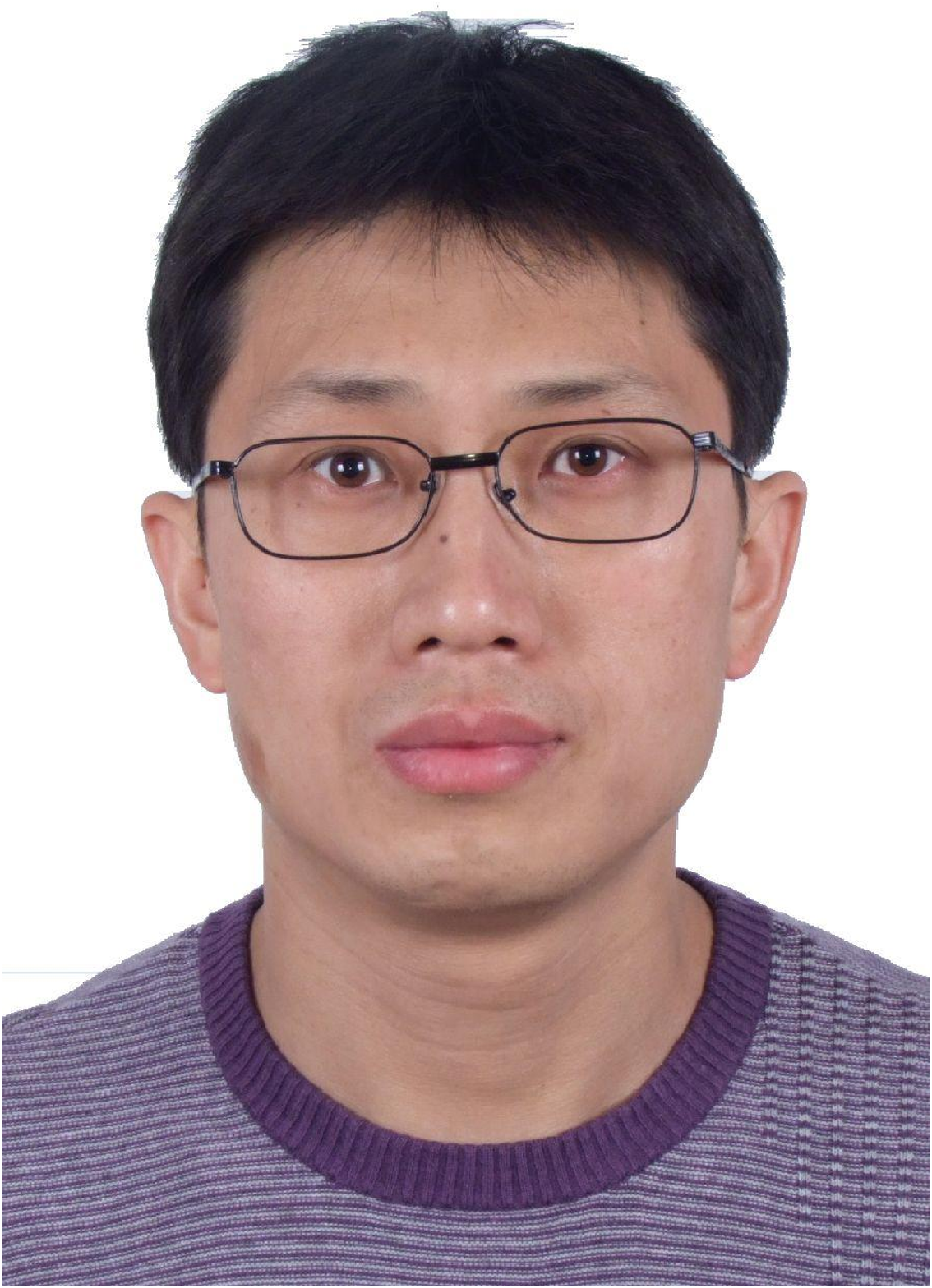}}]{Guo Cao}
	received the Ph.D. degree in pattern recognition and intelligent system from Shanghai Jiaotong University, Shanghai, China, in 2006.
	
	He is a Professor with the School of Computer Science and Engineering, Nanjing University of Science and Technology, Nanjing, China. His current research interests include pattern recognition, machine learning and intelligent systems.
	\vspace{-.5cm}
\end{IEEEbiography}
\begin{IEEEbiography}[{\includegraphics[width=1in,height=1.25in,clip,keepaspectratio]{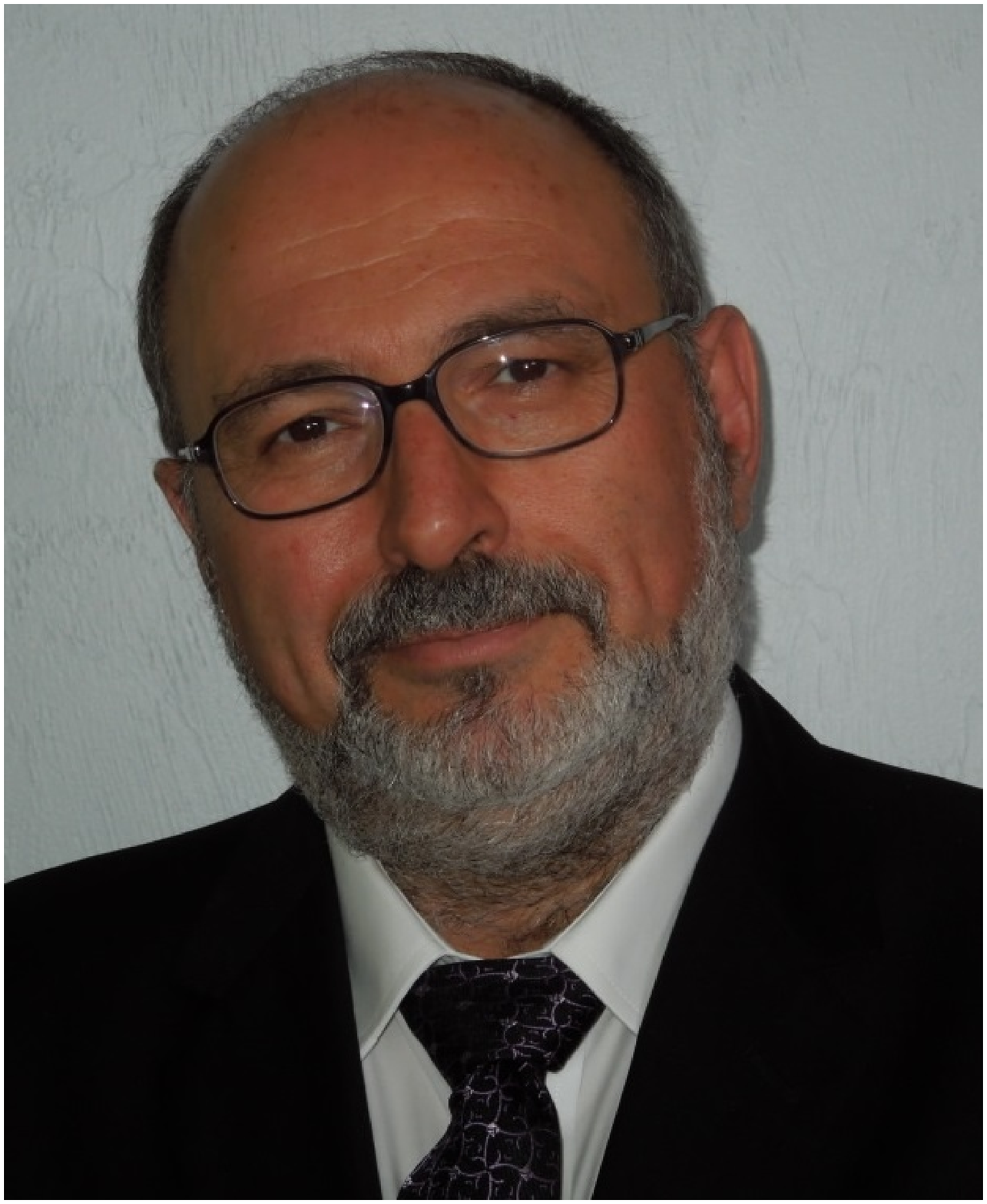}}]{Panos M. Pardalos}
	received the B.S. degree in mathematics from the Athens University, Athens, Greece, in 1977, the M.S. degree in mathematics and computer science from the Clarkson University, Potsdam, NY, USA, in 1978, and the Ph.D. degree in computer and information sciences from the University of Minnesota, Minneapolis, MN, USA, in 1985.
	
	He is a Distinguished Professor and the Paul and Heidi Brown Preeminent Professor of Industrial and Systems Engineering at the University of Florida, and a World Renowned Leader in Global Optimization, Mathematical Modeling, and Data Sciences. He has published over 500 papers, edited/authored more than 200 books, and organized over 80 conferences.
	
	Prof. Pardalos was a recipient of the 2018 Humboldt Research Award, the 2013 Constantin Caratheodory Prize of the International Society of Global Optimization, and the 2013 EURO Gold Medal Prize bestowed by the Association for European Operational Research Societies. He is the Founding Editor of Optimization Letters and Energy Systems, and the Co-Founder of Journal of Global Optimization. He is also a Foreign Member of the Lithuanian Academy of Sciences, the Royal Academy of Spain, and the National Academy of Sciences of Ukraine. He is a fellow of AAAS, AIMBE, and INFORMS.
\end{IEEEbiography}

\end{document}